
\input harvmac
\catcode`\:=11
\def\latex:adjust{\expandafter\ifx\the\textfont0\csname twlrm\endcsname
                                           \def\bm:scale{1200}%
                              \else\expandafter\ifx
                                 \the\textfont0\csname elvrm\endcsname
                                                 \def\bm:scale{1095}%
                                               \else\def\bm:scale{1000}%
                                               \fi
                              \fi}%
\latex:adjust
\font\tenbf=cmbx10 scaled \bm:scale
\font\sevenbf=cmbx7 scaled \bm:scale
\font\fivebf=cmbx5 scaled \bm:scale
\textfont\bffam=\tenbf
\scriptfont\bffam=\sevenbf
\scriptscriptfont\bffam=\fivebf
\font\tenbmit=cmmib10 scaled \bm:scale
\font\sevenbmit=cmmib7 scaled \bm:scale
\font\fivebmit=cmmib5 scaled \bm:scale
\newfam\bmitfam
\def\bmit{\fam\bmitfam}
\textfont\bmitfam=\tenbmit
\scriptfont\bmitfam=\sevenbmit
\scriptscriptfont\bmitfam=\fivebmit
\font\tenbsy=cmbsy10 scaled \bm:scale
\font\sevenbsy=cmbsy7 scaled \bm:scale
\font\fivebsy=cmbsy5 scaled \bm:scale
\newfam\bsyfam

\textfont\bsyfam=\tenbsy
\scriptfont\bsyfam=\sevenbsy
\scriptscriptfont\bsyfam=\fivebsy
\let\bm=\bmit  \let\bcal=\bsy
\newcount\bm:counta \newcount\bm:countb 
\newcount\bm:countc \newcount\bm:countd
\newtoks\bm:tok \newtoks\bm:pretok
\newif\ifbm:delim
\def\thehex#1{\ifcase\the#1 0\or 1\or 2\or 3\or 4\or 5\or 6\or 7\or
8\or 9\or A \or B\or C\or D\or E\or F\fi}%
\def\test#1#2{\ifcat#1#2\message{True}\else\message{False}\fi}
\newif\ifboldwarning
\def\bm:message#1{{\newlinechar=`^^J \immediate\write16{#1^^J}}}
\def\bold#1{{%
             \ifx\mit#1 \aftergroup\bmit           
             \else\ifx\cal#1 \aftergroup\bcal      
                  \else\ifx\{#1 \bmlbrace\else\ifx\}#1 \bmrbrace\else
                       \ifx\left#1 \aftergroup\boldchar\aftergroup\left%
                       \else\ifx\right#1
                                \aftergroup\boldchar\aftergroup\right%
                            \else\ifcat\noexpand#1A\boldletter{#1}%
                                 \else\ifcat\noexpand#1>\aftergroup\boldchar
                                           \aftergroup\relax\aftergroup#1%
                                       \else\ifcat#1\alpha
                       \aftergroup\boldchar\aftergroup\relax\aftergroup#1%
                                                \else
                                        \aftergroup\boldchar\aftergroup#1%
                                                \fi
                                       \fi
                                  \fi
                              \fi
                        \fi\fi\fi
                   \fi
            \fi}}
%
%
\def\boldit#1{{%
               \ifx\mit#1 \aftergroup\bmit      
               \else\ifx\cal#1 \aftergroup\bcal 
                    \else\ifcat #1\alpha%
                                \ifnum\the#1>"6FFF
                                      \aftergroup{\aftergroup\bmit
                                             \aftergroup#1\aftergroup}%
                                \else\aftergroup\bold\aftergroup#1%
                                \fi
                          \else\ifcat\noexpand#1A\aftergroup{\aftergroup\bmit
                                             \aftergroup#1\aftergroup}%
                               \else\aftergroup\bold\aftergroup#1%
                               \fi
                          \fi
                    \fi
               \fi}}%
%
%
\def\boldletter#1{{\bf #1}}
%
%
\def\boldchar#1#2{
\begingroup
\bm:pretok={#1}\bm:tok={#2}%
\ifx#1\relax\relax \bm:delimfalse  
    \expandafter\ifx\csname bold\string#2\endcsname\relax
                    \bold:mathchardef{#2}%
                \fi
    \csname bold\string#2\endcsname
\endgroup        
\else\bm:delimtrue
     \expandafter\ifx\csname bold:delim\string#2\endcsname\relax
                 \bold:mathchardef{#2}%
                 \fi
     \aftergroup#1\expandafter\aftergroup\csname
                                         bold:delim\string#2\endcsname
\endgroup        
     \bm:delimfalse
\fi               }%
\def\bold:mathchardef#1{
\ifcat#1\alpha         
      \bm:countd=-1 \bm:counta=\the#1
\else \bm:counta=\the\mathcode\expandafter`#1
      \bm:countd=\the\delcode\expandafter`#1
\fi
\ifnum\the\bm:counta="8000
           \expandafter\xdef\csname bold\string#1\endcsname{#1}%
\else 
\bm:countc=\bm:counta
\divide\bm:countc by "1000
\advance\bm:counta by -\expandafter"\thehex\bm:countc 000
\ifbm:delim \ifnum\the\bm:countd>-1     
                       \begingroup \bm:counta=\bm:countd
                       \divide\bm:counta by "1000
                            \begingroup
                            \multiply\bm:counta by "1000
                            \global \advance\bm:countd by
                                                      -\the\bm:counta
                            \endgroup
                       \bold:mathrecode
                       \multiply\bm:counta by "1000
                            \begingroup
                            \bm:counta=\bm:countd
                            \bold:mathrecode
                            \global\bm:countd=\bm:counta
                            \endgroup
                       \global\advance\bm:countd by \the\bm:counta
                       \endgroup
              \else    \bm:message{Did you mean
                                      \the\bm:pretok\string\bold\space\the\bm:tok\space on line \the\inputlineno?}%
                       \bold:mathrecode
                       \bm:countd=\bm:counta
                       \multiply\bm:counta by "1000
                       \advance\bm:countd by \the\bm:counta
              \fi
              \advance\bm:countd by \expandafter"\thehex\bm:countc000000
              \expandafter\xdef\csname bold:delim\string#1\endcsname
                               {\delimiter\the\bm:countd}%
              \xdef\bm:lastcode{\the\bm:countd}%
\else         \bold:mathrecode
              \advance\bm:counta by \expandafter"\thehex\bm:countc 000
              \expandafter\global\expandafter\mathchardef\csname
                                bold\string#1\endcsname=\the\bm:counta
              \xdef\bm:lastcode{\the\bm:counta}%
\fi
\fi                     }
\def\bold:mathrecode{
      \bm:countb=\bm:counta
      \divide\bm:countb by "100
      \advance\bm:counta by -\expandafter"\thehex\bm:countb 00
      \ifcase\the\bm:countb
              \bm:countb=\the\bffam
      \or     \bm:countb=\the\bmitfam
      \or     \bm:countb=\the\bsyfam
      \else   \ifboldwarning \bm:message{\string\bold\space warning on
                                                   line \the\inputlineno!}
                             \ifbm:delim\bm:message{Lack of bold
extension fonts means \string\bold\the\bm:pretok\the\bm:tok\space may
not be bold.}%
                             \else\bm:message{Sorry, there just aren't the
                                   fonts for \string\bold\the\bm:tok.}%
                             \fi
              \fi 
      \fi
\advance\bm:counta by \expandafter"\thehex\bm:countb 00
                    }
\def\defboldmacro#1#2#3{%
{\bm:counta=#3 \bm:countc=#3 \bm:countd=#3
\divide\bm:counta by "1000 \divide\bm:countc by "1000000
\advance\bm:countd by -\expandafter"\thehex\bm:countc 000000
\ifnum\the\bm:counta<"10 \bm:counta=\bm:countd \bm:countd=-1 
      \global\bm:delimfalse
      \else \global\bm:delimtrue
\fi
\mathcode`?=\the\bm:counta \delcode`?=\the\bm:countd
\bm:tok={#2}
\bold:mathchardef{?}%
\expandafter\xdef\csname bold:delim?\endcsname{\relax}
\xdef#1{#2\bm:lastcode}%
}%
                        }%
\defboldmacro{\bmlbrace}{\delimiter}{"4266308}
\defboldmacro{\bmrbrace}{\delimiter}{"5267309}
\let\int=\intop
\let\oint=\ointop
\boldwarningtrue
\catcode`\:=12

CHANGES MADE SINCE 14/10/94

17/10/94

1) \defboldmacro modified to reset \bold:delim? to \relax after each use.

2) \ifboldwarning introduced to allow suppression of warnings during
   input of boldmath.tex and at discretion of user.

26/10/94

3) Warnings improved to give input line number.

1/2/95

4) \latex:adjust introduced for compatibility with LaTeX2.09 font
   selection at \magstep0.

\input amssym.tex
\input amssym
\baselineskip 14pt 
\magnification\magstep1 
\font\bigcmsy=cmsy10 scaled 1500
\parskip 6pt
\newdimen\itemindent \itemindent=32pt
\def\textindent#1{\parindent=\itemindent\let\par=\resetpar%
\indent\llap{#1\enspace}\ignorespaces}

\let\oldpar=\par
\def\resetpar{\oldpar\parindent=20pt\let\par=\oldpar}

\font\ninerm=cmr9 \font\ninesy=cmsy9 \font\eightrm=cmr8
\font\sixrm=cmr6 \font\eighti=cmmi8 \font\sixi=cmmi6
\font\eightsy=cmsy8 \font\sixsy=cmsy6 \font\eightbf=cmbx8
\font\sixbf=cmbx6 \font\eightit=cmti8
\def\eightpoint{\def\rm{\fam0\eightrm}
  \textfont0=\eightrm \scriptfont0=\sixrm \scriptscriptfont0=\fiverm
  \textfont1=\eighti  \scriptfont1=\sixi  \scriptscriptfont1=\fivei
  \textfont2=\eightsy \scriptfont2=\sixsy \scriptscriptfont2=\fivesy
  \textfont3=\tenex   \scriptfont3=\tenex \scriptscriptfont3=\tenex
  \textfont\itfam=\eightit  \def\it{\fam\itfam\eightit}%
  \textfont\bffam=\eightbf  \scriptfont\bffam=\sixbf
  \scriptscriptfont\bffam=\fivebf  \def\bf{\fam\bffam\eightbf}%
  \normalbaselineskip=9pt
  \setbox\strutbox=\hbox{\vrule height7pt depth2pt width0pt}%
  \let\big=\eightbig  \normalbaselines\rm}
\catcode`@=11 %
\def\eightbig#1{{\hbox{$\textfont0=\ninerm\textfont2=\ninesy
  \left#1\vbox to6.5pt{}\right.\n@@space$}}}
\def\vfootnote#1{\insert\footins\bgroup\eightpoint
  \interlinepenalty=\interfootnotelinepenalty
  \splittopskip=\ht\strutbox %
  \splitmaxdepth=\dp\strutbox %
  \leftskip=0pt \rightskip=0pt \spaceskip=0pt \xspaceskip=0pt
  \textindent{#1}\footstrut\futurelet\next\fo@t}
\catcode`@=12 %

\def\ux{u}

\def \de{\delta}
\def \De{\Delta}
\def \si{\sigma}
\def \bsi{\bar \sigma}

\def \ga{\gamma}

\def \al{\alpha}
\def \be{\beta}
\def \pr{\partial}

\def \rO{{\rm O}}

\def \ep{\epsilon}
\def \vep{\varepsilon}
\def \half{{\textstyle {1 \over 2}}}
\def \thir{{\textstyle {1 \over 3}}}
\def \quar{{\textstyle {1 \over 4}}}
\def \eight{{\textstyle {1 \over 8}}}
\def \dps{\displaystyle}
\def \ts{\textstyle}
\def \ss{\scriptstyle}
\def \rC{{\cal R}}
\def \A{{\cal A}}
\def \B{{\cal B}}
\def \C{{\cal C}}
\def \D{{\cal D}}

\def \F{{\cal F}}
\def \G{{\cal G}}
\def \H{{\cal H}}
\def \I{{\cal I}}

\def \M{{\cal M}}
\def \N{{\cal N}}
\def \O{{\cal O}}
\def \P{{\cal P}}
\def \Q{{\cal Q}}
\def \S{{\cal S}}
\def \R{{\cal R}}

\def \W{{\cal W}}
\def \V{{\cal V}}
\def \Z{{\cal Z}}

\def \ba{\bar a}
\def \bh{\bar h}
\def \by{\bar x}

\def \bJ{{\bar J}}
\def \bQ{\bar Q}
\def \bS{\bar S}
\def \vphi{{\varphi}}
\def \hD{{\hat D}}
\def \hf{{\hat f}}
\def \hg{{\hat g}}
\def \hh{{\hat h}}
\def \hs{{\hat s}}
\def \hu{{\hat u}}
\def \hchi{{\hat \chi}}

\def \bj{\bar \jmath}
\def \al{\alpha}

\def \uv{{\underline{\rm v}}}

\font \bigbf=cmbx10 scaled \magstep1

\def \Bsw{\!\mathrel{\hbox{\bigcmsy\char'056}}\!}
\def \Bse{\!\mathrel{\hbox{\bigcmsy\char'046}}\!}

\def \leq{\leqslant}
\def \geq{\geqslant}

\def \dal{{\dot \alpha}}
\def \dbe{{\dot \beta}}
\def \dga{{\dot \gamma}}
\def \dde{{\dot \delta}}
\def \uu{{\rm u}}
\def \oD{\overline{\D}}

\def \ua{\underline\alpha}

\def \uH{\underline H}
\def \rH{{\sl H}}
\def \bfe{{\rm e}}
\def \uL{\underline \Lambda}
\def \ul{\underline \lambda}
\def \uv{\underline v}
\def \uw{\underline w}
\def \urho{\underline \rho}
\def \ll{\underline \lambda}
\def \uxx{\underline x}
\def \ttr{\widetilde{\tr}}
\def \lam{\lambda}
\def \blam{{\bar\lambda}}
\def \bl{\bold \ell}
\def \brho{\bold \rho}
\def \bt{\bar t}

\lref\malda{J. Maldacena, {\it The Large $N$ Limit of
Superconformal Field Theories and Supergravity},
Adv. Theor. Math. Phys. 2 (1998) 231, hep-th/9711200.}

\lref\fadho{F.A. Dolan and H. Osborn, {\it On short and semi-short
representations for four-dimensional superconformal symmetry},
Ann. Phys. 307 (2003) 41.}

\lref\BMStao{
 M.~Bianchi, J.F.~Morales and H.~Samtleben,
{\it On stringy $AdS_5 \times S^5$ and higher spin holography},
JHEP 0307 (2003) 062, hep-th/0305052.}

\lref\BeisertTE{
N.~Beisert, M.~Bianchi, J.F.~Morales and H.~Samtleben,
{\it On the spectrum of AdS/CFT beyond supergravity},
JHEP {0402} (2004) 001, hep-th/0310292.}

\lref\beis{N.~Beisert, M.~Bianchi, J.F.~Morales and H.~Samtleben,
{\it Higher spin symmetry and $N = 4$ SYM},
  JHEP 0407 (2004) 058, hep-th/0405057.}

\lref\mald{
J.~Kinney, J.M.~Maldacena, S.~Minwalla and S.~Raju,
{\it An index for $4$ dimensional super conformal theories},
hep-th/0510251.}

\lref\fadl{
F.A.~Dolan, {\it Character formulae and partition functions in higher
dimensional conformal field theory}, 
J. Math. Phys. 47 (2006) 062303, hep-th/0508031.}

\lref\and{G.E. Andrews, R. Askey and R. Roy,
{\it Special Functions}, Cambridge University Press, Cambridge, 1999.}

\lref\dunkl{C.F. Dunkl and Y. Xu, `{\it Orthogonal polynomials of several
variables}', Encyclopedia of Mathematics and its Applications, Vol. 81,
Cambridge University Press, 2001.}

\lref\biswas{
I.~Biswas, D.~Gaiotto, S.~Lahiri and S.~Minwalla,
{\it Supersymmetric states of $N=4$ Yang-Mills from giant gravitons},
hep-th/0606087.
}

\lref\Chiral{F. Cachazo, M.R. Douglas, N. Seiberg and E. Witten, {\it
Chiral Rings and Anomalies in Supersymmetric Gauge Theory}, JHEP 0212 (2002) 071,
hep-th/0211170.}

\lref\opemix{
M.~Bianchi, B.~Eden, G.~Rossi and Y.S.~Stanev,
{\it On operator mixing in $N = 4$ SYM}, Nucl.\ Phys.\ B646 (2002) 69,
hep-th/0205321.
}

\lref\Moraletal{
S.~Bellucci, P.Y.~Casteill, J.F.~Morales and C.~Sochichiu,
{\it Spin bit models from non-planar $N = 4$ SYM},
Nucl.\ Phys.\ B699 (2004) 151, hep-th/0404066.
}

\lref\HesH{
P.J.~Heslop and P.S.~Howe,
{\it A note on composite operators in $N = 4$ SYM},
Phys.\ Lett.\ B516 (2001) 367, hep-th/0106238.
}

\lref\Aspects{
P.J.~Heslop and P.S.~Howe,
{\it Aspects of $N = 4$ SYM}, JHEP 0401 (2004) 058, hep-th/0307210.
}

\lref\sok{
B.~Eden and E.~Sokatchev,
{\it On the OPE of $1/2$ BPS short operators in $N = 4$ SCFT($4$)},
Nucl.\ Phys.\ B618 (2001) 259, hep-th/0106249.
}

\lref\Dobrev{
V.K.~Dobrev, {\it Characters of the positive energy UIRs of $D = 4$
conformal  supersymmetry}, hep-th/0406154.
}

\lref\Dob{V.K. Dobrev and V.B. Petkova, {\it All positive energy unitary
irreducible representations of extended conformal symmetry}, Phys. Lett.
B162 (1985) 127.}

\lref\har{
T. Harmark and M. Orselli, {\it Quantum mechanical sectors
in thermal $\N=4$ super Yang Mills on $\Bbb {R}\times S^3$},
hep-th/0605234.}

\lref\pendant{M. Spradlin and A. Volovich, {\it A pendant for 
P\/olya: The One-Loop Partition Function of $\N=4$ SYM on $\Bbb {R} \times S^3$},
Nucl. Phys. B711 (2005) 199, hep-th/0408178.}

\lref\BRSta{ M.~Bianchi, G.~Rossi and Y.S.~Stanev,
{\it Surprises from the resolution of operator mixing in $N = 4$ SYM},
Nucl.\ Phys.\ B685 (2004) 65, hep-th/0312228.}

\lref\BHR{M.~Bianchi, P.J.~Heslop and F.~Riccioni,
{\it More on la grande bouffe},
  JHEP 0508 (2005) 088,  hep-th/0504156.}

\lref\Bianchi{
M.~Bianchi, {\it Higher spins and stringy $AdS_5 \times  S^5$},
Fortsch.\ Phys.\  53 (2005) 665, hep-th/0409304.}

\lref\beist{N.~Beisert, {\it Spin chain for quantum strings},
  Fortsch.\ Phys.\  53 (2005) 852, hep-th/0409054.}

\lref\kaza{
V.A.~Kazakov, A.~Marshakov, J.A.~Minahan and K.~Zarembo,
  {\it Classical / quantum integrability in AdS/CFT},
  JHEP 0405 (2004) 024 hep-th/0402207.}

\lref\Doker{
E.~D'Hoker, P.~Heslop, P.~Howe and A.V.~Ryzhov,
{\it Systematics of quarter BPS operators in $N = 4$ SYM},
JHEP 0304 (2003) 038, hep-th/0301104.}

\lref\Morales{
J.F.  Morales, {private communication}.
}

\lref\howe{
A.~Galperin, E.~Ivanov, S.~Kalitsyn, V.~Ogievetsky and E.~Sokatchev,
{\it Unconstrained $N=2$ matter, Yang-Mills and supergravity theories in
harmonic  superspace},
Class.\ Quant.\ Grav.\  {1} (1984) 469\semi
P.S.~Howe and G.G.~Hartwell,
{\it A superspace survey}, Class.\ Quant.\ Grav.\  {12} (1995) 1823.}

\lref\unconstrained{
P.~Heslop and P.S.~Howe,
{\it Harmonic superspaces and superconformal fields},
hep-th/0009217\semi
P.J.~Heslop,
{\it Superfield representations of superconformal groups},
Class.\ Quant.\ Grav.\  {19} (2002) 303,
  hep-th/0108235.
}

\lref\fourpoints{
P.J.~Heslop and P.S.~Howe,
{\it Four-point functions in $N = 4$ SYM},
JHEP {0301} (2003) 043, hep-th/0211252.}

\lref\stau{
N.~Beisert, C.~Kristjansen and M.~Staudacher,
{\it The dilatation operator of $N = 4$ super Yang-Mills theory},
Nucl.\ Phys.\ B664 (2003) 131, hep-th/0303060.}

\lref\beiserto{
N.~Beisert,
{\it The complete one-loop dilatation operator of $N = 4$ super Yang-Mills theory},
Nucl.\ Phys.\ B676 (2004) 3, hep-th/0307015.}

\lref\RSpeis{J. Fuchs and C Schweigert, {\it Symmetries, Lie Algebras
and Representations}, Cambridge University Press, Cambridge, 1997.}

\lref\phd{F.A. Dolan, {\it Aspects of Superconformal Quantum Field Theory},
University of Cambridge PhD thesis (2003).}

\lref\NO{M. Nirschl and H. Osborn, {\it Superconformal Ward Identities and
their Solution}, Nucl. Phys. B711 (2005) 409, hep-th/0407060.}

\lref\Superchar{
J. Van der Jeugt, J.W.B. Hughes, R.C. King and J. Thierry-Mieg, {\it Character
formulas for irreducible modules of the Lie superalgebras $sl(m|n)$}, J. Math.
Phys. 31 (1990) 2278\semi
J. Van der Jeugt and R.B. Zhang, {\it Character and Composition Factor
Multiplicities for the Lie superalgebras $sl(m|n)$}, Lett. Math. Physics 47
(1999) 49\semi
Y. Su and R.B. Zhang, {\it Character and Dimension Formulae for General
Linear Superalgebra}, math.QA/0403315\semi
S-J. Cheng, W. Wang and R.B. Zhang, {\it A Super Duality and Kazhdan-Lustztig
Polynomials}, math.RT/0409016.}

\lref\Gwyn{R. de Mello Koch and R. Gwyn, {\it Giant Graviton Correlators from
Dual $SU(n)$ super Yang-Mills theory}, JHEP 0411 (2004) 081, hep-th/0410236.}

\lref\OPEN{G. Arutyunov, S. Frolov and A.C. Petkou, {\it Operator Product
Expansion of the Lowest Weight CPOs in $\N=4$ SYM${}_4$ at Strong Coupling},
Nucl. Phys. B586 (2000) 547, hep-th/0005182;
(E) Nucl. Phys. B609 (2001) 539\semi
G. Arutyunov, S. Frolov and A.C. Petkou, {\it Perturbative and
instanton corrections to the OPE of CPOs in $\N=4$ SYM${}_4$}, Nucl.
Phys. B602 (2001) 238, hep-th/0010137; (E) Nucl. Phys. B609 (2001) 540.}

\lref\OPEexp{F.A. Dolan and H. Osborn, {\it Conformal Partial Wave Expansions for
$\N=4$ Chiral Four Point Functions}, Ann. Phys. 321 (2006) 581, hep-th/0412335.}

\lref\NDO{F.A. Dolan, M. Nirschl and H. Osborn, {\it Conjectures for Large $N$ 
$\N=4$ Superconformal  Chiral Four Point Functions}.
Nuclear Physics, {\bf B749} 109-152 (2006), hep-th/0601148.}

\lref\Sundborg{
P.~Haggi-Mani and B.~Sundborg,
{\it Free large N supersymmetric Yang-Mills theory as a string theory}, 
JHEP {0004} (2000) 031 (2000), hep-th/0002189\semi
B.~Sundborg, 
{\it Stringy gravity, interacting tensionless strings and massless higher spins},
Nucl.\ Phys.\ Proc.\ Suppl.\  {102} (2001) 113 (2001), hep-th/0103247.
}

\lref\SezginS{
E.~Sezgin and P.~Sundell, {\it Doubletons and 5D higher spin gauge theory},
JHEP {0109} (2001) 036, hep-th/0105001\semi
E.~Sezgin and P.~Sundell,
{\it Towards massless higher spin extension of D = 5, N = 8 gauged supergravity},
JHEP {0109} (2001) 025, hep-th/0107186.
}

{\nopagenumbers
\rightline{DAMTP/06-73}
\rightline{ROM2F/2006/21}
\rightline{NSF-KITP-06-80}
\rightline{hep-th/0609179}
\vskip 1.5truecm
\centerline {\bigbf $\N=4$ Superconformal Characters and Partition Functions}
\vskip  6pt
\vskip 2.0 true cm
\centerline {M. Bianchi,${}^\dagger$
F.A. Dolan,${}^\natural$ P.J. Heslop${}^\natural$ and H. Osborn${}^\natural$}

\vskip 12pt
\centerline {${}^\dagger$Dipartimento di Fisica,
Universit\'a di Roma `Tor Vergata',}
\centerline{I.N.F.N. - Sezione di Roma `Tor Vergata',}
\centerline{
Via della Ricerca Scientifica, 1-00133 Roma, Italia}
\vskip  6pt
\centerline {${}^\natural$Department of
Applied Mathematics and Theoretical Physics,}
\centerline {Wilberforce Road, Cambridge CB3 0WA, England}
\vskip 1.5 true cm

{\eightpoint
\parindent 1.5cm{
\noindent

{\narrower\smallskip\parindent 0pt

Character formulae for positive energy unitary representations of
the $\N=4$ superconformal group are obtained through use of reduced
Verma modules and Weyl group symmetry.  Expansions of these are given which
determine the particular representations present and results such as dimensions of
superconformal multiplets. By restriction of variables  various `blind' characters 
are also obtained. Limits, corresponding to reduction to particular subgroups, in 
the characters isolate contributions from particular subsets of multiplets and 
in many cases simplify the results considerably. As a 
special case, the index counting short and semi-short multiplets which do not 
form long multiplets found recently is shown to be related to particular cases 
of reduced characters.  Partition functions of $\N=4$ super Yang Mills are 
investigated.  Through analysis of these, exact formulae are obtained 
for counting $\half$ and some $\quar$-BPS operators in the free case.
Similarly, partial results for the counting of semi-short operators are given.  
It is also shown in particular examples how certain short operators which one 
might combine to form long multiplets due to group theoretic considerations
may be protected dynamically.

Keywords: Superconformal Symmetry,  $N=4$ Quantum Field Theory, Characters,
Partition Functions.

\narrower}}

\vfill
\line{\hskip0.2cm E-mails:
{{\tt Massimo.Bianchi@roma2.infn.it,
F.A.Dolan@damtp.cam.ac.uk}}\hfil}
{\line{{\tt P.J.Heslop@damtp.cam.ac.uk,
H.Osborn@damtp.cam.ac.uk}}\hfill}
}

\eject}
\pageno=1
\newsec{Introduction}

The dynamics of $\N=4$ super-Yang Mills theories is highly
constrained by the large supersymmetry group present in this
case. Invariance of the classical
action under $\N=4$ superconformal transformations persists as a
symmetry of these theories at the quantum level both
perturbatively and non-perturbatively
thanks to their remarkable ultra-violet properties.

Moreover, with a $U(N)$ or $SU(N)$ gauge group, in the large $N$ limit the
theory should admit an effective string description that has found
a concrete realisation in the holographic AdS/CFT correspondence
conjectured by Maldacena some time ago \malda. 
More recently there has been much interest in the potential
integrability of $\N=4$ super-Yang Mills theories at the planar
level, which is related to the existence of an infinite number of
non-local classically conserved charges for type IIB superstrings
on $AdS_5\times S^5$. This has found applications in the study of
states with large $R$-charge or spin in terms of various spin
chains that are expected to be the holographic counterpart of
various string solitons (for recent reviews of  spin chains and
integrability see \refs{\beist, \kaza}).

In terms of determining the full spectrum of operators in general
unfortunately at
weak 't Hooft coupling, the regime amenable to a (perturbative)
superconformal field theory analysis, the $AdS_5\times S^5$
background where the string propagates is highly curved and the
spectrum of its excitations can only be extrapolated from large
radius that corresponds to large 't Hooft coupling, where an
effective supergravity description is available. Nevertheless,
invoking the higher spin symmetry enhancement present for zero coupling
\refs{\Sundborg,\SezginS}  allows the precise
matching of the spectrum of single particle states of the type IIB
superstring on $AdS_5\times S^5$ with the spectrum of single-trace
gauge invariant operators in $\N=4$ super-Yang Mills. Turning on
interactions with a non zero coupling, all but the $\half$-BPS multiplets 
combine into long
multiplets of the $\N=4$ superconformal group $PSU(2,2|4)$ and
acquire anomalous dimensions. The bulk counterpart is a Higgs
mechanism, termed {\it La Grande Bouffe}, whereby higher spin
fields absorb lower spin Goldstone fields and become massive - see
\refs{\BMStao, \BeisertTE, \beis, \BHR},  and  also the review \Bianchi.

A similar analysis for the spectrum of multi-particle states of supergravity
or more generally string theory on $AdS_5\times S^5$, that
appear in various guises as multi-graviton states, giant
gravitons, wrapped D3-branes, and AdS black holes, and are
expected to be dual to multi-trace gauge invariant operators, is
still in its infancy. 

The purpose of the present investigation is
to develop a framework which allows the decomposition of the spectrum of $\N=4$
super-Yang Mills theories into positive energy unitary representations of
the $\N=4$ superconformal group. The large superconformal symmetry
allows for a rich variety of BPS and short/semi-short multiplets. An
index proposed recently should count the `unpaired' short
multiplets that cannot combine into long ones. The index is
`topological' in the sense that it is invariant under continuous
deformations that preserve $\N=4$ superconformal symmetry such as
turning on or switching off interactions. It can be thus reliably
computed at weak coupling.
The information stored in the index is however limited since, as we will
show, there exist certain short operators which would be allowed to
combine into long multiplets due to purely representation considerations, but
which, according to the analysis of~\refs{\HesH,\Aspects}, are in fact prevented from 
doing so dynamically. Indeed Morales has shown that these operators are protected 
in some cases at one loop~\Morales.  A mismatch of the macroscopic entropy of some 
class of $\eight$-BPS and ${1\over 16}$-BPS AdS black holes with the 
counting of corresponding operators in the gauge theory as revealed by
the index \mald\ is also in contradiction with the naive expectation 
that this counts all such operators and that all potential long multiplets
gain anomalous dimensions for non zero coupling.

The spectrum of operators, either as single trace or multi-trace operators,
in $\N=4$ superconformal theories may be determined in terms of various
partition functions involving operators constructed from the fundamental
fields. For particular classes of short or semi-short supermultiplets
these may be restricted to particular subsectors. For single trace
operators these are equivalent to various spin chains on which 
the dilatation operator $D$, whose eigenvalues determine the scaling
dimensions, takes an integrable form.  For the interacting 
large $N$ superconformal theory there should then perhaps be the possibility
of determining the partition function for non zero coupling, but such issues 
are not addressed here. In any event in order to disentangle the
particular supermultiplets present it is necessary to consider
expansions in terms of the characters for the different irreducible
representations. Only genuine unconstrained long multiplets may develop
anomalous dimensions with non zero coupling.

In this paper we attempt to construct explicit compact, as far as possible, 
formulae for the characters for the various short and semi-short multiplets of
$\N=4$ superconformal symmetry. The results are obtained by application 
of a similar technique to that outlined in \RSpeis\ for obtaining the Weyl 
character formula for compact Lie groups by using Verma modules. 
Verma modules are vector spaces constructed by
the action of lowering operators in the Lie algebra on highest weight states.
The representation space is then formed by acting on the Verma modules
with the elements of the Weyl symmetry group. This procedure is extended here to
the $\N=4$ superconformal group. 
The analysis depends only on the algebraic properties of the $\N=4$
superconformal group. Various appropriate shortening conditions are
applied on the highest weight state, at particular values of the conformal scale
dimension $\De$, and these lead to the different short and semi-short
supermultiplets for which characters are obtained.\foot{Mathematical results 
\Superchar\ for
characters for atypical representations, when shortening conditions apply,
of superalgebras are not straightforward. Our procedure is not valid in all
cases, but it has been carefully checked in applications here.} A more general, 
although in some respects similar, discussion is found in \Dobrev.

The plan of the paper is as follows. In section 2 we describe standard
mathematical results for the characters of Lie algebras as obtained from
Verma modules in a fashion which allows generalisation to the superconformal
case subsequently. Although this introduces some convenient notation this
is not essential for later applications. 
In section 3 we then describe how  character
formulae for positive energy unitary representations of
the $\N=4$ superconformal group may be obtained, taking account of various possible
shortening conditions for supermultiplets. It is shown how to use these results
to obtain the decomposition of long multiplets in terms of semi-short multiplets
in a simple fashion.
In section 4 the character formulae are applied
in particular cases for unitary supermultiplets  of interest and, for special 
choices of variables (`fugacities') we obtain the `blind' characters that encode 
less information, eventually only keep track of the scaling
dimensions of the components of the supermultiplets.
In section 5 we show that limits exist which reduce the characters 
to those for various subgroups of the superconformal group $PSU(2,2|4)$
and isolate contributions from subsets of multiplets of the full spectrum allowed by
superconformal symmetry.  In particular, we identify a limit that exposes
the characters contributing to the index in \mald. 
In section 6 we describe the basic supermultiplet formed by the 
fundamental fields and how this may be reduced to different subsectors
corresponding to particular subgroups of $PSU(2,2|4)$ which involve subsets of
the elementary fields. 
In section 7 we investigate partition functions of $\N=4$
super-Yang Mills in different sectors and obtain exact formulae for counting
$\half$ and some $\quar$-BPS operators in the free case
and partial results for the counting of semi-short operators. 
These are applied both for general multi-trace operators and also
for those dual to supergravity fields which are relevant in the strong
coupling limit. Results for the index are also obtained in this context. 
Finally in section 8 we discuss
further when semi-short operators are protected in the interacting theory, 
making connections with specific examples. The index is applied in particular
cases and it is argued that all semi-short operators dual to supergravity
fields should be protected although in some cases they may form long multiplets.

Some details are referred to four appendices.
Appendix A describes the Lie algebra for $PSU(2,2|4)$ and lists 
its relevant subgroups related to different shortening conditions.
Appendix B contains formulae for the expansion of infinite products in 
terms of Schur polynomials that are useful in the analysis of partition 
functions. Appendix C contains various tables of the semi-short operators that 
are required by the expansion of partition functions for the first few levels, 
as discussed in the main text. In addition appendix D
analyses the product of two characters for the fundamental representation
and obtains a decomposition into irreducible multiplets.

\newsec{Verma Modules and Characters}

We here outline, without any proofs, how characters are obtained from
Verma modules for Lie algebras in a form which is appropriate for
our later discussion of superconformal groups.

For $Sl(2)$, with generators $J_\pm , J_3$ having standard commutation 
relations, the Verma module $\V_j$
is spanned by states $(J_-)^N|j\rangle^{\rm h.w.}$, for $N=0,1,2,\dots $,
where the highest weight state satisfies
$J_+|j\rangle^{\rm h.w.}=0$, $J_3|j\rangle^{\rm h.w.}=j|j\rangle^{\rm h.w.} $.
The corresponding character for $\V_j$  may then be expressed, for general $j$, as
a formal trace involving a sum over all eigenvalues for $J_3$ in $\V_j$
\eqn\vermasut{
C_j(x)=\ttr_{\V_j} ( x^{2J_3} ) =\sum_{N=0}^\infty x^{2j-2N}
={x^{2j+2}\over x^2-1}\,.
}
For  $2j\in \Bbb{N}$, $\V_{-j-1} \subset \V_j$ and, with the $SU(2)$ 
hermiticity requirements for $J_\pm , J_3$,  $\V_{-j-1}$ contains null states.
A finite dimensional space is obtained by considering $V_j = \V_j / \V_{-j-1}$,
removing also null states, and the associated representation of $Sl(2)$
corresponds to the standard spin-$j$ unitary irreducible
representation for $SU(2)$.
The corresponding character for this representation on $V_j$ is given by
a well defined trace
\eqn\charj{
\chi_j(x)=  \tr_{V_j} ( x^{2J_3} ) =
C_j(x)-C_{-j-1} (x)={x^{2j+1}-x^{-2j-1}\over x-x^{-1}}\,.
}
The Weyl group $\W$ in this case is $\S_2 \simeq \Bbb Z_2$ with elements $\{e,\sigma\}$
where $\sigma^2 = e$. Defining $\sigma f(x) =f(x^{-1}) $ for any function $f$
then the character for the irreducible spin $j$
representation may also be written as
\eqn\weylt{
\chi_j(x) = \frak{W}^{\S_2} C_j(x) = C_j(x)+C_j(x^{-1}) \, , \qquad
\frak{W}^{\S_2}=e+\sigma \,.
}
Of course it is easy to see that ${\rm dim}(V_j)= \chi_j(1)=2j+1$, the standard
dimension of the spin $j$ representation.

A general simple Lie algebra is decomposed into an abelian Cartan subalgebra
with generators  $\uH=(H_1,H_2,\dots,H_r)$ and remaining generators
$\{E_{\ua}\}$ defined by roots $\ua \in  \Phi$
where $[\uH , E_{\ua} ] = \ua \, E_{\ua}$. We require $[E_{\ua}, E_{-\ua} ] =
\ua{}^{\vee} \cdot \uH$ with $\ua{}^{\vee} = 2\ua / \ua^2$ the coroots. 
Dividing the root space into positive and negative roots,
$\Phi = \Phi_+ \cup \Phi_-$, a highest weight
state $| \uL \rangle^{\rm h.w.}$ satisfies $E_{\ua} | \uL \rangle^{\rm h.w.}
= 0$, $\ua \in \Phi_+$ and
$ \uH | \uL \rangle^{\rm h.w.} = \uL | \uL \rangle^{\rm h.w.}$.  The associated 
Verma module $\V_{\uL}$ is then defined in terms of the basis of states
\eqn\VermL{
\prod_{\ua \in \Phi_+} (E_{-\ua})^{N_{\ua}} | \uL \rangle^{\rm h.w.} \, , \quad
N_{\ua}= 0,1,2,\dots , \, ,
}
for a particular choice of ordering of $E_{-\ua}$.
For simple roots $\ua{}_i$ and corresponding fundamental weights $\uw_i$, where
$\ua{}_i{\!}^{\vee} \cdot \uw{}_j = \de_{ij}$, we can expand
$\uL = \sum_i \lambda_i \uw{}_i$. For ${\tilde H}_i =
\ua{}_i{\!}^{\vee} \cdot \uH$, which with $E_{\pm \ua_i}$ forms a standard $Sl(2)$ 
algebra, we define the character for the Verma module \VermL\ by
\eqn\CVerm{\eqalign{
C_{\uL}(\uxx) = {}& \ttr_{\V_{\uL}}\big ( {\ts \prod_i} x_i^{{\tilde H}_i} \big )
= \sum_{\{ N_{\ua} \}} \prod_{\ua \in \Phi_+} \prod_{i=1}^r x_i^{\lambda_i
- N_{\ua} \sum_{j} n_{{\ua},j} K_{ji}} \cr
= {}& \prod_{i=1}^r x_i^{\lambda_i} \prod_{\ua \in \Phi_+} \bigg ( 1 - {\ts \prod_j}
x_j^{-\sum_k n_{\ua,k} K_{kj} } \bigg )^{-1} \, , \cr}
}
where $\ua = \sum_i n_{\ua,i} \, \ua{}_i$ for $\ua \in \Phi_+$ and
$K_{ji}= \ua{}_j \cdot \ua{}_i{\!}^{\vee}$ are the elements of the Cartan 
matrix.

In a similar fashion to the $Sl(2)$ case for  $\lambda_i \in \Bbb{N}$ the
Verma module is reducible since there are states in the Verma module $\V_{\uL}$ also
satisfying the highest weight conditions. Such states arise when $\uL^\sigma = \uL -
\sum_{\ua \in \Phi_+} N_{\ua} \, \ua$ for some $N_{\ua}$ and where $\uL^\sigma$
denotes the action of an element $\si$ of the Weyl group $\W$ on $\uL$.
$\W$ is generated by reflections corresponding to the simple roots,
$\sigma_i \uv = \uv - \ua_i \ \ua_i{\!}^{\vee} \cdot \uv$, $\si_i{\!}^2=e$, and is 
determined by the relations, for $i\ne j$, $(\si_i \si_j)^{2 + K_{ij}K_{ji}} =e$ 
(excluding the group $G_2$). For any $\si$ we define
$\uL^\sigma = \sigma(\uL + \urho) - \urho$,
with $\urho = \half \sum_{\ua \in \Phi_+} \ua = \sum_{i=1}^r \uw{}_i$.
Hence, as for $Sl(2)$, $\V_{\uL^\sigma} \subset \V_{\uL}$.
A vector space $V_{\uL}$ for a finite dimensional representation may be
formed from $\V_{\uL}$ by an extension of the procedure for $Sl(2)$. 
For any $\si\in \W$  the length $\ell(\si)$ is defined as the
minimal number of elementary reflections $\si_i$ in a product necessary to
generate $\sigma$, there is further a unique element $\si_{\rm max}$ such that
$\ell(\si_{\rm max}) = \ell_{\rm max}$.
Defining $V_{\uL}{\!}^{(p)} = (\oplus_{\si, \ell(\si)=p}
\V_{\uL^\sigma})/V_{\uL}{\!}^{(p+1)}$, with $V_{\uL}{\!}^{(\ell_{\rm max})} =
\V_{\uL^{\sigma_{\rm max}}}$, then $V_{\uL} = V_{\uL}{\!}^{(0)}$.
The associated character for this representation is
\eqn\charV{
\chi_{\uL}^{\vphantom \sigma}(\uxx) =
\tr_{V_{\uL}}\big ( {\ts \prod_i} \, x_i^{{\tilde H}_i} \big ) =
\sum_{\sigma \in \W} (-1)^{\ell(\si)} \, C_{\uL^\sigma} (\uxx) \, ,
}
which is the standard Weyl character formula for the representation with
Dynkin labels $\ul = [\lambda_1,\dots , \lambda_r]$.

For later application we obtain more explicit expressions for the characters 
of $Gl(n)$ and then $Sl(n)$.  In an orthonormal basis the generators are 
$R^i{}_{\! j}$, with $1\le i,j\le n$, and satisfy the Lie algebra
$[ R^i{}_{\! j} , R^k{}_{\! l}] = \de^k{}_{\! j}R^i{}_{\! l} -
\de^i{}_{\! l}R^k{}_{\! j}$. The Cartan subalgebra generators for 
$Gl(n)$ are then $\rH=(R^1{}_{\! 1}, \dots , R^n{}_{\! n})$ and
$[ \rH , R^i{}_{\!j}] = (\bfe_i - \bfe_j) R^i{}_{\! j}$
with $\{\bfe_1, \dots ,\bfe_n\}$ a set of $n$-dimensional
orthonormal unit vectors given by $(\bfe_i)_j = \de_{ij}$.
In this case $\bfe_i - \bfe_j \in \Phi_+$ for $1\le i<j \le n$
and a corresponding set of simple roots is given by $\bfe_i - \bfe_{i+1}$
for $i=1,\dots , {n-1}$.  Representations are obtained from
highest weight states $|\ell_1,\ell_2,\dots,\ell_n\rangle^{\rm h.w.}$
satisfying  $R^i{}_{\! j} |\ell_1,\dots,\ell_n\rangle^{\rm h.w.}=0$, $i<j$, 
and $ \rH |\ell_1,\dots,\ell_n\rangle^{\rm h.w.}= \bl
|\ell_1,\dots,\ell_n\rangle^{\rm h.w.}$ for $\bl=(\ell_1,\dots , \ell_n)$. 
The corresponding Verma module $\V_{\bl}$
is spanned by $\prod_{i>j} (R^i{}_{\! j})^{N_{ij}}
|\ell_1,\ell_2,\dots,\ell_n\rangle^{\rm h.w.}$, $N_{ij}=0,1,2,\dots$.
For $\uu=(u_1,\dots , u_n)$ the Verma module character may be defined
by a sum over all eigenvalues of $\rH$
\eqn\vermagln{\eqalign{
C_{\bl}(\uu)= {}& \sum_{\{N_{ij}\}} \prod_{i=1}^n
\bigg ( u_i^{\ \ell_i - \sum_{j>i} N_{ij} + \sum_{j<i} N_{ij}} \bigg )
= \prod_{i=1}^n u_i^{\ell_i} \prod_{j<k} \sum_{N_{jk}=0}^\infty \Big (
{u_k\over u_j} \Big )^{N_{jk}} \cr
= {}& \prod_{i=1}^n \ux_i{}^{\ell_i+n-i} \Delta(\uu)^{-1}\,, \cr}}
which converges for $u_{i+1}<u_i$ and 
where $\Delta(\uu)$ is the Vandermonde determinant,
\eqn\vander{ \Delta(\uu)= {\det} \big [\ux_i{}^{j-1} \big ]
=\prod_{1\leqslant i<j\leqslant n}(\ux_i-\ux_j)\,. }
Under a rescaling $C_{\bl}(\lambda \uu) = \lambda^{\sum_i \ell_i} C_{\bl}(\uu)$.

For $Sl(n)$ the generators satisfy $\sum_i R^i{}_{\! i}=0$ and a basis for the 
Cartan subalgebra
is given by $H_i = R^i{}_{\! i} - R^{i+1}{}_{\!\! i+1}$, $i=1,\dots , n-1$. 
Since the Verma module character \vermagln\ satisfies $C_{\bl}(\uu) \to 
(u_1 \dots u_n)^c \C_{\bl}(\uu)$ for  $\bl \to \bl + c \sum_i \bfe_i $, 
for any constant $c$, we may impose for the $Sl(n)$ case $\prod_{i=1}^{n}\ux_i=1$,
so that $\bl$ is then arbitrary up to $\bl \to \bl + c \sum_i \bfe_i $.
This condition on $\uu$ may be realised by letting 
$u_1=x_1, \, u_n=1/x_{n-1}$ and $u_i = x_i/x_{i-1}$ for $i=2,\dots ,n-1$ and 
then $C_{\bl}(\uu) = C_{\ll}(\uxx)$
where $\lambda_i = \ell_{i}-\ell_{i+1},\,i=1,\dots,n-1$. 

For the simple roots considered above $\bfe_i-\bfe_{i+1}=(\bfe_i-\bfe_{i+1})^{\vee}$
and it is easy to see that $\sigma_i \bfe_i =
\bfe_{i+1}, \, \sigma_i \bfe_{i+1} = \bfe_i$ and $\si_i \bfe_j = \bfe_j$ 
otherwise, so that the Weyl group $\W \simeq \S_n$ the usual permutation 
group. Taking
$\brho = (n-1,n-2,\dots,0)$ and $\bl^\sigma = \sigma(\bl + \brho) - \brho$
it is straightforward to show that $C_{\bl^\sigma}(\uu) = \hbox{sign}(\si) \,
C_{\bl}(\sigma\uu)$ where
$\si (u_1,\dots,u_n)=(u_{\si(1)},\dots,u_{\si(n)})$ is a simple  permutation
and $\hbox{sign}(\si)=\pm 1$ for even, odd permutations, so that
$\Delta(\sigma \uu) = \hbox{sign}(\sigma) \, \Delta (\uu)$. Hence
applying \charV\ in this case the character for the finite dimensional
representation obtained when we take $\ell_i - \ell_{i+1} \in \Bbb{N}$ is,
\eqn\chargln{
\chi_{\bl}(\uu) = \sum_{\sigma\in \S_n}\hbox{sign}(\sigma) \, C_{\bl^\sigma}(\uu)
=\sum_{\sigma\in \S_n}C_{\bl}(\sigma \uu)
={\det}\big [\ux_i{}^{\ell_j+n-j} \big ]\, \Delta(\uu)^{-1}\, ,}
which are Schur polynomials, expressing compactly the characters for $Gl(n)$ and also 
$Sl(n)$ with $\prod_i u_i =1$.
The result \chargln\ may also be expressed, if $\si f(\uu)= f(\si \uu)$,  as
\eqn\weylsym{
\chi_{\bl}(\uu) =  \frak{W}^{\S_n} C_{\bl}(\uu) \, , \qquad
\frak{W}^{\S_n}=\sum_{\si\in \S_n}\si\, .
}
A particular example which is useful later is
\eqn\pzerorep{
\chi_{(p,0,\dots,0)}(\uu)=\sum_{a_i \geqslant 0 \atop a_1 + \dots + a_n =p}
\!\!\!\! {\ts \prod}_{i=1}^n \, \ux_i{\!}^{a_i} \, . }

For calculating products involving  $Sl(2)$ and $Gl(n)$ or $Sl(n)$
characters we may use
\eqn\chartwo{
(x^n+x^{-n}) \chi_j(x) = \chi_{j+{1\over 2}n}(x) + \chi_{j-{1\over 2}n}(x) \, ,
\qquad \chi_j(x) = - \chi_{-j-1}(x) \, ,
}
and results such as
\eqn\charfour{
\sum_{\si \in \S_n/\S} \! u_{\si(1)}{}^{\!r_1} \dots u_{\si(n)}{}^{\!r_n} \,
\chi_{\bl}(\uu) = \sum_{\si \in \S_n/\S} \!
\chi_{\bl + r_1 \bfe_{\si(1)} + \dots  + r_n \bfe_{\si(n)} }(\uu) \, , \quad
r_i = 0,1,2,\dots \, , 
}
with $\S \subset \S_n$ defined by $u_{\si(1)}{}^{\!r_1} \dots u_{\si(n)}{}^{\!r_n} 
= u_{1}{}^{\!r_1} \dots u_{n}{}^{\!r_n}$.
With the relations 
\eqn\charW{
\chi_{\bl}(\uu) = (-1)^{{\rm sign}(\si)} \chi_{\bl^\si}(\uu) \, , \quad 
\si \in \S_n \, ,}
and also for $\chi_j$ under $j\to -j-1$ in \chartwo, we
may  ensure all resulting characters in the expansion
have $j\ge 0$, $\ell_i\ge \ell_{i+1}$, requiring also $\chi_{-{1\over 2}}=0$
and $\chi_{\bl}=0$ if $\bl=\bl^\si$ for ${\rm sign}(\si)=-1$.
For $Gl(4)$, which is relevant later, the Weyl reflections are generated by
\eqn\reflect{
\bl^{\si_1} = (\ell_2-1,\ell_1+1,\ell_3,\ell_4) \, , \ \
\bl^{\si_2} = (\ell_1,\ell_3-1,\ell_2+1,\ell_4) \, , \ \
\bl^{\si_3} = (\ell_1,\ell_2,\ell_4-1,\ell_3+1) \, .
}

For positive energy unitary representations of non compact groups, such as the
conformal group $SO(d,2)$, in order to obtain the corresponding characters
it is necessary to restrict the action of elements of the Weyl group to
those corresponding to a maximal semi-simple compact subgroup, for the conformal 
group  this is $SO(d)$. The Verma module in this
case is obtained from a highest weight state for $O(d) \otimes O(2)$ which is
an eigenvector of the dilatation operator $D$ with eigenvalue $\Delta$
and is annihilated by the generators for special conformal transformations $K_\mu$.
A basis for the Verma module is then constructed by the action of arbitrary
products of the momentum operators $\prod_\mu (P_\mu)^{N_\mu}$,
as well lowering operators for the Lie algebra of $SO(d)$. A detailed discussion
was given in \fadl. There are various possible shortening conditions
when appropriate subsets of certain combinations of the
generators $P_{\mu}$ annihilate the highest weight state for particular
$\Delta$ which lead to reduced Verma modules. The characters for the
corresponding unitary representations are then obtained by acting with
the reduced Weyl symmetry group on the reduced Verma module characters.

\newsec{Characters for unitary irreducible representations of the
$\N=4$ superconformal group}

The conformal group $SU(2,2)$ algebra consists of
translation generators $P_{\al\dal}$,
special conformal generators ${\tilde K}^{\dal\al}$,
dilatations $D$ and $SU(2)_J\otimes SU(2)_{\bar J}$ spin generators
$J_3,J_{\pm},{\bar J}_3,{\bar J}_{\pm}$, for $\al,\dal=1,2$ spinor indices.
The superconformal extension $PSU(2,2|4)$ has in addition
supercharges $Q^i{}_\al,{\bar Q}_{i\dal}$,
along with their superconformal extensions $S_i{}^{\al},{\bar S}^{i\dal}$,
for $SU(4)$ indices $i=1,\dots,4$, and also $SU(4)_R$ $R$-symmetry generators
$R^i{}_{\! j}$.
The details of the  conformal algebra are given in appendix A.\foot{As compared
to \fadho\ 
$D \to i D $ so that $D$ here, although anti-hermitian, has real eigenvalues.
\vskip -6pt}

A generic highest weight primary state for this superalgebra
$|\De;k,p,q;j,\bj\rangle^{\rm h.w.}$ has conformal
dimension $\De$ belongs to the
spin $SU(2)\otimes SU(2)$ representation $(j,\bj)$ and
the $SU(4)$ representation with Dynkin labels $[k,p,q]$ and satisfies
\eqn\hws{\eqalign{
& (K_{\al\dal},S_i{}^{\al},{\bar S}^{i\dal},J_{+},{\bar J}_+,R^i{}_{\! i+1})
|\De;k,p,q;j,\bj\rangle^{\rm h.w.}=0\, , \cr
(D; H_1,H_2,H_3; & J_3,{\bar J}_3) |\De;k,p,q;j,\bj\rangle^{\rm h.w.}
= (\De ; k,p,q ; j , \bj ) |\De;k,p,q;j,\bj\rangle^{\rm h.w.} \, , \cr}
}
with $H_i$ the Cartan generators for $SU(4)$. 
The corresponding Verma module $\V_{(\De;k,p,q;j,\bj)}$ is then
spanned by the states
\eqn\vermam{
\prod_{i,j,k,l=1,\dots,4,k>l\atop
\al,\dal,\beta,\dbe=1,2}(P_{\al\dal})^{N_{\al\dal}}
(Q^i{}_{\beta})^{n_{i\beta}^{\vphantom g}}
({\bar Q}_{j\dbe})^{{\bar n}_{j\dbe}}(J_-)^{N}({\bar J}_-)^{{\bar N}}
(R^k{}_{\! l})^{N_{kl}}|\De;k,p,q;j,\bj\rangle^{\rm h.w.}\, , }
for $N_{\al\dal},N,\,{\bar N},\,{N_{kl}} = 0,1,2, \dots$ and
$n_{i\beta}^{\vphantom g},\,{\bar n}_{j\dbe} =0,1$.
The character for the Verma module  $\V_{(\De;k,p,q;j,\bj)}$
is expressed in terms of variables $s,\ux_1,\ux_2,\ux_3,\ux_4,x,\by$,
with $\prod_i u_i =1$, so that in the series expansion of the character then
$s^{2\De}\ux_1{}^{\ell_1}\ux_2{}^{\ell_2}\ux_3{}^{\ell_3}
\ux_4{}^{\ell_4} x^{2j}\by^{2\bj}$, for
$k=\ell_1-\ell_2,p=\ell_2-\ell_3,q=\ell_3-\ell_4$, corresponds to
the highest weight state.
In general in translating between orthonormal basis labels $(\ell_1,\dots,\ell_4)$
and Dynkin labels $[k,p,q]$ for $SU(4)$ representations then we may take 
\eqn\takevar{
\ell_1=k+p+q\,,\qquad \ell_2=p+q\,,\qquad \ell_3=q\,,
\qquad \ell_4=0\, ,
}
without loss of generality.

The action of the generators on the highest
weight state, as in \vermam, introduces further factors according to
$ P_{\al\dal} \rightarrow s^2 x^{\pm 1} \by^{\pm 1} $, along with
\eqn\wsq{
Q^{1}{}_\al \to s \, u_1 \, x^{\pm1} \, , \quad
Q^{2}{}_\al \to s \, u_3 \, x^{\pm1} \, , \quad
Q^{3}{}_\al \to s \, u_3 \, x^{\pm1} \, , \quad
Q^{4}{}_\al \to s \, u_4 \, x^{\pm1} \, , 
}
and,{\foot{More generally
for the conjugate fundamental representation of $Gl(4)$  we should take
$ \ux_i{}^{-1} \to \prod_{j\ne i}^4 \ux_j$.}}
\eqn\wsqq{ 
{\bar Q}_{4\dal} \to s \, u_4{}^{-1} \by^{\pm 1} \, , \quad
{\bar Q}_{3\dal} \to s \, u_3{}^{-1} \by^{\pm 1}\, , \quad
{\bar Q}_{2\dal} \to s \, u_2{}^{-1} \by^{\pm 1}\, , \quad
{\bar Q}_{1\dal} \to s \, u_1{}^{-1} \by^{\pm 1}\, . 
}
In the above $\al=1,2$ correspond to $x,x^{-1}$ and
$\dal=1,2$ to $\by^{-1},\by$ respectively.

By using \vermasut, \vermagln\ we may easily determine 
the Verma module character, which is given by a formal trace,
\eqn\vermasup{\eqalign{ C_{(\De;k,p,q;j,\bj)}& (s;\uu;x,\by)
= \ttr_{\V_{(\De;k,p,q;j,\bj)}}
\big ( s^{2D} \, u_1{}^{H_1+H_2+H_3} u_2{}^{H_2+H_3} u_3{}^{H_3}
\, x^{2J_3} \, \by^{2{\bar J}_3} \big ) \cr
\noalign{\vskip 2pt}
={} & s^{2\De}
C_{(\ell_1,\ell_2,\ell_3,\ell_4)}(\uu)C_{j}(x) C_{\bj}(\by)\cr & {} \times
\!\!\! \sum_{n_{\vep\eta}=0,1,2,\dots,\atop \vep,\eta=\pm 1}\!\!\! 
(s^2 x^\vep \by^\eta)^{n_{\vep\eta}} 
\!\!\! \sum_{i,j=1,\dots 4,\,\vep,\eta=\pm 1 \atop
n_{i\vep},{\bar n}_{j\eta}=0,1}\!\!\! (s\, \ux_i\, x^\vep)^{n_{i\vep}}(s\,
\ux_j{}^{-1}\, \by^\eta)^{{\bar n}_{j\eta}} \, ,} }
where $\uu=(u_1,u_2,u_3,u_4)$ with $u_1u_2u_3u_4=1$,
For a general long multiplet,
with all possibilities for $n_{\vep\eta}, n_{i\vep},{\bar n}_{i\eta}$ as in
\vermasup\ included, we have assuming \takevar
\eqn\vermasupp{ C_{(\De;k,p,q;j,\bj)}(s;\uu;x,\by)=
s^{2\De}C_{(\ell_1,\ell_2,\ell_3,\ell_4)}(\uu)C_{j}(x)
C_{\bj}(\by)P(s,x,\by)\Q(s\uu,x){\bar \Q}(s^{-1}\uu,\by)\,, } 
where
\eqn\vermasuppo{ \!\!\!\!P(s,x,\by)=\!\!\!\prod_{\vep,\eta=\pm
1}\!\!\! (1-s^2 x^\vep \by^\eta)^{-1}\,,\quad \!\!
\Q(\uu,x)=\!\!\!\prod_{i=1\atop \vep=\pm 1}^4\!\! (1+ \ux_i\,
x^\vep)\,,\quad \!\! {\bar \Q}(\uu,\by)=\!\!\!\prod_{i=1\atop \eta=\pm
1}^4\!\!(1+ \ux_i{}^{\! {-}1} \,\by^\eta)\,. }
According to our  prescription
the character of the irreducible long representation is then given by
a trace over a representation space $V_{(\De;k,p,q;j,\bj)}$
\eqnn\charlong
$$\eqalignno{
\chi_{(\De;k,p,q;j,\bj)}(s,\uu,x,\by) 
={} & \tr_{V_{(\De;k,p,q;j,\bj)}}
\big ( s^{2D} \, u_1{}^{H_1+H_2+H_3} u_2{}^{H_2+H_3} u_3{}^{H_3}
\, x^{2J_3} \, \by^{2{\bar J}_3} \big ) \cr
={}& \frak{W}^{\S_4}\frak{W}^{\S_2}{\overline \frak{W}}{}^{\S_2} 
\big (C_{(\De;k,p,q;j,\bj)}(s;\uu;x,\by)\big ) & \charlong \cr
= {}& s^{2\De} \chi_{(\ell_1,\ell_2,\ell_3,\ell_4)}(\uu)
\chi_j(x)\chi_{\bj}(\by)P(s,x,\by) \Q(s\uu,x){\bar \Q}(s^{-1}\uu,\by)\, , \cr}
$$
since $P(s,x,\by),\Q(s\uu,x),{\bar{\Q}}(s^{-1}\uu,\by)$ are
invariant under the action of the Weyl symmetriser. Here $\frak{W}^{\S_2}$
imposes symmetry under $x\to x^{-1}$ and ${\overline \frak{W}}{}^{\S_2}$
under $\by \to \by^{-1}$. 
Factoring off $P(s,x,\by)$ in \charlong\ and
setting $s,\ux_i,x,\by=1$ then this gives the usual dimension formula
for the conformal primary states in a long multiplet as
$2^{16} (2j+1)(2\bj+1)d_{[k,p,q]}$ where $d_{[k,p,q]}$ is the dimension of
the $SU(4)$ irreducible representation labelled by $[k,p,q]$,
\eqn\dimsuf{
d_{[k,p,q]}=\chi_{(\ell_1,\dots,\ell_4)}(1,1,1,1)={\textstyle
{1\over 12}}(k+1)(p+1)(q+1)(k+p+2)(p+q+2)(k+p+q+3)\, .
}
In \charlong\ the factors $\Q,{\bar \Q},P$ can be decomposed into 
$SU(4)_R \otimes SU(2)_J \otimes SU(2)_{\bar J}$ characters according to
\eqn\PQQ{\eqalign{
\Q(s\uu,x) = {}& \sum_{r=0}^4 s^r \chi_{(1^r 0^{4-r})}(\uu)\, \chi_{j_r}(x)\, , \quad
{\bar \Q}(s^{-1}\uu,\by) = \sum_{r=0}^4 s^r \chi_{(1^{4-r} 0^{r})}(\uu) \,
\chi_{j_r}(\by)\ ,  \cr
P(s,x,\by) = {}& \sum_{r=0}^\infty {s^{2r} \over 1- s^4} \, 
\chi_{{1\over 2}r}(x)\chi_{{1\over 2}r}(\by) \, , \qquad j_0=j_4=0\, , \,
j_1=j_3= \half \, , \, j_2=1 \, . }
}

There are various possible shortening conditions \refs{\Dob,\fadho} which fall into
essentially three classes for unitary representations of $PSU(2,2|4)$
which we label by $t,\bt$, the fraction of the $Q,{\bar Q}$ supercharges
which are eliminated from the Verma module in the generic case.
For the semi-short conditions from \fadho\ we have
\eqn\semis{\eqalign{
\Big ( Q^1{}_2 - {1\over 2j+1}J_- Q^1{}_1 \Big ) 
|\De;k,p,q;j,\bj\rangle^{\rm h.w.}  = {}& 0 \, , \quad
\Delta = 2+2j+ \half (3k+2p+q) \, , \quad   t = \eight \, , \cr
\Big ( {\bar Q}_{41} + {1\over 2\bj+1} \bJ_- {\bar Q}_{4 2} \Big ) 
|\De;k,p,q;j,\bj\rangle^{\rm h.w.} = {}& 0 \, , \quad
\Delta = 2+2\bj+ \half (k+2p+3q) \, , \quad   \bt = \eight \, . \cr}
}
The conditions \semis\ may be extended to $Q^i{}_2$
for $i=1,2$ if $k=0$, $i=1,2,3$ if $k=p=0$ and conversely to ${\bar Q}_{i1}$
for $i=3,4$ if $q=0$ and $i=2,3,4$ if $p=q=0$. For short multiplets there
are two cases given by
\eqn\quars{\eqalign{
Q^1{}_\alpha |\De;k,p,q;0,\bj\rangle^{\rm h.w.}  = {}& 0 \, , \quad
\Delta = \half (3k+2p+q) \, , \qquad   t = \quar \, , \cr
{\bar Q}_{4\dal} |\De;k,p,q;j,0\rangle^{\rm h.w.}  =  {}& 0 \, , \quad
\Delta = \half (k+2p+3q) \, , \qquad   \bt = \quar \, , \cr}
}
and
\eqn\halfs{\eqalign{
Q^i{}_\alpha |\De;0,p,q;0,\bj\rangle^{\rm h.w.}  = {}& 0 \, , \quad i=1,2, \
\Delta = \half (2p+q) \, , \qquad   t = \half \, , \cr
{\bar Q}_{j\dal} |\De;k,p,0;j,0\rangle^{\rm h.w.} = {}& 0 \, , \quad j=3,4, \
\Delta = \half (k+2p) \, , \qquad   \bt = \half \, . \cr}
}
The conditions listed in \semis\ become lower bounds for $\De$ for unitary
long representations where $\Delta$ is not determined in terms of
other parameters. There are also conditions for which both $t,\bt$ are non
zero, for $t=\bt={1\over 8}$ then $j-\bj = \half (q-k)$, for $t=\bt=\quar$
and $t=\bt=\half$ then $j=\bj=0$ with $k=q$ and $k=q=0$ respectively.
We may also impose conditions which involve $Q^i{}_\al$
and ${\bar Q}_{j\dal}$ for some $\al,\dal$ and $i=j$
but since $\{ Q^i{}_{\! \alpha} , {\bar Q}_{j\dal} \} =  2 \de^i{}_{\! j}
P_{\alpha\dal}$ there are then constraints on the highest weight state
involving various components of the momentum operator as well. 
The unitary cases of interest in this paper, which cover all gauge invariant
operators,  are given by
$(t,{\bar t})=(0,0)$, $(\quar,0)$, $(\quar,\quar)$, $(\half,\half)$,
$(\eight,0)$, $(\eight,\eight)$, $(\quar,\eight)$ along with
conjugate representations $(0,\quar)$, $(0,\eight)$, $(\eight,\quar)$.
In terms of the notation in \fadho\ for various supermultiplets we
have for the significant cases\foot{For those cases where only one condition 
was imposed there are also
$(\eight,0)\leftrightarrow \C^{{1\over 4},0}$,
$(0,\eight)\leftrightarrow {\C}^{0,{1\over 4}}$ (which are sometimes
referred to as ${1\over 16}$-BPS),
and also $(\quar,0)\leftrightarrow \B^{{1\over 4},0}$,
$(0,\quar)\leftrightarrow \B^{0,{1\over 4}}$. Multiplets
which were both short and semi-short are denoted by
$(\quar,\eight)\leftrightarrow {\D}^{{1\over 4},{1\over 4}}$,
$(\eight,\quar)\leftrightarrow {\bar \D}^{{1\over 4},{1\over 4}}$.\vskip -8pt}
\eqn\multiplets{\eqalign{
&\hbox{long}: \ \ \qquad t=\bt=0 \, , \quad \ \, \A^\De_{[k,p,q](j,\bj)} \, , \cr
&\hbox{short}: 
\qquad \, t=\bt=\quar \, , \quad \ \, 
\B^{{1\over 4},{1\over 4}}_{[q,p,q](0,0)} \, ,
\quad t=\bt=\half \, , \quad \B^{{1\over 2},{1\over 2}}_{[0,p,0](0,0)} \, ,\cr
&\hbox{semi-short}: \, t=\bt={\ts{1\over 8}} \, , \quad \, 
\C^{{1\over 4},{1\over 4}}_{[k,p,q](j,\bj)} \, , \quad
\C^{{1\over 2},{1\over 2}}_{[0,p,0](j,j)} \, , \quad
\C^{1,1}_{[0,0,0](j,j)} \, ,  \cr}
}
where $\C^{{1\over 2},{1\over 2}}$ and $\C^{1,1}$ are semi-short
multiplets  with $Q^i{}_2\sim 0,\bQ_{k1}\sim 0$, as in \semis, with
$i=1,2,k=3,4$ and $i,k=1,2,3,4$ respectively. 
Other conditions than \semis, \quars\ and \halfs\ are possible but
they are not consistent with unitary representations,\foot{Non unitary
representation were considered in \refs{\phd,\NO}.} they are related to
the above by elements of the Weyl group, corresponding to the full superconformal
group, which do not belong to the Weyl group for the compact subgroup
used in our construction.

The prescription for obtaining the characters for unitary irreducible
representations of the $\N=4$ superconformal group corresponding to 
the various short and semi-short supermultiplets is an extension of
that used for the conformal group in \fadl. For a highest weight state
satisfying the conditions in \semis, \quars\ or \halfs\ a reduced Verma 
module $\V^{t,\bt} \subseteq \V_{(\De;k,p,q;j,\bj)}$ may be constructed,
similarly to \vermam, by the action of  $\{J_-,{\bar J}_-, R^i{}_{\! j}\}$, 
$1\leq j<i\leq 4$, along with an appropriate subset of
$\{P_{\al\dal},Q^{i}{}_{\beta},{\bar Q}_{i\dbe}\}$, $\al,\beta,\dal,\dbe=1,2$.
As a consequence of \semis\ the contribution of the supercharges 
$Q^1{}_2$ or $\bQ_{41}$ are expressible in terms  of other contributions 
present in the Verma module so that these operators may be removed from the 
basis in \vermam.
For \quars\ or \halfs\ the supercharges $Q^i{}_\alpha$ and/or
${\bar Q}_{j\dal}$ are removed either for $i=1$ or $i=1,2$ and/or
$j=4$ or $j=3,4$. The character for the reduced Verma module 
$\V^{t,\bt}$ is then directly obtained and 
the character of the corresponding unitary irreducible representation
is then given by the action of the Weyl symmetriser 
$\frak{W}^{\S_4}\frak{W}^{\S_2}{\overline \frak{W}}{}^{\S_2}$ associated
with the maximal compact subgroup $SU(4)_R\otimes SU(2)_J \otimes
SU(2)_{{\bar J}}$ on the character for the Verma module for $\V^{t,\bt}$. 
As in the usual Weyl character formula discussed in section 2 this
corresponds to a trace over the space $V^{t,\bt}_{(\De;k,p,q;j,\bj)}$
on which unitary representations are defined.

Previously \fadho\ those $SU(4)_R \otimes SU(2)_J \otimes SU(2)_{\bar J}$ 
representations
present in the full short or semi-short supermultiplet were determined by
using the Racah-Speiser algorithm for considering the tensor products
of the representations carried by the various supercharges and that
carried by the highest weight state where shortening conditions were
applied to remove particular sets of supercharges in accord with
\semis, \quars\ and \halfs. The present approach gives equivalent
results in terms of characters, which of course is as expected since the
Racah-Speiser algorithm may be 
proved using Weyl symmetrisers acting on Verma module characters.
Without attempting a more rigorous proof along the lines
of \fadl\ we endeavour to show in the next section and later
that the procedure yields characters which agree with the analysis
in \fadho\ and also with other results in the literature.

The expressions for characters found by the prescription described  may also
be derived using similar techniques on  analytic superspace~\howe\ on which
all positive energy irreducible  representations are carried by
unconstrained superfields~\unconstrained\ thus providing some simplification.
Consequently the  problem is to some extent reduced to that of finding
$Gl(2|2)$ characters as extended Schur polynomials, which were written out in an 
expanded form in~\fourpoints.

For various short and semi-short multiplets the character for the
associated reduced Verma module $\V^{t,\bt}$ is thus obtained by restricting
some $n_{\vep\eta},n_{i\vep},\bar{n}_{i\eta}$ in \vermasup\ to be zero
depending on which of $P_{\al\dal},Q^i{}_\al,{\bar Q}_{i\dal}$ are to be
omitted from the operators generating $\V_{(\De;k,p,q;j,\bj)}$. 
For example from \semis\ either $n_{12}=0$ or ${\bar n}_{41}=0$. 
This prescription then requires that all such unitary multiplets of
$PSU(2,2|4)$ have character expressible in the form
\eqn\charunit{\eqalign{ &\chi^{t,{\bar t}}_{(\De;k,p,q;j,\bj)}(s;\uu;x,\by)\cr
&{} = \tr_{V^{t,\bt}_{(\De;k,p,q;j,\bj)}}
\big ( s^{2D} \, u_1{}^{H_1+H_2+H_3} u_2{}^{H_2+H_3} u_3{}^{H_3}
\, x^{2J_3} \, \by^{2{\bar J}_3} \big ) \cr
&{} =s^{2\De}{\frak W}^{\S_4}{\frak W}^{\S_2}{\overline \frak{W}}{}^{\S_2}
\bigg(C_{(\ell_1,\ell_2,\ell_3,\ell_4)}(\uu)
C_j(x)C_{\bj}(\by)P(s,x,\by) {\Q(s\uu,x)\,{\bar \Q}(s^{-1}\uu,\by)\over
{\Q}_t(s\uu,x)\, {\bar \Q}_{\bar t}(s^{-1}\uu,\by)}\bigg)\,,} }
for appropriate $\De,k,p,q,j,\bj$ compatible with \semis, \quars\ and \halfs\
where
\eqn\defQ{ \Q_t(\uu,x)=\cases{1 \,,& if $t=0$, \cr
(1+ \ux_1\, x^{-1})\,, & if $t=\eight $,  \cr
(1+ \ux_1\, x)(1+ \ux_1 x^{-1})\,, & if $t=\quar$,\cr
(1+ \ux_1\, x)(1+ \ux_1\, x^{-1})
(1+ \ux_2\, x)(1+ \ux_2\, x^{-1})\,,& if $t=\half$, } }
and
\eqn\defQt{
{\bar \Q}_{\bar t}(\uu,\by)=\cases{1 \,,& if ${\bar t}=0$, \cr
(1+ \ux_4{}^{-1}\, \by^{-1})\,, & if ${\bar t}=\eight$,  \cr
(1+ \ux_4{}^{-1}\, \by)(1+ \ux_4{}^{-1} \by^{-1})\,, &
if ${\bar t}=\quar$,\cr
(1+ \ux_4{}^{-1}\, \by)(1+ \ux_4{}^{-1}\, \by^{-1})
(1+ \ux_3{}^{-1}\, \by)(1+ \ux_3{}^{-1}\, \by^{-1})\,,& if ${\bar t}=\half$. } }
Trivially of course $\chi^{0,0}_{(\De;k,p,q;j,\bj)}(s,\uu,x,\by)=
\chi_{(\De;k,p,q;j,\bj)}(s,\uu,x,\by)$ as in \charlong.

In the following it is sometimes convenient to allow for more general
$\De,\ell_1,\ell_2,\ell_3,\ell_4,j,\bj$ in most
character formulae so as to reveal simplifications but we stress that
for characters corresponding to positive energy unitary
representations then we must restrict the values appropriately.

Using the character formulae allows an easy derivation of various
identities for semi-short and short supermultiplets found earlier \fadho.
Since $C_{(\ell_1+1,\ell_2,\ell_3,\ell_4)}(\uu) = u_1
C_{(\ell_1,\ell_2,\ell_3,\ell_4)}(\uu)$ and $C_{j-{1\over 2}}(x) =
x^{-1} C_j(x)$ it is trivial to see from \charunit\ and \defQ\ that
\eqn\decomp{
\chi^{0,{\bar t}}_{(\De;k,p,q;j,\bj)}(s;\uu;x,\by) =
\chi^{{1\over 8},{\bar t}}_{(\De;k,p,q;j,\bj)}(s;\uu;x,\by)
+ \chi^{{1\over 8},{\bar t}}_{(\De+{1\over 2};k+1,p,q;j-{1\over 2},\bj)}
(s;\uu;x,\by) \, ,
}
since the Weyl symmetriser is linear. Similarly
\eqn\decompt{
\chi^{t,0}_{(\De;k,p,q;j,\bj)}(s;\uu;x,\by) =
\chi^{t,{1\over 8}}_{(\De;k,p,q;j,\bj)}(s;\uu;x,\by) + 
\chi^{t,{1\over 8}}_{(\De+{1\over 2};k,p,q+1;j,\bj-{1\over 2})}(s;\uu;x,\by)\, .
}
For a long multiplet 
at the unitarity threshold $\Delta = 2+k+p+q+j+\bj$ and also
$k-q=2(\bj-j)$ these results 
may be combined to give a decomposition of the character in terms of four
$\chi^{{1\over 8},{1\over 8}}$ characters,
\eqnn\fourch
$$\eqalignno{
\chi_{(\De;k,p,q;j,\bj)}(s;\uu;x,\by)  
= {}& \chi^{{1\over 8},{1\over 8}}_{(\De;k,p,q;j,\bj)}(s;\uu;x,\by) \cr
&{} + \chi^{{1\over 8},{1\over 8}}_{(\De+{1\over 2};k+1,p,q;j-{1\over 2},\bj)}
(s;\uu;x,\by) + 
\chi^{{1\over 8},{1\over 8}}_{(\De+{1\over 2};k,p,q+1;j,\bj-{1\over 2})}
(s;\uu;x,\by) \cr
&{} + \chi^{{1\over 8},{1\over 8}}_{(\De+1;k+1,p,q+1;j-{1\over 2},\bj-{1\over 2})}
(s;\uu;x,\by) \, .  & \fourch \cr}
$$

{}From \charunit\ and \defQ\ we may also write
\eqn\reduct{\eqalign{ &
\chi^{{1\over 8},{\bar t}}_{(\De;k,p,q;-{1\over 2},\bj)}(s;\uu;x,\by)
=s^{2\De} P(s,x,\by) \cr & \times {\frak W}^{\S_4}
{\frak W}^{\S_2}{\overline \frak{W}}{}^{\S_2}
\bigg(C_{(\ell_1,\ell_2,\ell_3,\ell_4)}(\uu)
(1+s u_1 x)C_{-{1\over 2}}(x)C_{\bj}(\by) {\Q(s\uu,x)\,{\bar \Q}(s^{-1}\uu,\by)
\over {\Q}_{1\over 4}(s\uu,x)\, {\bar \Q}_{\bar t}(s^{-1}\uu,\by)}\bigg)\, ,} }
since $P(s,x,\by)$ is invariant under 
${\frak W}^{\S_2}{\overline \frak{W}}{}^{\S_2}$.
The expressions involving $\Q(s\uu,x)$ and ${\Q}_{1\over 4}(s\uu,x)$ are
also invariant under ${\frak W}^{\S_2}$ and ${\frak W}^{\S_2} 
C_{-{1\over 2}}(x) =0$. Further the term $s u_1 x$ may be absorbed by 
$\ell_1 \to \ell_1+1$ and $x C_{-{1\over 2}}(x) = C_{0}(x)$ giving then
\eqn\red{
\chi^{{1\over 8},{\bar t}}_{(\De;k,p,q;-{1\over 2},\bj)}(s;\uu;x,\by)
= \chi^{{1\over 4},{\bar t}}_{(\De+{1\over 2};k+1,p,q;0,\bj)}(s;\uu;x,\by) \, .
}
Following a similar argument
\eqn\redt{
\chi^{t,{1\over 8},}_{(\De;k,p,q;j,-{1\over 2})}(s;\uu;x,\by)
= \chi^{t,{1\over 4}}_{(\De+{1\over 2};k,p,q+1;j,0)}(s;\uu;x,\by) \, .
}
and hence
\eqn\redtt{
\chi^{{1\over 8},{1\over 8},}_{(p+2q+1;q,p,q;-{1\over 2},-{1\over 2})}
(s;\uu;x,\by)
= \chi^{{1\over 4},{1\over 4}}_{(p+2q+2;q+1,p,q+1;0,0)}(s;\uu;x,\by) \, .
}
Hence we may identify 
\eqn\idCB{
\C^{{1\over4},{1\over 4}}_{[q,p,q](-{1\over 2},-{1\over 2})} \simeq
\B^{{1\over4},{1\over 4}}_{[q+1,p,q+1](0,0)} \, .
}
The results \red\ and \redt\ may be used in  \decomp\ and \decompt\ when 
$j=0$ and $\bj=0$ respectively. In general they show how $t,\bt=\quar$ 
characters are a special case of those for $t,\bt=\eight$.

As described in \fadho\ the result \fourch\ is a consequence of the 
representation for a long multiplet, at unitarity threshold, containing
four superconformal primary operators from which $(\eight,\eight)$
semi-short multiplets can be constructed. This may be described by
the diagrams
\eqn\diamond{
{\def\normalbaselines{\baselineskip20pt\lineskip3pt
\lineskiplimit3pt}\hskip-0cm
\matrix{
&&\hidewidth \O^{{1\over 8},{1\over 8}}_{[k,p,q](j,\bj)}\hidewidth&&\cr
&\Bsw&~~~~~~~~~~&\Bse&\cr
\hidewidth \O^{{1\over 8},{1\over 8}}_{[k+1,p,q](j-{1\over 2},\bj)}\hidewidth
&&&&
\hidewidth \O^{{1\over 8},{1\over 8}}_{[k,p,q+1](j,\bj-{1\over 2})}
\hidewidth\cr
&\Bse&&\Bsw&\cr
&&\hidewidth
\O^{{1\over 8},{1\over 8}}_{[k+1,p,q+1](j-{1\over 2},\bj-{1\over 2})}
\hidewidth
&&
}
\hskip 3cm
\matrix{
&&\hidewidth \O^{{1\over 8},{1\over 8}}_{[q,p,q](0,0)}\hidewidth&&\cr
&\Bsw&~~~~~~~~~~&\Bse&\cr
\hidewidth \O^{{1\over 4},{1\over 8}}_{[q+2,p,q](0,0)}\hidewidth
&&&&
\hidewidth \O^{{1\over 8},{1\over 4}}_{[q,p,q+2](0,0)}
\hidewidth\cr
&\Bse&&\Bsw&\cr
&&\hidewidth
\O^{{1\over 4},{1\over 4}}_{[q+2,p,q+2](0,0)}
\hidewidth
&&
}
}}
where $\swarrow$ in the first case represents the action of
the $Q^1{}_{\! 2}$ supercharge and  $\searrow$ that of the ${\bar Q}_{41}$
supercharge, in the second they represent $\vep^{\alpha\beta}
Q^1{}_{\! \alpha}Q^1{}_{\! \beta}$ and $\vep^{\dal\dbe}
{\bar Q}_{4\dal}{\bar Q}_{\smash{4\dbe}} $,
and we have ignored other descendant operators.
The $j,\bj=0$ case may be obtained by interpreting operators with
$j,\bj=-{1\over 2}$ as $t,\bt = \quar$-BPS operators, in accord with
the results \red, \redt\ and \redtt.

For later discussion it is convenient to adapt the treatment of the
semi-short conditions shown in \semis, following \refs{\beis,\mald},
so that the conditions are just $\de = {\bar \de}=0$ where
\eqn\defdd{
\de  = \De - 2j  - \half(3k+2p+q) \, , \qquad {\bar \de} =
\De - 2 \bj - \half (k+2p+3q) \, .
}
For any state satisfying \hws\ we  define a corresponding state
\eqn\hwsa{
|\De+\half;k+1,p,q;j+\half,\bj\rangle^{\widetilde{\rm h.w.}} =  
Q^1{}_{1} |\De;k,p,q;j,\bj\rangle^{\rm h.w.}\,,
}
which is easily seen to satisfy the same conditions as in \hws\
($S_i{}^\alpha |\De;k,p,q;j,\bj\rangle^{\widetilde{\rm h.w.}}=0$
for $i=2,3,4$ follows from $R^1{}_{\!i} |\De;k,p,q;j,\bj\rangle^{\rm h.w.}
=0$ for $i=2,3,4$, similarly for $S_1{}^2$),
except that $S_1{}^1$ is omitted and instead
\eqn\Qhws{
Q^1{}_{1} |\De;k,p,q;j,\bj\rangle^{\widetilde{\rm h.w.}} = 0 \, .
}
{}From \hwsa\ we may easily find an inverse relation,
\eqn\Invhw{
S_1{}^1 |\De;k,p,q;j,\bj\rangle^{\widetilde{\rm h.w.}} =
2 ( 4j+\de) |\De-\half;k-1,p,q;j-\half,\bj\rangle^{\rm h.w.}\, . 
}
For this state we may then impose
\eqn\short{
Q^1{}_2 |\De;k,p,q;j,\bj\rangle^{\widetilde{\rm h.w.}} = 0 \quad
\Rightarrow \quad \De = 2j+\half(3k+2p+q) \, ,
}
which, by considering $S_1{}^1 Q^1{}_2 Q^1{}_{1} 
|\De;k,p,q;j,\bj\rangle^{\rm h.w.} = 0$ and using $\{S_1{}^1, Q^1{}_2\} = 4J_-$,
$\{S_1{}^1,  Q^1{}_{1} \} = 4J_3 + 2D - 3H_1-2H_2-H_3$,
is equivalent to the $t={1\over 8}$ condition in \semis. Hence
for analysing $t={1\over 8}$ semi-short multiplets it is sufficient
to consider states satisfying $\de=0$. If $j=\de=0$ then
\Invhw\ shows that  $|\De;k,p,q;0,0\rangle^{\widetilde{\rm h.w.}}$
becomes a highest weight $t=\quar$ state.

For $\bt={1\over 8}$ a similar analysis is possible while in the
$t=\bt={1\over 8}$ case we may consider
\eqn\hwsaa{
|\De+1;k+1,p,q+1;j+\half,\bj+\half\rangle^{\widetilde{\rm h.w.}} =
\bQ_{42} Q^1{}_{1} |\De;k,p,q;j,\bj\rangle^{\rm h.w.}\,,
}
and we omit $S_1{}^1,\bS^{42}$ from the conditions in \hws\ and they are
replaced by $Q^1{}_1,\bQ_{42}$. Imposing $Q^1{}_2
|\De;k,p,q;j,\bj\rangle^{\widetilde{\rm h.w.}} 
= \bQ_{41} |\De;k,p,q;j,\bj\rangle^{\widetilde{\rm h.w.}} =0$,
which requires $\de={\bar \de}=0$ or $\Delta=j+\bj + k+p+q, j-\bj=\half (q-k)$,
is then equivalent to the combined $t=\bt={1\over 8}$ conditions in \semis.
For $j=\bj=0$, $k=q$ then  $|\De;q,p,q;0,0\rangle^{\widetilde{\rm h.w.}}$ 
becomes a $t=\bt=\quar$ highest weight state. If we denote 
${\tilde O}^{{1\over 8},{1\over 8}}_{[k,p,q](j,\bj)}$ as the operator
corresponding to these alternative highest weight conditions, with 
$\De = j + \bj + k+p+q, \, j-\bj = \half(q-k)$, then instead of \diamond\ we
have an equivalent diagram,
\eqn\diamondt{
{\def\normalbaselines{\baselineskip20pt\lineskip3pt
\lineskiplimit3pt}\hskip-0cm
\matrix{
&&\hidewidth 
{\tilde \O}^{{1\over 8},{1\over 8}}_{[k+1,p,q+1](j+{1\over 2},\bj+{1\over 2})}
\hidewidth&&\cr
&\Bsw&~~~~~~~~~~&\Bse&\cr
\hidewidth {\tilde \O}^{{1\over 8},{1\over 8}}_{[k+2,p,q+1](j,\bj+{1\over 2})}
\hidewidth &&&&
\hidewidth {\tilde \O}^{{1\over 8},{1\over 8}}_{[k+1,p,q+2](j+{1\over 2},\bj)}
\hidewidth\cr
&\Bse&&\Bsw&\cr
&&\hidewidth
{\tilde \O}^{{1\over 8},{1\over 8}}_{[k+2,p,q+2](j,\bj)}
\hidewidth
&&
}
}}
where ${\tilde \O}^{{1\over 8},{1\over 8}}_{[k,p,q](0,\bj)} =
{\tilde \O}^{{1\over 4},{1\over 8}}_{[k,p,q](0,\bj)}$, similarly
${\tilde \O}^{{1\over 8},{1\over 8}} = {\tilde \O}^{{1\over 8},{1\over 4}}$ for 
$\bj=0$, and ${\tilde \O}^{{1\over 8},{1\over 8}}_{[q,p,q](0,0)} =
\O^{{1\over 4},{1\over 4}}_{[q,p,q](0,0)}$.

\newsec{Characters in Special Cases}

In this section we consider in more detail characters for the various 
short and semi-short multiplets 
discussed in the previous section  and attempt, where possible, to simplify them and 
to write them in a more explicit fashion,  allowing 
particular cases to be considered subsequently.
This then allows us to determine `blind' characters (such as where the variables 
$\ux_i,x,\by=1$ so that there is then a dependence just on $s$).
With these results we are able to greatly
simplify dimension formula for multiplets given in \fadho\ and
to make manifest the identities  for semi-short multiplets in \multiplets,
$\C^{{1\over 4},{1\over 4}}_{[0,p,0](j,\bj)}
\simeq \C^{{1\over 2},{1\over 2}}_{[0,p,0](j,\bj)}$,
$\C^{{1\over 2},{1\over 2}}_{[0,0,0](j,\bj)}
\simeq \C^{{1,1}}_{[0,0,0](j,\bj)}$.

\noindent
{\bf Basic Supermultiplets}

Our results may be illustrated first by considering the multiplet for
the fundamental fields in $\N=4$ superconformal theories which
corresponds to $\B^{{1\over 2},{1\over 2}}_{[0,1,0](0,0)}$. The character is
\eqn\elemp{\eqalign{
\chi^{{1\over 2},{1\over 2}}_{(1;0,1,0;0,0)}(s;\uu;x,\by)
= {}& \D_0(s,x,\by)\, \chi_{(1,1,0,0)}(\uu) \cr
&{} +\D_{1\over 2}(s,x,\by)\, \chi_{(1,1,1,0)}(\uu)
+ \oD_{1\over 2}(s,x,\by)\, \chi_{(1,0,0,0)}(\uu)\cr
& {} + \D_{1}(s,x,\by)+\oD_1(s,x,\by) \, , \cr }
}
where
\eqn\charfree{\eqalign{
\D_j(s,x,\by)& =s^{2j+2}\big(\chi_j(x)-s^2 \chi_{j-{1\over 2}}(x)
\chi_{1\over 2}(\by)+s^4\chi_{j-1}(x)\big)P(s,x,\by)\,,\cr
\oD_{\bj}(s,x,\by)& =s^{2\bj+2}\big(\chi_{\bj}(\by)-s^2 \chi_{\bj-{1\over  2}}(\by)
\chi_{1\over 2}(x)+s^4\chi_{\bj-1}(\by)\big)P(s,x,\by)\,,}
}
are the characters corresponding to the
spin-$j$ chiral, respectively, spin-$\bj$ anti-chiral free field
representations of the conformal group in $4$ dimensions \fadl.
($\D_0$ corresponds to a free scalar field of conformal dimension 1,
$\D_{1\over 2}$ to a free spin-${1\over 2}$ chiral fermion of
dimension $3\over 2$, $\oD_{1\over 2}$ to
a free spin-${1\over 2}$ anti-chiral fermion of dimension $3\over 2$
and $\D_1$,$\oD_1$ to free chiral and
anti-chiral vector fields of conformal dimension $2$
so that the combination corresponds to a free Maxwell field).
We may also easily determine
\eqn\chis{\eqalign{
\chi_{(1,0,0,0)}(\uu)&=u_1+u_2+u_3+u_4\,, \qquad 
\chi_{(1,1,1,0)}(\uu)=u_1{}^{-1}+u_2{}^{-1}+u_3{}^{-1}+u_4{}^{-1}\, , \cr
\chi_{(1,1,0,0)}(\uu)&=u_1\,u_2+u_1\,u_3+u_1\,u_4+u_2\,u_3+u_2\,u_4+u_3\,u_4
\,,
}}
assuming $\prod_{i=1}^4 u_i=1$, corresponding to the $4$, $\overline 4$ and $6$
representations of $SU(4)$ respectively.

The simplest gauge invariant $\half$-BPS multiplet is 
$\B^{{1\over 2},{1\over 2}}_{[0,2,0](0,0)}$, 
when the chiral primary operators belong
to a 20 dimensional representation of $SU(4)_R$. This supermultiplet
contains the energy momentum tensor as well as the $SU(4)_R$ current.
To expand the character we use the characters for a long representation
of $SO(4,2)$
\eqn\long{
\A_{\De,j,\bj}(s,x,\by)= s^{2\De} \chi_j(x)\chi_{\bj}(\by)P(s,x,\by)
\,,}
with unitarity requiring $\De\geq j+\bj+2$, and also that for
conserved currents
\eqn\conserved{
\D_{j,\bj}(s,x,\by)=
s^{2j+2\bj+4}\big (\chi_j(x)\chi_{\bj}(\by)-s^2\,\chi_{j-{1\over
2}}(x)\chi_{\bj-{1\over 2}}(\by)\big )P(s,x,\by)\, .}
Then we may obtain for the superconformal character
\eqn\zerotwozero{\eqalign{
\chi^{{1\over 2},{1\over 2}}_{(2;0,2,0;0,0)}(s,\uu,x,\by)
={}& \A_{2,0,0}(s,x,\by)\chi_{(2,2,0,0)}(\uu)\cr
& +\A_{{5\over 2},{1\over 2},0}(s,x,\by)\chi_{(2,2,1,0)}(\uu)
+ \A_{{5\over 2},0,{1\over 2}}(s,x,\by)\chi_{(2,1,0,0)}(\uu)\cr & + 
\A_{3,0,0}(s,x,\by)\big(\chi_{(2,0,0,0)}(\uu)+\chi_{(2,2,2,0)}(\uu)\big)\cr
& +
\big(\A_{3,1,0}(s,x,\by)+\A_{3,0,1}(s,x,\by)\big)\chi_{(1,1,0,0)}(\uu)\cr
& + \D_{{1\over 2},{1\over 2}}(s,x,\by)\chi_{(2,1,1,0)}(\uu)\cr
& + \big ( \A_{{7\over 2},{1\over 2},0}(s,x,\by)
+ \D_{{1\over 2},1}(s,x,\by) \big ) \chi_{(1,1,1,0)}(\uu) \cr
& + \big ( \A_{{7\over 2},0,{1\over 2}}(s,x,\by)
+ \D_{1,{1\over 2}}(s,x,\by) \big ) \chi_{(1,0,0,0)}(\uu) \, . }
}

\noindent
{\bf Characters for short multiplets}

More generally we consider initially characters $t=\bt=\quar,\half$
requiring $k=q,j=\bj=0$, $\De=p+2q$, with $q=0$ when $t=\bt=\half$.
In this case in \charunit\ the action of 
$\frak{W}^{\S_2}{\overline \frak{W}}{}^{\S_2}$ becomes simple, since all
factors are invariant under $x\leftrightarrow x^{-1}, \by
\leftrightarrow \by^{-1}$. Hence the general result reduces to
\eqn\charshort{\eqalign{
\chi^{t,t}_{(p+2q;q,p,q;0,0)}(s;\uu;x,\by) = {}& s^{2p+4q}P(s,x,\by) \cr
&{} \times {\frak W}^{\S_4} \big (C_{(p+2q,p+q,q,0)}(\uu) \, \P_{N_t}(s\uu,x)
\, {\bar \P}{}_{{\bar N}_{t}} (s^{-1}\uu,\by) \big ) \, , \cr}
}
where
\eqn\defPP{
\P_N(\uu,x) = \prod_{i=N}^4
\prod_{\vep=\pm 1 }\!(1+ \ux_i x^\vep) \,, \quad
{\bar \P}_{{\bar N}} (\uu,\by) 
= \prod_{j=1}^{{\bar N}}
\prod_{\eta=\pm 1} \! (1+ \ux_j{}^{-1} \by^\eta) \,, 
}
are invariant under $\frak{W}^{\S_2},{\overline \frak{W}}{}^{\S_2}$ and
\eqn\defPN{
N_{1\over 4} = 2 \, , \quad N_{1\over 2} = 3 \, , \qquad
{\bar N}_{1\over 4} = 3 \, , \quad {\bar N}_{1\over 2} = 2 \, .
}

For $t=\half$ the character may be calculated by expanding the factors 
$\P_{3}(s\uu,x)$ and ${\bar \P}_{2} (s^{-1}\uu,\by)$ in \charshort\ giving
\eqn\charBPS{\eqalign{
& \chi^{{1\over 2},{1\over 2}}_{(p;0,p,0;0,0)}(s;\uu;x,\by) \cr
&{} = s^{2p}P(s,x,\by) \!\!\!\! \sum_{a,b,c,d=0}^2 \!\! s^{a+b+c+d} \,
\chi_{j_a}(\by)\chi_{j_b}(\by)\chi_{j_c}(x)\chi_{j_d}(x) \, 
\chi_{(p-a,p-b,c,d)}(\uu) \, , \cr}
}
with the notation here
\eqn\deffa{
j_0 = j_2 = 0 \, , \qquad j_1 = \half \, .
}
For $x=1$, $\chi_{j_a}(1)= {2\choose a}$.
With the aid of \charW\ and \reflect\ this may be simplified further
\eqnn\charhalf
$$\eqalignno{
\chi^{{1\over 2},{1\over 2}}_{(p;0,p,0;0,0)}(s;\uu;x,\by) = {}&
s^{2p}P(s,x,\by) \!\!\!\! \sum_{0\le a \le b \le 2 \atop 0 \le d \le c \le 2}
\!\! s^{a+b+c+d}
\chi_{j_{ba}}(\by)\chi_{j_{cd}}(x)\, \chi_{(p-a,p-b,c,d)}(\uu) \, , \cr
j_{00}=j_{22} = j_{20} = {}& 1 \, , \qquad 
j_{10}=j_{21} = \half \, , \qquad j_{11} = 1 \, . & \charhalf \cr}
$$
Noting that $\P_{3}(\uu,1) {\bar \P}_{2}(\uu,1) = (u_1u_2)^{-2}\prod_{i=1}^4 (1+u_i)^2$ 
we may also straightforwardly obtain
\eqn\quarzerftuthro{ 
\chi^{{1\over 2},{1\over 2}}_{(p;0,p,0;0,0)}(s;\uu;1,1)/P(s,1,1)
\big|_{s\to 1} = \prod_{i=1}^{4}(1+\ux_i)^2
\chi_{(p-2,p-2,0,0)}(\uu)\, ,}
which immediately gives $2^8 d_{[0,p-2,0]}$ for the number of conformal
primary operators as in \fadho.{\foot{For $p=1$, the singleton
representation, this is zero but the counting of physical degrees of freedom
for fields satisfying equations of motion is different in this case.}}

For the special cases $p=1,2$ applying \charW\ leads to further reductions and 
gives the results \elemp\ and \zerotwozero. For $p=0$ from \charhalf\ we get
\eqn\charzero{\eqalign{
\chi^{{1\over 2},{1\over 2}}_{(0;0,0,0;0,0)}(s;\uu;x,\by) = {}&
P(s,x,\by)\big ( 1 - (s^2+s^6)\chi_{{1\over 2}}(x)\chi_{{1\over 2}}(\by)
+s^4 ( \chi_1(x) + \chi_1(\by) \big ) \cr
= {}& 1 \, , \cr}
}
in accord with $\B^{{1\over 2},{1\over 2}}_{[0,0,0](0,0)} = \I$, the trivial 
representation.

For $t=\quar$ in \charshort\ we may expand $\P_{2}(s\uu,x){\bar \P}_{3}(s^{-1}\uu,\by)$
in  a similar fashion as above to obtain, with the definitions \deffa,
\eqnn\charquar
$$\eqalignno{ 
\chi^{{1\over 4},{1\over 4}}_{(p+2q;q,p,q;0,0)}(s;\uu;x,\by) = {}& 
s^{2p+4q }P(s,x,\by)\cr
\noalign{\vskip-2pt}
& {} \times \!\!\!\!\sum_{a,b,c,d,e,f=0 }^{2}\!\!\!\!
s^{a+b+c+d+e+f} \chi_{j_a}(\by)\chi_{j_b}(\by)\chi_{j_c}(\by)
\chi_{j_d}(x)\chi_{j_e}(x)\chi_{j_f}(x) \cr
\noalign{\vskip-6pt}
& \qquad \qquad \quad {}\times 
\chi_{(p+2q-a,p+q-b+d,q-c+e,f)} (\uu)\,. & \charquar} 
$$
Except for $q=1$ there are no additional simplifications as in \charhalf.
Furthermore using $\P_{2}(\uu,1) {\bar \P}_{3} (\uu,1)
= (u_1u_2u_3)^{-2}  (1+u_2)^2(1+u_3)^2\prod_{i=1}^4 (1+u_i)^2 $
we have for $s\to 1$
\eqn\quarzerftutho{\eqalign{  
\chi^{{1\over 4},{1\over 4}}_{(p+2q;q,p,q;0,0)}& (s;\uu;1,1)/
P(s,1,1)\big|_{s\to 1}\cr & =
\prod_{i=1}^{4}(1+\ux_i)^2\sum_{a,b=0}^{2}
\left({\textstyle{2\atop a}}\right)\left({\textstyle{2\atop b}}\right)
\chi_{(p+2q-2,p+q+a-2,q+b-2,0)}(\uu)\, .}}
For $\ux_i\to 1$ this gives $2^8 \sum_{a,b}
\left({\textstyle{2\atop a}}\right)\left({\textstyle{2\atop b}}\right)
d_{[q-a,p+a-b,q+b-2]}$ for the number of conformal primary operators, which
may be shown to agree with the corresponding formula
for the $\quar$-BPS multiplet given in \fadho. 

When $q=0$, $\chi^{{1\over 4},{1\over 4}}$ may be reduced to 
$\chi^{{1\over 2},{1\over 2}}$ characters. Using in 
\charshort, from \defPP,
$\P_2(\uu,x){\bar P}_3(\uu,\by) = 
\prod_{\vep = \pm 1} ( 1 + u_2 x^\vep) \prod_{\eta=\pm 1} (1+ u_3{}^{-1} \by^\eta)
\, \P_3(\uu,x){\bar P}_2(\uu,\by)$
the action of $\frak{W}^{\S_4}$ may be simplified for a 
subgroup $\S_2 \times \S_2 \subset \S_4$, generated by the
permutations $(12),(34)$, to
\eqn\simqz{\eqalign{
\frak{W}^{\S_2\times \S_2} \big ( & 
{\ts \prod}_{\vep = \pm 1} ( 1 + s u_2 x^\vep) {\ts \prod}_{\eta=\pm 1} 
(1+ s u_3{}^{-1} \by^\eta) \, C_{(p,p,0,0)}(\uu)  \big ) \cr
= {}& \frak{W}^{\S_2\times \S_2} 
\big ( {\ts \sum_{r=0}^2 {2\choose r}} (-1)^r s^{2r} C_{(p+r,p+r,0,0)}(\uu)  \big ) \, , \cr}
}
since other factors in \charshort\ are invariant under this $\S_2 \times \S_2$.
Hence we obtain
\eqn\Chalf{
\chi^{{1\over 4},{1\over 4}}_{(p;0,p,0;0,0)}(s;\uu;x,\by)=
{\ts \sum_{r=0}^2 {2\choose r}} (-1)^r
\chi^{{1\over 2},{1\over 2}}_{(p+r;0,p+r,0;0,0)}(s;\uu;x,\by) \, ,
}
with a corresponding decomposition for $\B^{{1\over 4},{1\over 4}}_{[0,p,0](0,0)}$.

For the case $t=\quar,\bt=0$, $j=0$ \charshort\ is modified to
\eqn\charquart{\eqalign{
\chi^{{1\over 4},0}_{({1\over 2}(3k+2p+q);k,p,q;0,\bj)}(s;\uu;x,\by)  
= {}& s^{3k+2p+q}P(s,x,\by)\, {\bar \Q}(s^{-1}\uu,\by) \chi_{\bj}(\by)  \cr
&{}\times 
{\frak W}^{\S_4} \big (C_{(k+p+q,p+q,q,0)}(\uu) \P_{2}(s,\uu,x) \big ) \, .}
}
As before we may obtain an expansion
\eqn\quarzefy{\eqalign{
&\chi^{{1\over 4},0}_{({1\over 2}(3k+2p+q);k,p,q;0,\bj)}(s;\uu;x,\by)
\cr 
& {} =  s^{3k+2p+q} P(s,x,\by)\, \bar{\Q}(s^{-1}\uu,\by)  \chi_{\bj}(\by)\!
\!\sum_{a,b,c=0}^2 \!\! s^{a+b+c} \chi_{j_a}(x)\chi_{j_b}(x)\chi_{j_c}(x)\cr
\noalign{\vskip-6pt}
& \hskip 7cm {}\times
\chi_{(k+p+q,p+q+a,q+b,c)}(\uu)\,.}
}
The dimension of conformal primary states for the $(\quar,0)$ multiplet 
is obtained by factoring $P(s,1,1)$ out of the last expression and then setting 
$s=1,u_i=1$, using \dimsuf.
The result agrees with a comparable result in \fadho.

The forms exhibited in \quarzefy, \charquar\ and
\charhalf, along with \dimsuf, allow for easy determination of
the `blind' partition functions when $\ux_i,x,\by=1$. 
For the $t=\quar$ case we find from \quarzefy\ that
\eqnn\squarzer
$$\eqalignno{ &
\chi^{{1\over 4},0}_{({1\over 2}(3k+2p+q);k,p,q;0,\bj)}
(s;1,1,1,1;1,1)={\textstyle{1\over 12}}
s^{3k+2p+q}{(1+s)^7\over
(1-s)^4}(2\bj{+}1)(p{+1})(q{+1})(p{+}q{+}2)\cr
&\qquad\qquad\times\Big((k{+1})(k{+}p{+}2)(k{+}p{+}q{+}3)\cr
&\qquad\qquad\qquad+\big(5 (k{+1})(k{+}p{+}2)(k{+}p{+}q{+}3)-2
(k{+2})(k{+}p{+}3)(k{+}p{+}q{+}4)\big)\,s\cr
&\qquad\qquad\qquad+\big(5 (k{-1})(k{+}p)(k{+}p{+}q{+}1)-2
(k{-2})(k{+}p{-}1)(k{+}p{+}q)\big)\,s^2\cr
&\qquad\qquad\qquad+(k{-1})(k{+}p)(k{+}p{+}q{+}1)\,s^4\Big)\,, & \squarzer }
$$
from \charquar\ we may find directly for the $\quar$-BPS multiplet that
\eqnn\squarquar
$$\eqalignno{
&\!\!\!\!\!\!\!\!\!\!\!\!\!\!\chi^{{1\over 4},{1\over 4}}
_{(p+2q;q,p,q;0,0)}(s;1,1,1,1;1,1)
={\textstyle{1\over 12}}
s^{2p+4q}{(1+s)^3\over (1-s)^4}(p+1)\cr
&\times
\Big((q{+}1)(p{+}q{+}2)+2\big(q(p{+}q{+}1)-2\big)s+(q{-}1)(p{+}q)s^2\Big)\cr
&\times\Big((q{+}1)(p{+}q{+}2)(p{+}2q{+}3)\cr
&\qquad+{\textstyle{1\over 9}}\big(
2(q{+}2)(3p{+}3q{+}1)(3p{+}6q{+}5)+(q{-}3)(3p{+}3q{+}10)(3p{+}6q{+}5)
-140\big)s\cr
&\qquad+{\textstyle{1\over 9}}\big(
2(q{-}2)(3p{+}3q{+}5)(3p{+}6q{+}1)+(q{+}3)(3p{+}3q{-}4)(3p{+}6q{+}1)+140
\big)s^2\cr
&\qquad+(q{-}1)(p{+}q)(p{+}2q{-}1)s^3
\Big)\,, & \squarquar }
$$
and, finally, we have for the $\half$-BPS multiplet
from \charhalf\ that
\eqn\shalfhalf{\eqalign{
\chi^{{1\over 2},{1\over 2}}_{(p;0,p,0;0,0)}(s;1,1,1,1;1,1)
={}& {\textstyle{1\over 12}} {s^{2p}\over (1-s)^4} \big(p+2 + (p-2)s\big)\cr
&{} \times\big ((p{+}1)(p{+}2)(p{+}3)+3(p{-}2)(p{+}1)(p{+}3)s \cr
&\quad {}+ 3(p{-}3)(p{-}1)(p{+}2)s^2+(p{-}3)(p{-}2)(p{-}1)s^3 \big)\,.
}}
For $p=1$, $\chi^{{1\over 2},{1\over 2}}_{(1;0,1,0;0,0)}(s;1,1,1,1;1,1)
= 2s^2(3-s)/(1-s)^3$ which is the standard result for the fundamental
multiplet.

\noindent
{\bf Characters for semi-short multiplets}

Corresponding to the $(\eight,0)$ case for which
$\De=2+2j+{1\over 2}(3k+2p+q)$ then in \charunit\ we may obtain
an expansion in terms of $(\quar,0)$ characters by reducing the
dependence on $\uu$ to expressions of the form \charquart\
\eqnn\quarzerss
$$\eqalignno{
&\!\!\!\!\!\chi^{{1\over 8},0}_{(\De;k,p,q;j,\bj)}(s;\uu;x,\by) \cr
&{} =s^{2\De}P(s,x,\by){\bar \Q}(s^{-1}\uu,\by)\cr 
&\quad {}\times {\frak W}^{\S_4}{\frak W}^{\S_2}{\overline \frak{W}}{}^{\S_2}
\big((1+su_1x)C_{(k+p+q,p+q,q,0)}(\uu)
C_j(x)C_{\bj}(\by)\P_2(s\uu,x) \big)\cr
&{} = \chi^{{1\over 4},0}_{(\De;k,p,q;0,\bj)}(s;\ux;x,\by)\chi_j(x)+
\chi^{{1\over 4},0}_{(\De+{1\over 2};k+1,p,q;0,\bj)}(s;\uu;x,\by)
\chi_{j+{1\over 2}}(x) \,, & \quarzerss  }
$$
so that previous results \quarzefy\ for 
$\chi^{{1\over 4},0}_{(\De;k,p,q;0,\bj)}$
may be used although with different $\De$. 
Setting $j=-{1\over 2}$ gives \red\ for $\bt=0$.

For the $(\eight,\eight)$ semi-short multiplet with
$\De=2+j+\bj+k+p+q$ and $k-q=2(\bj-j)$ then a similar expansion
may be obtained from \charunit\ in terms of $(\quar,\quar)$ characters
for more general $\De,k,p,q$,

\vbox{
\eqnn\quarquarss
$$\eqalignno{ \!\!\!\!\!\!\!\!\!\!\!\!\!\!\!\!\!\!\!\!
\chi^{{1\over 8},{1\over 8}}_{(\De;k,p,q;j,\bj)}(s;\uu;x,\by)
= {} & \chi^{{1\over 4},{1\over 4}}_{(\De;k,p,q;0,0)}(s;\uu;x,\by)
\chi_j(x)\chi_{\bj}(\by)\cr
&{} + \chi^{{1\over 4},{1\over 4}}_{(\De+{1\over 2};k+1,p,q;0,0)}(s;\uu;x,\by)
\chi_{j+{1\over 2}}(x)\chi_{\bj}(\by) \cr
& {} + \chi^{{1\over 4},{1\over 4}}_{(\De+{1\over 2};k,p,q+1;0,0)}
(s;\uu;x,\by)\chi_j(x) \chi_{\bj+{1\over 2}}(\by) \cr &{} +
\chi^{{1\over 4},{1\over 4}}_{(\De+1;k{+1},p,q{+1};0,0)}(s;\uu;x,\by)
\chi_{j{+}{1\over 2}}(x)\chi_{\bj{+}{1\over 2}}(\by) \, . & \quarquarss }
$$}
For $j=\bj=-\half$ in \quarquarss\ we simply obtain \redtt.

Other cases considered in \fadho\ are for the $\C^{{1\over 2},{1\over 2}}$, 
$\C^{1,1}$ semi-short multiplets, which we now show
reduce to the $(\eight,\eight)$ case above.

For the $\C^{{1\over 2},{1\over 2}}$ semi-short multiplet with
$\De=2+2j+p$, $j=\bj$ and $k,q=0$ then $Q^{1}{}_{2},Q^{2}{}_{2},
{\bar Q}_{31},{\bar Q}_{41}$ are missing from the full Verma module \vermam\ 
so that the character formula becomes
\eqn\halfhalfssexp{\eqalign{
& \chi^{\C^{{1\over 2},{1\over 2}}}_{(\De;0,p,0;j,j)} (s;\uu;x,\by) 
=  s^{2\De}P(s,x,\by) \, \frak{W}^{\S_2}{\overline \frak{W}}{}^{\S_2} 
\Big ( C_j(x)C_{j}(\by) \, \frak{W}^{\S_4}\big ( X_p(s;\uu;x,\by) \big ) \Big ) \, , \cr
& X_p(s;\uu;x,\by) = C_{(p,p,0,0)}(\uu)\, {\ts \prod_{i=1}^2}
(1+s \ux_i x){\ts \prod_{k=3}^4} (1+s \ux_k{}^{-1}\by)\,
\P_3(s\uu,x) {\bar \P}_2(s^{-1}\uu,\by) \, . \cr}
}
The corresponding expression for 
$\chi^{{1\over 8},{1\over 8}}_{(\De;0,p,0;j,j)}(s;\uu;x,\by)$ obtained from 
\charunit\ is identical with \halfhalfssexp\ except that it involves
\eqn\symM{
\frak{W}^{\S_4}\big ( (1+s u_2 x^{-1})(1+s u_3{}^{-1} \by^{-1})
X_p(s;\uu;x,\by) \big )  \, .
}
Nevertheless this leads to the same result as \halfhalfssexp\ as a consequence of
\eqn\redM{\eqalign{
\frak{W}^{\S_2\times \S_2} \big ( & (1+s u_2 x^{-1})(1+s u_3{}^{-1} \by^{-1})
C_{(p,p,0,0)}(\uu)  \big ) = \frak{W}^{\S_2\times \S_2} 
\big ( C_{(p,p,0,0)}(\uu)  \big ) \cr
&{} = { (u_1 u_2)^{p+2} \over
(u_1-u_3)(u_1-u_4)(u_2- u_3)(u_2-u_4)} \, , \cr}
}
for $\S_2\times \S_2$ defined in the same fashion as in  \simqz. This then implies 
\eqn\Chalf{
\chi^{{1\over 8},{1\over 8}}_{(2+2j+p;0,p,0;j,j)}(s;\uu;x,\by) =
\chi^{\C^{{1\over 2},{1\over 2}}}_{(2+2j+p;0,p,0;j,j)}(s;\uu;x,\by) \, , 
}
demonstrating $\C^{{1\over 4},{1\over 4}}_{[0,p,0](j,\bj)}
\simeq \C^{{1\over 2},{1\over 2}}_{[0,p,0](j,\bj)}$, as was also noted in \fadho.

For the $\C^{1,1}$ multiplet, for which $\De=2+2j$ and $k,p,q=0$,
then $Q^i{}_2,\bQ_{k1}$ for all $i,k$ are removed from the operators
generating the Verma module and hence we must also drop the
contribution of $P_{21}$ arising from their anticommutator in \vermam. 
Thus \charunit\ must be modified in this case and takes the form
\eqn\oneoness{\eqalign{
& \chi^{\C^{1,1}}_{(2+2j;0,0,0;j,j)}(s;\uu;x,\by)\cr 
&{} = s^{4+4j}P(s,x,\by) \, \frak{W}^{\S_2}{\overline \frak{W}}{}^{\S_2}
\bigg ( \prod_{i=1}^4 (1+s \ux_i x)(1+s \ux_i{}^{-1}\by) 
( 1 -s^2x^{-1}\by^{-1}) C_j(x)C_j(\by)\bigg )  \,. \cr }
}
Expanding in powers of $x,\by$ and using the definitions \chis\ and
\conserved\ we easily obtain the succinct form
\eqn\oneone{
\chi^{\C^{1,1}}_{(2+2j;0,0,0;j,j)}(s;\uu;x,\by)  = \!\!
\sum_{0\le m,n\le4} \!\! \D_{j+{1\over 2}m,j+{1\over 2}n}(s,x,\by) \,
\chi_{(1^m 0^{4-m})}(\uu)\chi_{(1^{4-n}0^n)}(\uu)\, ,
}
which involves solely contributions corresponding to
conserved currents.
As a particular limit from \oneoness\ we have
\eqn\oneonesss{\eqalign{
\chi^{\C^{1,1}}_{(2+2j;0,0,0;j,j)}(1;1,1,1,1;x,\by)
={} &P(1,x,\by) (1+x)^2(1+x^{-1})^2(1+\by)^2(1+\by^{-1})^2\cr &
\times\big(\chi_{j{+1}}(x)\chi_{j{+1}}(\by) -\chi_{j+{1\over
2}}(x)\chi_{j+{1\over 2}}(\by)\big) \,, }}
so that, factoring
off $P(1,x,\by)$ and setting $x=\by=1$, the dimension of the
corresponding multiplet is $2^8(4j+5)$ which agrees with \fadho.
We also have
\eqn\oneones{
\chi^{\C^{1,1}}_{(2+2j;0,0,0;j,j)}(s;1,1,1,1;1,1) 
= s^{4+4j} \, {(1+s)^3\over (1-s)^4} \big ( 2j+1+(2j+4)s \big ) 
\big ( 2j+1 + 5s - (2j+4)s^2 \big ) \, .
}
To verify $\C^{{1\over 2},{1\over 2}}_{[0,0,0](j,j)} \simeq 
\C^{1,1}_{[0,0,0](j,j)}$, \fadho, we set $p=0$ in \halfhalfssexp\
and use
\eqn\Sym{
\frak{W}^{\S_4}\big ( C_{(0,0,0,0)}(\uu)\,  {\ts \prod_{i=3}^4}
(1+s \ux_i x^{-1})\, {\ts \prod_{k=1}^2} (1+s \ux_k{}^{-1}\by^{-1} )
\big ) = 1 - s^2 x^{-1}\by^{-1} \, ,
}
to ensure that it is then identical with \oneoness\ and so,
\eqn\Chalfone{
\chi^{\C^{{1\over 2},{1\over 2}}}_{(2+2j;0,0,0;j,j)}(s;\uu;x,\by)=
\chi^{\C^{{1},{1}}}_{(2+2j;0,0,0;j,j)}(s;\uu;x,\by) \, .
}

Other characters which are readily obtainable are those for
multiplets with mixtures of shortening and semi-shortening
conditions, denoted $\D^{s,\bar{t}},{\bar \D}^{t,{\bar s}}$ in
\fadho.  For the $(\quar,\eight)$ semi-short multiplet for which
$\Delta={1\over 2}(3k+2p+q)$ and $j=0,\,\bj={1\over 2}(k-q)-1$ then
the general formula gives,
\eqn\charmixed{\eqalign{
& \chi^{{1\over 4},{1\over 8}}_{(\De;k,p,q;0,\bj)}(s;\uu;x,\by) \cr
&{} =  
s^{2\De}P(s,x,\by) \, {\frak W}^{\S_4} {\overline \frak{W}}{}^{\S_2} 
\big ( (1+su_4{}^{-1} \by) C_{(k+p+q,p+q,q,0)}(\uu) C_{\bj}(\by) \, \P_2(s\uu,x)
{\bar \P}{}_3 (s^{-1}\uu,\by) \big ) \, . \cr}
}
This can be expanded as
\eqn\dquarquarss{\eqalign{
&\chi^{{1\over 4},{1\over 8}}_{(\De;k,p,q;0,\bj)}(s;\uu;x,\by) \cr
&{} =
\chi^{{1\over 4},{1\over 4}}_{(\De;k,p,q;0,0)}(s;\uu;x,\by)\chi_{\bj}(\by)+
\chi^{{1\over 4},{1\over 4}}_{(\De+{1\over 2};k,p,q+1;0,0)}(s;\uu;x,\by)
\chi_{\bj{+{1\over 2}}}(\by) \,, } }
where \charquar\ may be  used for $\chi^{{1\over 4},{1\over 4}}$ with an
extension to more general $\De,k,p,q$. 
For $\bj=-\half$ \dquarquarss\ reproduces \redt.

\newsec{Reduction of Characters in Different Limits}

Here we consider various limits for the $PSU(2,2|4)$ characters,
as obtained in sections 3 and 4, which restrict them to different
subgroups. These restrictions ensure that particular short and semi-short
multiplet characters survive to give non-zero contribution and also
that in most cases the expressions for the characters greatly simplify.
Furthermore the expression for the general long multiplet character in 
\charlong\ has factors $1+su_1 x^{-1}$ and $1+s u_4{}^{-1}\by^{-1}$ which are not
present in the characters for the different short multiplets that are given
by \charunit\ with \defQ\ when $t,\bt$ are non zero. These factors, which 
survive in particular limits, may then be set to zero to eliminate contributions
for multiplets which have $t$ and/or $\bt$ zero, and hence remove all long
multiplets. The result corresponds to the index in \mald.

We consider the different cases in turn making use of the analysis in
appendix A of the relevant subgroups of $\G_{t,\bt} \subset PSU(2,2|4)$,
which correspond to the various shortening conditions in section 
3 and are the symmetry groups for differing reduced sectors of the theory.
The limits are obtained by re-expressing the basic trace
$\tr ( s^{2D} \, u_1{}^{H_1+H_2+H_3} u_2{}^{H_2+H_3} u_3{}^{H_3}
\, x^{2J_3} \, \by^{2{\bar J}_3})$ in terms of different 
linear combinations of operators $\hD, \dots $ with corresponding variables 
$\hs,\dots$ and then requiring  appropriate variables to vanish 
when the character reduces to one for the reduced group $\G_{t,\bt}$.
For a variable $h$ which contributes to the trace in the form $h^{2\H}$,
where $\H$ has a positive semi-definite spectrum, the limit $h\to 0$
ensures that the trace is reduced to a subspace on which $\H$ has zero
eigenvalue. Furthermore for a long multiplet,
which is reducible to semi-short multiplets, the limit for suitable
cases gives zero, realising the index defined in \mald. In 
each case considered we first list $(t,\bt)$ and the residual group
$\G_{t,\bt}$ which is relevant for the particular limit. The limits 
are applied to $\chi^{t,\bt}$ in this case and also for 
$\chi^{{1\over 2},{1\over 2}}_{(p;0,p,0;0,0)}$ which is then restricted to 
$p=1$, corresponding to the fundamental or singleton representation 
$\F = \B^{{1\over 2},{1\over 2}}_{[0,1,0](0,0)}$.

\noindent
{\bf{(i)}} $t=\bt=\half$, $\G_{{1\over 2},{1\over 2}}=U(1)_D$

In this case we write the trace in the form
\eqn\halftrace{
\tr \big (h^{2\tilde \H} \, \hs^{\, 2D} \, \hu_1{}^{H_1} \hu_3{}^{H_3}
\, x^{2J_3} \, \by^{2{\bar J}_3} \big )\, , \quad 
{\tilde \H} = D - \half ( H_1+ 2H_2+H_3) \, ,
}
and then consider the limit $h\to 0$. Applying this to the 
character for the $\half$-BPS multiplet,
from \charshort\ with $t=\bt=\half$, we obtain
\eqn\halfhalfshv{\eqalign{ 
\lim_{h\to 0}& \chi^{{1\over 2},{1\over 2}}_{(p;0,p,0;0,0)} 
\big (h\, \hs;\uu_h;x,\by \big ) = \chi^{U(1)}_{p}(\hs)= \hs^{\, 2p}\,, \cr
&  \uu_h = ( h^{-1}\hu_1,h^{-1}\hu_2,h \hu_3,h \hu_4) \, , \ 
\hu_1\hu_2 = \hu_3\hu_4 = 1 \, .}  
}
Trivially we have, when $p=1$,
\eqn\fundhalf{
\chi^{U(1)}_{\F}(\hs)= \hs^{\, 2} \, .
}

\noindent
{\bf{(ii)}} $t=\bt=\quar$, $\G_{{1\over 4},{1\over 4}}= SU(2) \otimes U(1)_{H_+}$

The trace is now rewritten as
\eqn\quartrace{\eqalign{
& \tr \big (h^{2\H_0} \, \bh^{\, 2{\bar \H}_0} \, \hu_2{}^{H_2} \, u^{H_+} 
\, x^{2J_3} \, \by^{2{\bar J}_3} \big )\, , \quad H_+ = H_1+H_2+H_3 \, , \cr
& {\H}_0 = D - \half ( 3H_1+ 2H_2+H_3) \, ,\quad  
{\bar \H}_0 = D - \half ( H_1+ 2H_2+ 3H_3) \, , \cr}
}
and we now consider the limit $h,\bh \to 0$. For the $\quar$-BPS multiplet
character
\eqn\quarquarshv{\eqalign{
\lim_{h,\bh\to 0}
& \chi^{{1\over 4},{1\over 4}}_{(p+2q;q,p,q;0,0)} 
\big ( h\bh\, \hs ; \uu_{h,\bh} ; x,\by \big ) 
= \chi^{U(2)}_{(p,q)}(u;\hu_2,\hu_3) = u^{p+2q} \, \chi_{(p,0)}(\hu_2,\hu_3)  \,, \cr
& \uu_{h,\bh} = \big (h^{-3}\bh^{-1}u ,h\bh^{-1} \hu_2, h\bh^{-1} \hu_3, 
h \bh^3 u^{-1} \big )\, , \quad \hu_2\hu_3 = 1 \, , \cr}
}
which is a $U(2)$ character. Applying this limit to
$\chi^{{1\over 2},{1\over 2}}_{(p;0,p,0;0,0)}$ gives the same result with $q=0$.
In particular for $p=1$
\eqn\quarF{
\chi^{U(2)}_{\F}(u,\hu_2,\hu_3) = u ( \hu_2 + \hu_3 ) \, .
}

\noindent
{\bf(iii)} $t=\quar, \, \bt=0$, $\G_{{1\over 4},0}= SU(2|3)$

The trace is re-expressed as
\eqn\qztrace{
\tr \big (h^{2\H_0} \, \hs^{\, 2\hD}\, \hu_2{}^{H_2+H_3} \, \hu_3{}^{H_3}  
\, x^{2J_3} \, \by^{2{\bar J}_3} \big )\, ,  \quad
\hD = {\ts {3\over 2}} D -  {\ts{1\over 4}} ( 3H_1+ 2H_2+H_3) \, ,
}
with $\H_0$ as in \quartrace. The limit $h\to 0$ reduces this trace to a $SU(2|3)$ 
character. Applying this limit to $\chi^{{1\over 4},0}$ as in \charquart\ gives
\eqn\quarzeshv{\eqalign{ 
\lim_{h \to 0}&\chi^{{1\over 4},0}_{({1\over 2}(3k+2p+q);k,p,q;0,\bj)}
\big (h\, \hs^{3\over 2} ;\uu_h ;x,\by \big ) = 
\chi^{SU(2|3)}_{({1\over 2}(3k+2p+q),p,q,\bj)}(\hs;{\hat \uu};\by) \, , \cr
& \uu_h = \big ((h \hs^{1\over 2})^{-3}, h \hs^{1\over 2} {\hat \uu} \big ) \, , 
\quad {\hat \uu} = (\hu_2,\hu_3,\hu_4) \, , \quad \hu_2\hu_3\hu_4 = 1 \, , 
} }
where
\eqn\charsu{
\chi^{SU(2|3)}_{(\kappa,p,q,\bj)}(\hs;{\hat \uu};\by)
= \hs^{2\kappa}\, \chi_{(p+q,q,0)}({\hat \uu} )\, \chi_{\bj}(\by) \, 
{\ts \prod}_{i=2}^4 {\ts \prod}_{\eta=\pm 1} 
\big (1+ \hs\, \hu_i{}^{-1}\by^\eta \big )\,.
}

The same limit for $\chi^{{1\over 2},{1\over 2}}_{(p;0,p,0;0,0)}$ gives
\eqn\limchartt{
\lim_{h\to 0}\chi^{{1\over 2},{1\over 2}}_{(p;0,p,0;0,0)} 
\big (h\, \hs^{3\over 2} ;\uu_h ;x,\by \big ) =  
\chi^{SU(2|3)}_{{\rm short}, p}(\hs;{\hat \uu};\by) \, ,
}
where
\eqn\quarquarshvot{\eqalign{ 
\chi^{SU(2|3)}_{{\rm short}, p}(\hs;{\hat \uu};\by)
= {}& \hs^{2p}\, {\frak W}^{\S_3}
\big(C_{(p,0,0)}({\hat \uu})\, {\ts \prod}_{\eta=\pm 1 }
(1+ \hs \, \hu _2{}^{-1}\by^\eta)\big) \cr
= {}&  \hs^{\, 2p}\, \big(\chi_{(p,0,0)}({\hat \uu})+
\hs\, \chi_{(p-1,0,0)}({\hat \uu}) \, \chi_{1\over 2}(\by) + \hs^{\,2}
\chi_{(p-2,0,0)}({\hat \uu}  )\big) \, , \cr}
}
and the $SU(3)$ characters are all given by \pzerorep. 
Here the associated $SU(2|3)$ representation satisfies a shortening condition.
In the particular case of  $p=1$
\eqn\quarone{
\chi^{SU(2|3)}_{\F}(\hs;{\hat \uu};\by) =  
\hs^{2}\, \big( \hu_2 + \hu_3 + \hu_4 + \hs \, ( \by + \by^{-1}) \big) \, .
}

\noindent
{\bf(iv)} $t=\quar, \, \bt = {1\over 8}$, $\G_{{1\over 4},{1\over 8}}= 
SU(1|2)\otimes U(1)_{H_+}$

The trace is now rewritten as
\eqn\qetrace{\eqalign{
& \tr \big (h^{2\H_0} \, \bh^{\, 2{\bar \H}} \, \hs^{\,2\hD}\, \hu_2{}^{H_2} \, u^{H_+} 
\, x^{2J_3} \big )\, , \quad \hD = D + 2\bJ_3 - \half (H_1-H_3) \, , \cr
& {\bar \H} = D - 2 \bJ_3 - \half ( H_1+ 2H_2+ 3H_3) \, , \cr}
}
with $\H_0,H_+$ defined in \quartrace. With the result in \charmixed,
and requiring the necessary conditions on $\De$ from \semis\ and \quars, we have
\eqn\quarquarshvo{\eqalign{
\lim_{h,\bh\to 0} & \chi^{{1\over 4},{1\over 8}}_{({1\over 2}(3k+2p+q);k,p,q;0,\bj)}
\big (h\bh\, \hs;\uu_{h,\bh} ;x,\bh{}^{-2} \hs^2 \big ) = u^{k+p+q+1} 
\chi^{SU(1|2)}_{(2k+p,p)}(\hs;\hu_2,\hu_3) \, , \cr
& \quad \uu_{h,\bh} = \big ( h^{-3}\bh^{-1} \hs^{-1} u , h\bh^{-1} 
\hs \hu_2 , h \bh^{-1} \hs \hu_3,
h \bh^3 \hs^{-1} u^{-1} \big ) \, , \quad \hu_2\hu_3 = 1 \, ,  }
}
with
\eqn\charlim{
\chi^{SU(1|2)}_{(\kappa,p)}(\hs;\hu_2,\hu_3)  
= \hs^{\, 2\kappa} \, \chi_{(p,0)} (\hu_2,\hu_3) \, {\ts \prod}_{i=2,3} 
( 1+ \hs^2 \hu_i{}^{-1} ) \, .
}
For $\chi^{{1\over 4},0}$ as $\bh\to 0$, with $\bar \de$ as in \defdd,
\eqn\quarzero{
\chi^{{1\over 4},0}_{({1\over 2}(3k+2p+q);k,p,q;0,\bj)}
\big (h\bh\, \hs;\uu_{h,\bh} ;x,\bh{}^{-2} \hs^{\, 2} \big )
\sim {\bh}^{2\bar \de} (1+u)\,  u^{k+p+q+1} 
\chi^{SU(1|2)}_{(2k+p,p)}(\hs;\hu_2,\hu_3) \, . 
}
Even for ${\bar \de}=0$ we may set $u=-1$ to remove all $\chi^{{1\over 4},0}$ 
contributions.

Applying this limit for $\chi^{{1\over 2},{1\over 2}}_{(p;0,p,0;0,0)}$ gives
\eqn\quarei{\eqalign{
\lim_{h,\bh\to 0} \chi^{{1\over 2},{1\over 2}}_{(p;0,p,0;0,0)}
\big (h\bh \, \hs;\uu_{h,\bh} ;x,\bh{}^{-2} \hs^{\,2} \big ) 
= {} & u^p \, \chi^{SU(1|2)}_{{\rm short}, p}(\hs;\hu_2,\hu_3) \, , \cr
\chi^{SU(1|2)}_{{\rm short},p}(\hs;\hu_2,\hu_3) =  
\hs^{\, 2p} \big ( \chi_{(p,0)}(\hu_2,\hu_3)&{} + \hs^{\, 2}\chi_{(p-1,0)}(\hu_2,\hu_3)
\big ) \, . \cr}
}
For $p=1$
\eqn\fund{
\chi^{SU(1|2)}_\F(\hs;\hu_2,\hu_3) = \hs^{\,2} ( \hu_2 + \hu_3 + \hs^{\,2} ) \, .
}

\noindent
{\bf(v)} $t=\bt=\eight$, $\G_{{1\over 8},{1\over 8}}= PSU(1,1|2) \ltimes U(1)_{H_-}
\otimes U(1)_{H_+}$

The trace here has the form
\eqn\eighttrace{\eqalign{
& \tr \big (h^{2\H} \, \bh^{\, 2{\bar \H}} \, \hs^{\,2\hD} \, \hu_2{}^{H_2} \, 
u^{\, H_+} \, v^{\, H_-} \big )\, , \quad H_+ = H_1+H_2+H_3 \, , \  
H_- = H_1 - H_3 \, , \cr
& {\H} = D - 2J_3 - \half ( 3H_1+ 2H_2+H_3) \, ,\quad  
{\bar \H}_0 = D - 2 \bJ_3 - \half ( H_1+ 2H_2+ 3H_3) \, , \cr
& \hskip 3cm \hD = D + J_3 + \bJ_3 \, , \cr}
}
and we consider the double limit $h,\bh\to 0$.
Assuming both conditions on $\De$ in \semis\ are satisfied then
the result for $\chi^{{1\over 8},{1\over 8}}$ is
\eqnn\quarquarshvn
$$\eqalignno{
& \lim_{h,\bh\to 0} \chi^{{1\over 8},{1\over 8}}_{(\De;k,p,q;j,\bj)}
\big (h\bh\, \hs;\uu_{h,\bh} ;h^{-2}\hs,\bh{}^{-2} \hs \big ) = u^{k+p+q+2} 
\chi^{PSU(1,1|2)\ltimes U(1)}_{(k+p+q+2j+2\bj+4,p,k-q)}(\hs;\hu_2,\hu_3;v) \, , \cr
&\ \  \uu_{h,\bh} = \big ( h^{-3}\bh^{-1} u v , h\bh^{-1} 
v^{-1} \hu_2 , h \bh^{-1} v^{-1} \hu_3, h \bh^3 u^{-1} v \big ) \, , 
\quad \hu_2\hu_3 = 1 \, ,  & \quarquarshvn }
$$
where\foot{This result is equivalent to a combination of formulae in \beis.}
\eqn\shorchar{
\chi^{PSU(1,1|2)\ltimes U(1)}_{(\kappa,p,m)}(\hs;\hu_2,\hu_3;v) = v^m \, 
{\hs^{\, 2\kappa } \over 1 - \hs^{\, 4}} \, \chi_{(p,0)}(\hu_2,\hu_3) \, 
\prod_{i=2}^3 \prod_{\ep=\pm 1} ( 1 + \hs^2 v^\ep \hu_i ) \, .
}
For a general long multiplet as $h,\bh\to 0$
\eqn\longeight{\eqalign{
& \chi_{(\De;k,p,q;j,\bj)}
\big (h\bh\, \hs;\uu_{h,\bh} ;h^{-2}\hs,\bh{}^{-2} \hs \big ) \cr
&{} \sim
h^{2\de} \bh{}^{2\bar \de} \, (1+uv) (1+uv^{-1}) \, u^{k+p+q+2}
\chi^{PSU(1,1|2)\ltimes U(1)}_{(\De + j+\bj+2,p,k-q)}(\hs;\hu_2,\hu_3;v) \, , \cr}
}
with $\de,{\bar \de}$ defined in \defdd. At the unitarity threshold,
$\de={\bar \de}=0$, the long multiplet character in this case vanishes if
$uv=-1$ or $uv^{-1}=-1$. If both $uv=uv^{-1}=-1$ then $ \chi^{{1\over 8},0}$
and $ \chi^{0,{1\over 8}}$ characters vanish in this limit for $\de = 0$ and
${\bar \de} = 0$ respectively.

The $\half$-BPS multiplet character in this limit satisfies
\eqn\quarquarshvnp{
\lim_{h,\bh\to 0}
\chi^{{1\over 2},{1\over 2}}_{(p;0,p,0;0,0)}
\big (h\bh \, \hs;\uu_{h,\bh} ;h^{-2}\hs,\bh{}^{-2} \hs \big ) 
= u^p \, \chi^{PSU(1,1|2)\ltimes U(1)}_{{\rm short},p}(\hs;\hu_2,\hu_3;v) \, ,
}
where
\eqn\chish{\eqalign{
\chi^{PSU(1,1|2)\ltimes U(1)}_{{\rm short},p}(\hs;\hu_2,\hu_3;v) = 
{\hs^{\, 2p} \over 1 - \hs^4} \, & \big ( \chi_{(p,0)}(\hu_2,\hu_3)  + \hs^{\, 2}
\chi_{{1\over 2}}(v) \,\chi_{(p-1,0)}(\hu_2,\hu_3) \cr
\noalign{\vskip -3pt}
&\quad {} + \hs^{\,4} \chi_{(p-2,0)}(\hu_2,\hu_3) \big ) \, . \cr}
}
For $p=1$
\eqn\pone{
\chi^{PSU(1,1|2)\ltimes U(1)}_{\F}(\hs;\hu_2,\hu_3;v) =
{\hs^{\, 2} \over 1 - \hs^{\,4}} \big ( \hu_2 + \hu_3 + \hs^{\, 2}
(v + v^{-1} ) \big ) \, .
}

\noindent
{\bf(vi)} $t=\eight$, $\bt=0$, $\G_{{1\over 8},0}= PSU(1,2|3) \ltimes U(1)_{R}$

The trace here has the form
\eqn\eightztrace{
\tr \big (h^{2\H} \, \hs^{\,2\hD} \, \hu_2{}^{H_2+H_3} \, \hu_3{}^{H_3} 
\by^{2\bar J_3} \, u^R \big )\, , \quad R = 3H_1+2H_2+H_3 \, , \  \hD = D + J_3 \, ,
}
with $\H$ as defined in \eighttrace.
Applying the limit $h\to 0$ to $\chi^{{1\over 8},0}$ gives
\eqn\quarquarshvnew{ \eqalign{ 
\lim_{h \to 0}&\chi^{{1\over 8},0}_{(2+2j+{1\over 2}(3k+2p+q);k,p,q;j,\bj)}
\big (h \, \hs  ;\uu_h ; h^{-2} \hs,\by \big ) \cr
&{} = 
\chi^{PSU(1,2|3)\ltimes U(1)}_{({1\over 2}(3k+2p+q)+3j+3,3k+2p+q+3;p,q;\bj)}
(\hs,u;{\hat \uu};\by) \, , \cr
& \uu_h = \big (h^{-3} u^3, h u^{-1} {\hat \uu} \big ) \, , 
\quad {\hat \uu} = (\hu_2,\hu_3,\hu_4) \, , \quad \hu_2\hu_3\hu_4 = 1 \, ,}
}
where
\eqn\chared{\eqalign{
\chi^{PSU(1,2|3)\ltimes U(1)}_{(\kappa,r;p,q;\bj)}(\hs,u;{\hat \uu};\by)
= {}& {\hs^{\, 2\kappa} \, u^r\over (1 - \hs^{\,3} \by ) (1 - \hs^{\,3} \by^{-1} )}\, 
\chi_{(p+q,q,0)}({\hat \uu})\, \chi_{\bj}( \by)\cr
& {} \times
\prod_{i=2}^4 \Big ( \big (1+ \hs^{\, 2} u^{-1} \hu_i \big ) \, 
{\ts \prod}_{\eta=\pm 1} \big (1+ \hs \,u\, \hu_i{}^{-1}\by^\eta\big ) \Big ) \, ,
} }
where $\chi_{(p+q,q,0)}({\hat \uu})$ is a $SU(3)$ character. For a long 
multiplet for $h\to 0$
\eqn\eightlong{
\chi_{(\De;k,p,q;j,\bj)}
\big (h \, \hs  ;\uu_h ; h^{-2} \hs,\by \big ) 
\sim h^{2\de} \, (1+u^3) \,
\chi^{PSU(1,2|3)\ltimes U(1)}_{(\De+j+1,3k+2p+q+3;p,q;\bj)}
(\hs,u;{\hat \uu};\by) \, , 
}
so that $u=-1$ removes such contributions even at threshold $\de=0$.

For the $\half$-BPS character we find in this limit
\eqn\halfred{
\lim_{h \to 0}\chi^{{1\over 2},{1\over 2}}_{(p;0,p,0;0)}
\big (h \, \hs  ;\uu_h ; h^{-2} \hs ,\by \big ) 
= \chi^{PSU(1,2|3)\ltimes U(1)}_{{\rm short},p}(\hs,u;{\hat \uu};\by) \, ,
}
for
\eqnn\halfhalfshvnew
$$\eqalignno{
& \chi^{PSU(1,2|3)\ltimes U(1)}_{{\rm short},p}(\hs,u;{\hat \uu};\by) \cr
&\quad {} =  {(\hs \, u)^{2p}\over (1-\hs^{\, 3} \by) (1-\hs^{\, 3} \by^{-1} )}
\, {\frak W}^{\S_3}
\bigg(C_{(p,0,0)}({\hat \uu})\prod_{j=3,4}
(1+ \hs^{\, 2} u^{-1} {\hat \ux}_j ) 
\prod_{\eta=\pm 1} (1+ \hs\ u\,\ux_2{}^{-1}\by^\eta)\bigg) \cr
&\quad {}=  {(\hs \, u)^{2p}\over (1-\hs^{\, 3} \by) (1-\hs^{\, 3} \by^{-1} )}\,
\sum_{a=0}^2 \ (\hs \, u)^a \, \chi_{j_a}(\by) \cr
& \qquad \qquad \qquad {} \times \big ( \chi_{(p-a,0,0)}({\hat \uu})
+ \hs^{\, 2} u^{-1} \chi_{(p-a,1,0)}({\hat \uu}) 
+ \hs^{\, 4} u^{-2} \chi_{(p-a-1,0,0)}({\hat \uu}) \big ) \, , & \halfhalfshvnew \cr}
$$
using the notation in \deffa.
To calculate the $SU(3)$ characters in \halfhalfshvnew\ we may use \pzerorep\ and 
\eqn\ponesrep{
\chi_{(p,1,0)}({\hat \uu})= {\ts \sum}_{i=2}^4 {\hat \ux}_i{}^{-1} \,
\chi_{(p{-1},0,0)}({\hat \uu} ) - \chi_{(p{-2},0,0)}({\hat \uu} )\, .
}
When $p=1$, using $\chi_{(-1,0,0)}=\chi_{(-2,0,0)}, \ \chi_{(-1,1,0)} = -1$,
we obtain for the reduced character for the fundamental representation,
\eqn\fundea{\eqalign{
& \chi^{PSU(1,2|3)\ltimes U(1)}_{\F}(\hs,u;{\hat \uu};\by) \cr
&\ \ {} = { (\hs u)^2 \big ( {\ts \sum}_{i=2}^4 {\hat \ux}_i  + \hs^{\,2} u^{-1}
{\ts \sum}_{i=2}^4 {\hat \ux}_i{}^{-1}  + \hs \, u ( \by + \by^{-1} ) 
+ \hs^{\,4} u^{-2} ( 1 - u^3) \big ) \over 
(1-\hs^{\, 3} \by)\,  (1-\hs^{\, 3} \by^{-1} )}  \, . }
}
For $u=-1$, corresponding to the index, and also taking ${\hat \ux}_2 = v, \,
{\hat \ux}_3=w/v, \, {\hat \ux}_4=1/w$, this is identical with the result 
in \mald\ obtained by direct calculation of all contributions to the trace.\foot{
Comparing with (5.3) of \mald\ we should take $\hs^{\, 3} = x^2$, 
$\by =e^\zeta$, $u^3=x \, e^{{1\over 2}(\mu_1+\mu_2+\mu_3)}$, 
$\hu_2= e^{{1\over 3}(2\mu_1-\mu_2-\mu_3)}$, 
$\hu_3= e^{{1\over 3}(2\mu_2-\mu_1-\mu_3)}$,
$\hu_4= e^{{1\over 3}(2\mu_3-\mu_1-\mu_2)}$.}

\newsec{Free Fields}

The fundamental fields for $\N=4$ supersymmetric Yang Mills are
six scalars $X_r$, four chiral fermions (gauginos) $\lambda_{i\alpha},
{\bar \lambda}^{i}{}_{\! \dal}$ and gauge field strengths 
$F_{\alpha\beta}=F_{\beta\alpha}, {\bar F}_{\smash{\dal\dbe}}=
{\bar F}_{\smash{\dbe\dal}}$, all belonging to the adjoint representation
of the relevant gauge group and to the $[0,1,0],[1,0,0],[0,0,1]$
and trivial $SU(4)_R$ representations respectively.
By using $SU(4)$ gamma matrices it is convenient to let $X_r \to \vphi_{ij}
= - \vphi_{ji}$. The character for this representation is given by \elemp\
and the non zero action of the supercharges on these fields, at the origin $x=0$,
is given by
\eqn\QSfree{\eqalign{
\big [ Q^i{}_{\! \alpha} , \vphi_{kl} \big ]  = {}& 2\de^i{}_{\! [k}
\lambda_{l]\alpha} \, , \qquad
\big [ \bQ_{\smash {i\dal}}  , \vphi_{kl} \big ]  = \vep_{ijkl}
{\bar \lambda}^{j}{}_{\! \dal} \, , \qquad \ \ \, 
\big \{ Q^i{}_{\! \alpha} , \lambda_{j\beta} \big \} = \de^i{}_{\! j} 
F_{\alpha\beta} \, , \cr
\big \{ \bQ_{\smash {i\dal}} , \lambda_{j\al} \big \} = {}& 2i \,\pr_{\al\dal}
\vphi_{ij} \, , \quad \ \big \{ Q^i{}_{\! \alpha} , 
{\bar \lambda}^{j}{}_{\! \smash \dal} \big \} = \vep^{ijkl} i 
\pr_{\al\dal} \vphi_{kl} \, , \quad
\big \{ \bQ_{\smash {j\dal}} , {\bar \lambda}^i{}_{\! \smash \dbe} \big \}
= \de^i{}_{\! j} {\bar F}_{\smash{\dal\dbe}} \, , \cr
\big \{ S_i{}^{\!\beta} , \lambda_{j\al} \big \} ={}& 4 \de_{\al}{}^{\!\be}
\vphi_{ij} \, , \qquad \ \big \{ \bS{}^{i\dal} ,
{\bar \lambda}^{j}{}_{\! \smash \dbe} \big \} =
- 2 \de^\dal{}_{\! \smash \dbe}\vep^{ijkl} \vphi_{kl}  \, , \cr
\big [ \bQ_{\smash {i\dal}}  , F_{\al\be} \big ] = {}& 2 i \, 
\pr_{(\al \dal} \lambda_{i \be)} \, , \quad \big [ Q^i{}_{\! \alpha} ,
{\bar F}_{\smash{\dal\dbe}} \big ] = 2 i \, \pr_{\al( \dal} 
{\bar \lambda}^{i}{}_{\! \smash \dbe)} \, ,\cr
\big [  S_i{}^{\!\alpha} , F_{\beta\ga} \big ] = {}& 
8 \de_{(\beta}{}^{\!\al}  \lambda_{i \gamma)} \qquad  \ \, 
\big [ \bS{}^{i\dal} , {\bar F}_{\smash{\dbe\dga}} \big ]
= - 8\de^\dal{}_{\! \smash {(\dbe}} {\bar \lambda}^{i}{}_{\! \dga)} \, .  \cr}
}
As is well known the algebra closes subject to the equations of motion
$\pr^{\dal\al}\lambda_{i\alpha}=0, \pr^{\dal\al}F_{\alpha\beta}=0$, 
and their conjugates, which also imply $\pr^2 \vphi_{kl} = 0$.

The basic gauge singlet operators are formed by products of
multiple traces over adjoint indices of products of the fundamental fields 
with derivatives. These form the various possible supermultiplets
each of which arises from a unique highest weight operator 
together with descendants. For the various short multiplets it
is possible to restrict to just those elementary fields 
annihilated by the relevant supercharges as described in section 3.
Below we list in turn the crucial cases for different possible $t,\bt$ the
associated supercharges which have a trivial action and then
the remaining subgroup $\G_{t,\bt}\subset PSU(2,2|4)$. 
Following this the corresponding  kernel in terms
of the fundamental fields, which is easily determined from \QSfree\ and forms
a representation of $\G_{t,\bt}$, is displayed. 

\noindent
$t={1\over 8}$, $Q^1{}_{\! 2}, S_1{}^{\!2}$, $PSU(1,2|3) \ltimes U(1)_{R}$;
\eqn\ZYX{
\pr_{11}^{n_1} \pr_{12}^{n_2}(Z, \ Y, \ X, \ \lambda_{i 1}, \
{\bar \lambda}^1{}_{\! \smash \dal}, \ F_{11} ) \, , \quad
i=2,3,4, \, \dal=1,2, \, n_1,n_2 = 0,1,2,\dots \, ,
}
where
\eqn\defZYX{
Z= \vphi_{34}\, , \ Y=\vphi_{42} \, , \ X= \vphi_{23} \, , \quad
\pr_{11} {\bar \lambda}^1{}_{\! 2} = \pr_{12} {\bar \lambda}^1{}_{\! 1} \, .
}
Derivatives are present in \ZYX\ since
$[Q^1{}_{\! 2}, P_{1\dal}] = [S_1{}^{\!2}, P_{1\dal}] =0$. The weights
associated with each field are listed in Table 1, along with their elementary
contribution to the character which is identified with a particular letter,
up to identifications, for later application in partition functions.
The Cartan subalgebra for $PSU(1,2|3)$ is $(\hD;H_2,H_3;\bJ_3)$, with 
$\hD=D+J_3$ and the $U(1)$ generator $R = 3H_1+2H_2+H_3$ and the associated
supercharges are $Q^i{}_{\! 1}, S_j{}^{\!1}, \bQ_{{\smash {j\dbe}}}, 
\bS^{i\dal}$, $i,j=2,3,4$.
$Z,Y,X$ form a $3$ representation and $\lambda_{i1}, \, i=2,3,4$ a 
$\bar 3$ representation of $SU(3)$, their $SU(3)$ weights $(H_2,H_3)$ can
be read off from Table 1.

\noindent
$t=\bt={1\over 8}$, $Q^1{}_{\! 2}, S_1{}^{\!2}, \bQ_{41}, \bS^{41}$,
$PSU(1,1|2)\ltimes U(1)_{H_-} \otimes U(1)_{H_+}$;
\eqn\ZYd{
\pr_{12}^n ( Z, \ Y, \ \lambda_{4 1}, \ {\bar \lambda}^1{}_{\! 2} ) \, ,
\quad n =0,1,2, \dots \, .
}
The Cartan generators for $PSU(1,1|2)$ are here $\hD=D+J_3+\bJ_3, H_2$
and $H_- = H_1-H_3$, $H_+ = H_1 + H_2 +H_3$, and the algebra contains supercharges
$Q^i{}_{\! 1}, S_j{}^{\!1}, \bQ_{j2}, \bS^{i2}$, $i,j=2,3$.

\noindent
$t={1\over 4}$, $Q^1{}_{\! \alpha}, S_1{}^{\! \beta}$, $SU(2|3)$;
\eqn\Zquar{
Z \, , \ \ Y \, , \ \ X \, , \ \ {\bar \lambda}^1{}_{\! \dal} \, .
}
The Cartan generators for $SU(2|3)$ are $J_3,H_2,H_3$ and
$\hD= {3\over 2} D - {1\over 4}(3H_1+2H_2+H_3)$ with supercharges
$\bQ_{\smash{j\dbe}}, \bS^{i \dal}$, $i,j=2,3,4$.

\noindent
$t={1\over 4}, \bt ={1\over 8}$, $Q^1{}_{\! \alpha}, S_1{}^{\! \beta}, \bQ_{41}, 
\bS^{41}$, $SU(1|2) \otimes U(1)_{H_+}$;
\eqn\ZYfe{
Z \, , \ \  Y \, , \ \ {\bar \lambda}^1{}_{\! 2} \, .
}
The Cartan generator for $SU(1|2)$ are $H_2$, $\hD = D + 2 \bJ_3 - \half H_-$,
and $H_\pm$ are  as previously and the supercharges are
$\bQ_{\smash{j2}}, \bS^{i2}$, $i,j=2,3$.
 
\noindent
$t=\bt={1\over 4}$, $Q^1{}_{\! \alpha}, S_1{}^{\! \beta}, \bQ_{4\dal}, 
\bS^{4\dbe}$, $SU(2) \otimes U(1)_{H_+}$;
\eqn\ZY{
Z \, , \ \  Y \, .
}
The Cartan generator for $SU(2)$, for which $(Z,Y)$ form doublet, is $H_2$
and $H_+$ is as previously. 

\noindent
$t=\bt={1\over 2}$, $Q^i{}_{\! \alpha}, S_j{}^{\! \beta}, \bQ_{k\dal},
\bS^{l\dbe}$, $i,j=1,2, \ k,l=3,4$, $U(1)_D$;
\eqn\ZZZ{
Z \, .
}

\vbox{
\noindent Table 1
\nobreak
\vskip 6pt

\vbox{\tabskip=0pt \offinterlineskip
\halign{&\vrule# &\strut \ \hfil#\  \cr
\noalign{\vskip 3pt}
&\multispan7\hrulefill& \cr
height2pt&\omit&&\omit&&\omit&&\omit&\cr
&\ Field  \ \hfil   && \ \ $(\Delta;H_1,H_2,H_3;J_3,{\bar J}_3)$ \ \ &&
$s^{2\Delta}u_1^{H_1+H_1+H_3}u_2^{H_2+H_3}u_3^{H_3}x^{2J_3}\by^{2{\bar J}_3}$\hfil
&& \ Letter \ \hfil &\cr
height2pt&\omit&&\omit&&\omit&&\omit&\cr
\noalign{\hrule}
height2pt&\omit&&\omit&&\omit&&\omit&\cr
& \ $Z$  \ \hfil && \ $(1;0,1,0;0,0)$ \hfil &
& \ $s^2u_1u_2$ \ \hfil &&\ $z$ \ \hfil &\cr
& \ $Y$  \ \hfil&& \ $(1;1,-1,1;0,0)$ \hfil &
& \ $s^2u_1u_3$ \ \hfil &&\ $y$ \ \hfil &\cr
& \ $X$  \ \hfil && \ $(1;1,0,-1;0,0)$ \hfil &
& \ $s^2 u_1u_4 $ \hfil &&\ $x$ \ \hfil &\cr
height2pt&\omit&&\omit&&\omit&&\omit&\cr
& \ $\lambda_{41}$ \ \hfil&& \ $({3\over 2};0,0,1;{1\over 2},0)$ 
\hfil  & & \ $s^3u_4{}^{-1} x$  \ \hfil &&\ $b$ \ \hfil &\cr
& \ $\lambda_{31}$ \ \hfil&& \ $({3\over 2};0,1,-1;{1\over 2},0)$ 
\hfil  & & \ $s^3u_3{}^{-1} x$  \ \hfil &&\ $c$ \ \hfil &\cr
& \ $\lambda_{21}$ \ \hfil&& \ $({3\over 2};1,-1,0;{1\over 2},0)$ 
\hfil  & & \ $s^3u_2{}^{-1} x$  \ \hfil &&\ $d$ \ \hfil &\cr
height2pt&\omit&&\omit&&\omit&&\omit&\cr
& \ ${\bar \lambda}{}^1{}_{2}$ \ \hfil&& \ $({3\over 2};1,0,0;0,{1\over 2})$ 
\hfil  & & \ $s^3u_1 \by$  \ \hfil &&\ $a$ \ \hfil &\cr
& \ ${\bar \lambda}{}^1{}_{1}$ \ \hfil&& \ $({3\over 2};1,0,0;0,-{1\over 2})$ 
\hfil  & & \ $s^3u_1 \by^{-1}$  \ \hfil &&\ $\bar a$ \ \hfil &\cr
height2pt&\omit&&\omit&&\omit&&\omit&\cr
& \ $F_{11}$  \ \hfil&& \ $(2;0,0,0;1,0)$ \hfil &
& \ $s^4x^2$  \ \hfil &&\ $a\,{\bar a}\,b^2\,y^{-2}z^{-2}$\ \hfil &\cr
& \ $\partial_{12}$  \  \hfil && \ $(1;0,0,0;{1\over 2},{1\over 2})$ \hfil &
& \ $s^2 x \by $ \ \hfil &&\ $a\, b\, y^{-1}z^{-1}$ \ \hfil &\cr
& \ $\partial_{11}$  \  \hfil && \ $(1;0,0,0;{1\over 2},-{1\over 2})$ \hfil &
& \ $s^2 x \by^{-1} $ \ \hfil &&\ ${\bar a}\,b \,y^{-1}z^{-1}$ \ \hfil &\cr
height2pt&\omit&&\omit&&\omit&&\omit&\cr
}
\hrule}

\noindent Relations: $xyz=a{\bar a}$, $xb=yc=zd$.
}

Each case listed above defines a sector of operators 
formed by multiple traces over gauge indices of products of the fields
in each set, with appropriate derivatives in the 
$({1\over 8},0)$ and $({1\over 8},{1\over 8})$ cases.
With an $SU(N)$ gauge group then in each trace with $n$ field operators we 
require $n\ge 2$ and also $n$ is restricted by removal trace identities 
present for finite $N$, thus $\tr(Z^n)$ for $n>N$ is expressible in terms of
products of traces with $\sum_i n_i = n$, $n_i\le N$.
Restricting to multi-trace operators of the fields listed above
in each case provides a basis for constructing the various potential
short and semi-short supermultiplets.
In the $(\half,\half)$ sector a basis for $k$-trace, $k=1,2,\dots$, 
operators is given by
\eqn\hbpsz{
\tr \big (Z^{n_1}\big )\cdots \tr\big (Z^{n_k}\big )\, , \quad 
{\ts \sum_i} n_i = n \, .
}
These are all $\half$-BPS operators belonging to the $[0,n,0]$
representation. 
In the  $t=\bt = \quar$ sector a basis of $k$-trace operators is
\eqn\quarbpsz{
\tr\big({\ts \prod_{j}}Z^{n_{1j}}Y^{m_{1j}}
\big )\cdots \tr\big ({\ts \prod_{j}}
Z^{n_{kj}}Y^{m_{kj}}\big )\,,  \quad {\ts \sum_j} (n_{ij}+m_{ij} )
= n_i \, , \ {\ts \sum_{ij}} m_{ij} = m \, ,
}
where there is now a choice of ordering within each trace. Those
related by application of the  $SU(2)$ lowering operator to the operators
given \hbpsz\ are 
part of $\half$-BPS multiplets, the rest are potential superconformal
primary operators for $\quar$-BPS multiplets belonging to the
$[m,n-2m,m]$ representation. They may become long multiplets with
anomalous dimensions if joined with other semi-short multiplets as
in \diamond. The $m=1$ case is always protected (for a single trace
this is part of the $\half$-BPS multiplet).

Other examples are similarly constructed for the
$t=\quar$, $t=\bt ={1\over8}$ and $t={1\over 8}$ sectors.  

In the various limits described in section 5 it is important to recognise
that only those letters $z,y,\dots $ listed in table 1 survive which correspond
to operators which are present in each sector. We here rephrase some of
the previous results in terms of these letters.

\noindent
$t=\bt=\half$. Here only $z=s^2 u_1u_2 = \hs^{\, 2}$ survives and we have from 
\halfhalfshv\
\eqn\hhph{
\hchi^{\vphantom X}_p{\!\!}^{{1\over 2},{1\over 2}}(z) = 
\chi^{U(1)}_p(\hs) = z^p \, , \qquad 
\hchi^{\vphantom X}_{1{\vphantom d}}{\!\!}^{{1\over 2},{1\over 2}}(z) = z \, .
}

\noindent
$t=\bt=\quar$. Here $(z,y)= u (\hu_2,\hu_3) $ survive and we have from 
\quarquarshv\
\eqn\qqph{
\hchi^{\vphantom g}_p{\!\!}^{{1\over 4},{1\over 4}}(z,y) = \chi^{U(2)}_{(p,0)}(\hs;\hu_2,\hu_3) = 
\chi_{(p,0)}(z,y)  \, , \qquad 
\hchi^{\vphantom g}_1{\!\!}^{{1\over 4},{1\over 4}}(z,y) = z+y  \, .
}

\noindent
$t=\quar, \ \bt =0 $. Here $(z,y,x)= \hs^{\, 2} {\hat \uu}$, 
$(a,{\bar a})= \hs^{\, 3}(\by,\by^{-1})$, $zyx=a\ba$. From  \quarquarshvot\
\eqnn\qzph
$$\eqalignno{
\hchi^{\vphantom g}_p{\!\!}^{{1\over 4},0}(z,y,x;a,\ba) = {}& 
\chi^{SU(2|3)}_{{\rm short},p}(\hs;{\hat\uu};\by)\cr
= {}&  \chi_{(p,0,0)}(z,y,x) + (a + \ba) \chi_{(p-1,0,0)}(z,y,x) + 
\chi_{(p-1,1,1)}(z,y,x) \, ,\cr 
\hchi^{\vphantom g}_1{\!\!}^{{1\over 4},0}(z,y,x;a,\ba) = {}& z+y+x+a+\ba  \, . 
& \qzph \cr}
$$

\noindent
$t=\quar, \ \bt =\eight $. Here $(z,y)= u \hs^{\, 2}({\hat u}_2,{\hat u}_3)$, $a =
u \hs^{\, 4}$. From  \quarei\
\eqn\qeph{\eqalign{
\hchi^{\vphantom g}_p{\!\!}^{{1\over 4},{1\over 8}}(z,y;a) = {}& u^p 
\chi^{SU(1|2)}_{{\rm short},p}(\hs;{\hat u}_2,{\hat u}_3)  
= \chi_{(p,0)}(z,y) + a \, \chi_{(p-1,0)}(z,y)  \, ,\cr 
\hchi^{\vphantom g}_1{\!\!}^{{1\over 4},{1\over 8}}(z,y;a) = {}& z+y+a \, . \cr}
}
For $u=-1$, appropriate for the index, $a=-zy$.

\noindent
$t= \bt =\eight $. Here $(z,y)= u \hs^{\, 2}({\hat u}_2,{\hat u}_3)$, $a =
u v \hs^{\, 4}$, $b = u v^{-1} \hs^{\, 4}$. From  \chish\
\eqn\eeph{\eqalign{
\hchi^{\vphantom g}_p{\!\!}^{{1\over 8},{1\over 8}}(z,y;a,b) = {}& u^p 
\chi^{PSU(1,1|2)\ltimes U(1)}_{{\rm short},p}(\hs;{\hat u}_2,{\hat u}_3;v) \cr 
= {}& { 1 \over 1 - \si } \, \big ( \chi_{(p,0)}(z,y) + 
(a+b) \chi_{(p-1,0)}(z,y) + ab\, \chi_{(p-2,0)}(z,y) \big )  \, ,\cr 
\hchi^{\vphantom g}_1{\!\!}^{{1\over 8},{1\over 8}}(z,y;a,b) = {}& 
{ 1 \over 1 - \si} \,( z+y+a+b) \, , \qquad \si={ab\over zy} \, . \cr}
}
For $uv^{\pm 1} =-1$, appropriate for the index, $a=b=-zy$.

\noindent
$t=\eight, \ \bt =0 $. Here $(z,y,x)= u^2 \hs^{\, 2} {\hat \uu}$,
$(a,{\bar a})= u^3 \hs^{\, 3}(\by,\by^{-1})$, $(d,c,b)=u\hs^{\, 4}({\hat u}_2{\!}^{-1},
{\hat u}_3{\!}^{-1},{\hat u}_4{\!}^{-1})$. From  \halfhalfshvnew\ with
$\lam = u^{-3} = b/yz$
\eqnn\ezph
$$\eqalignno{
& \hchi^{\vphantom g}_p{\!\!}^{{1\over 8},0}(z,y,x;a,\ba,b)  = 
\chi^{PSU(1,2|3)\ltimes U(1)}_{{\rm short},p}(\hs,u;{\hat \uu};\by) \cr 
&{} =  { 1 \over (1 - \lam a)(1-\lam \ba)}  \Big ( \chi_{(p,0,0)}(z,y,x) + 
\lam \chi_{(p,1,0)}(z,y,x) +\lam^2 \chi_{(p,1,1)}(z,y,x) \cr
&\hskip 3.2cm{}+ (a+\ba)\big ( \chi_{(p-1,0,0)}(z,y,x) + 
\lam \chi_{(p-1,1,0)}(z,y,x) + \lam^2 \chi_{(p-1,1,1)}(z,y,x) \big ) \cr
&\hskip 3.2cm{} + \chi_{(p-1,1,1)}(z,y,x) + \lam \chi_{(p-1,2,1)}(z,y,x) +
\lam^2 \chi_{(p-1,2,2)}(z,y,x)  \Big ) \, ,\cr 
& \hchi_{\lower 1pt \hbox{$\ss 1$}}^{{1\over 8},0}(z,y,x;a,\ba,b) = 
{ 1 \over (1 - \lam a)(1-\lam \ba)} \, \big ( z+y+x + a + \ba \cr
\noalign{\vskip -6pt}
& \hskip 6cm{}+ \lam(yz+zx+xy) + xyz(\lam^2-\lam) \big ) \, . & \ezph\cr}
$$
For $u =-1$, appropriate for the index, $\lam =-1$.

\newsec{Partition functions}

Here we analyse partition functions at large $N$ for free $\N=4$ super
Yang-Mills making use of results from the discussion of characters in previous
sections.
We endeavour to decompose them into linear sums of characters in order to 
identify the operator content of the theory.

In the general case these partition functions{\foot{Note that what we 
refer to as a  `partition function' is defined by a supertrace, as appropriate
for determining an index.  
This is reflected by the sign change $(x,\by)\to -(x,\by)$
in the characters which introduces a sign $(-1)^F$ for
fermion number $F$.}} involve the single particle partition function
which is given by,
\eqn\singp{Z(s;\uu;x,\by) =
\chi^{{1\over 2},{1\over 2}}_{(1;0,1,0;0,0)}(s;\uu;-x,-\by)
}
and is expanded in detail in \elemp. The partition function for single 
trace operators, for $SU(N)$ theories as $N\to \infty$, is then given by
\eqn\singt{\eqalign{
Z_{\rm s.t.}(s;\uu;x,\by) = {}&
\sum_{n=2}^\infty {1\over n}\sum_{d|n}\phi(d) \, 
Z(s^d;\uu^d;x^d,\by^d)^{n/d}\cr
= {}& -\sum_{d=1}^{\infty}{\phi(d)\over d}\log\big(1-Z(s^d;
\uu{}^d;x^d,\by^d)\big)-Z(s;\uu;x,\by)\,, \cr 
\uu^d = {}& (u_1{}^d,u_2{}^d,u_3{}^d,u_4{}^d) \, ,
}}
where $\phi(n)$ is the Euler totient function, being the number of
integers relatively prime to and smaller than $n$ with $\phi(1)=1$. This
formula reflects the modification of counting as a consequence of cyclic
symmetry of the trace. For $U(N)$ theories
the lower limit of the first sum in \singt\ becomes 1 so that the 
last term in the second line is missing.

The multi-trace partition function is expressed in terms of this by
\eqn\multitr{\eqalign{
Z_{\rm m.t.}(s;\uu;x,\by) = {}&
\exp\Big(\sum_{n=1}^\infty
{1\over n}Z_{\rm s.t.}(s^n;\uu^n;x^n,\by^n)\Big)\cr
= {}& \exp\Big(-\sum_{n=1}^\infty{1\over n}Z(s^n;\uu^n;x^n,\by^n) \Big)
\prod_{m=1}^\infty{1\over 1-Z(s^m;\uu{}^m;x^m,\by^m)}\,,
}}
which is achieved using $\sum_{d|n}{\phi(d)}=n$.
The modification for $U(N)$ theories is simply that
the pre-factor in the last line of \multitr\ is missing.

We will also be interested in the partition function over operators
whose fundamental fields are completely symmetrised within the trace.
The single trace partition function over such completely symmetrised operators
is given by
\eqn\singtsym{
Z_{\rm s.t., sym.}(s;\uu;x,\by) = \exp\Big(\sum_{n=1}^\infty
{1\over n}Z(s^n;\uu^n;x^n,\by^n)\Big)
-Z(s;\uu;x,\by) -1\,,
}
and the multi-trace partition function over such operators is
\eqn\multitrsym{
Z_{\rm m.t., sym.}(s;\uu;x,\by) =
\exp\Big(\sum_{n=1}^\infty
{1\over n}Z_{\rm s.t., sym.}(s^n;\uu^n;x^n,\by^n)\Big)\,.
}
For $U(N)$ theories the $Z(s;\uu;x,\by)$ term
in \singtsym\ is not subtracted.

We also consider the single trace partition function over $\half$-BPS
operators, given by,
\eqn\singthbps{
Z_{\rm s.t.,{1\over 2}-BPS}(s;\uu;x,\by) =\sum_{p=2}^\infty
\chi^{{1\over 2},{1\over 2}}_{(p;0,p,0;0,0)}(s;\uu;-x,-\by)\,,
}
along with the multi-trace partition function over operators
formed from these ({\it i.e.} multi-trace operators constructed 
by multiplying together single-trace $\half$-BPS operators and descendants).
In the large $N$, large coupling limit, these multi-trace operators correspond 
to supergravity multi-particle states via the AdS/CFT correspondence, and hence 
we label the partition function in this case by `sugra'.
The resulting partition function is then given by
\eqn\multigrav{
 Z_{\rm sugra}(s;\uu;x,\by) = \exp\Big(\sum_{n=1}^\infty
{1\over n}Z_{\rm s.t.,{1\over 2}-BPS}(s^n;\uu^n;x^n,\by^n)\Big)\,.
}

The operator content in the different sectors labelled by $t,\bt$ is determined
by using the reduced characters discussed in section 5. 
The relevant partition functions $Z^{t,\bt}$ can be
obtained from the general partition function by  taking the
appropriate limits described there. In each case we seek expansions of the form
\eqn\expart{
Z^{t,\bt} = \sum_{\M} N_\M^{\vphantom g} \, {\hchi}^{t,\bt}_\M \, ,
}
in terms of the appropriate reduced characters ${\hchi}^{t,\bt}_\M$ corresponding
to different supermultiplets $\M$, the integers $N_\M$ then determine the numbers
of such multiplets in each sector of the theory. The supermultiplets which are
accessible in the expansion \expart\ depend on $t,\bt$, it is of course crucial
that the resulting $N_\M^{\vphantom g}$ are consistent between different sectors. 
This follows quite simply since both $Z^{t,\bt}$ and ${\hchi}^{t,\bt}_\M$
are related by setting various letters to zero.

\noindent
{\bf $t=\bt = \half$ operators}

The operators here are constructed from the single field $Z$.
The  single particle partition function is determined in \hhph,
\eqn\halfsingpart{ Z^{{1\over 2},{1\over 2}}(z)= 
\hchi_{\lower 1pt \hbox{$\ss 1$}}^{{1\over 2},{1\over 2}}(z) =  z\,.  } 
In this sector it is easy to see from \singt, since $\sum_{k=1}^\infty
{\phi(k)\over k} \, \ln(1-x^k) = -x/(1-x)$, \singtsym\ and \singthbps\ 
\eqn\iden{
Z^{{1\over 2},{1\over 2}}_{\rm s.t.}(z)=
Z^{{1\over 2},{1\over 2}}_{\rm s.t., sym.}(z)=
Z^{{1\over 2},{1\over 2}}_{\rm s.t., {1\over 2}-BPS}(z) = {z^2 \over 1-z} 
= \sum_{n=2}^\infty z^p \, .
}
Trivially for single trace operators there is just one symmetric $1\over 2$-BPS 
operator represented by $\tr(Z^n)$ for $n=2,3,\dots$.

The multi-trace partition function is then simply given from \multitr\ by
\eqn\halfmt{ 
Z^{{1\over 2},{1\over 2}}_{\rm m.t.}(z)=\prod_{m=2}^\infty{1\over 1-z^m}\,,  }
and it follows immediately from \multitrsym\ and \multigrav, using \iden, that
\eqn\idenmt{
Z^{{1\over 2},{1\over 2}}_{\rm m.t.}(z)= Z^{{1\over 2},{1\over 2}}_{\rm m.t., sym.}(z)
= Z^{{1\over 2},{1\over 2}}_{\rm sugra}(z)\,.}
The character expansion is here
\eqn\halfmts{ Z^{{1\over 2},{1\over 2}}_{\rm m.t.}(z)
=\sum_{n=0}^{\infty} N^{{1\over 2}{\rm -BPS}}_{{\rm m.t.},n} \, z^n \,. } 
Using
\eqn\serhalf{
\prod_{k=1}^{\infty}{1\over 1-z^k}=\sum_{n=0}^{\infty}p(n)\, z^n\,, }
where $p(n)$ is the number of unrestricted partitions of the
positive integer $n$ into a sum of strictly positive integers
where the order of summands is irrelevant, with $p(0)=1$,
$p(-1)=0$, we have \Gwyn
\eqn\halfbps{
N^{{1\over 2}{\rm -BPS}}_{{\rm m.t.},n}  = p(n)-p(n-1) \, , \quad n=2,3,\dots \, ,
}
which counts the number of $\half$-BPS operators of conformal dimension $n$,
$N^{{1\over 2}-{\rm BPS}}_0=1$ corresponds to the identity operator. 
The corresponding operators for the first few cases
are summarised in the Table 2.

\medskip
\vbox{
\hskip4cm Table 2
\nobreak

\hskip4cm
\vbox{\tabskip=0pt \offinterlineskip
\hrule
\halign{&\vrule# &\strut \ \hfil#\  \cr
height2pt &\multispan3 &\cr
&\multispan3\ \ \ \ \ \ $(\half,\half)$ primary operators \ \ \hfil &\cr
height2pt& \multispan3 &\cr
&\multispan3\hrulefill& \cr
height2pt&\omit&&\omit&\cr
& \ $\De$ \ \hfil  && \ Operators \ \ \hfil  &\cr
height2pt&\omit&&\omit&\cr
\noalign{\hrule}
height2pt&\omit&&\omit&\cr
&  2 \hfil  &&  \quad ${\R}_{[0,2,0]}^{(0,0)}\,$ \hfil &\cr
&  3 \hfil  &&  \quad ${\R}_{[0,3,0]}^{(0,0)}\,$ \hfil &\cr
&  4 \hfil  &&  \quad 2 ${\R}_{[0,4,0]}^{(0,0)}$ \hfil &\cr
&  5 \hfil  &&  \quad 2 ${\R}_{[0,5,0]}^{(0,0)}$ \hfil &\cr
&  6 \hfil  &&  \quad 4 ${\R}_{[0,6,0]}^{(0,0)}$ \hfil &\cr
&  7 \hfil  &&  \quad 4 ${\R}_{[0,7,0]}^{(0,0)}$ \hfil &\cr
&  8 \hfil  &&  \quad 7 ${\R}_{[0,8,0]}^{(0,0)}$ \hfil &\cr
height2pt&\omit&&\omit&\cr
}
\hrule}

{\eightpoint
{\parindent 1.5cm{\narrower
\noindent
$\half$-BPS primary operators with conformal dimensions $\De$ 
and belonging to $SU(4)_R \otimes SU(2)_J \otimes SU(2)_{\bar J}$
representations $\R^{(j,\bj)}_{[k,p,q]}$ obtained from expansion of partition function. 
For each $\R^{(0,0)}_{[0,p,0]}$, $p\ge 2$, there is one single trace operator.

}}}}

\noindent
{\bf $t=\bt =\quar$ operators}

The relevant fields are the scalars $Z,Y$ with associated letters $z,y$.
Here the single particle partition function from \qqph\ is
\eqn\quarsingpart{ 
Z^{{1\over 4},{1\over 4}}(z,y)=
\hchi^{{1\over 4},{1\over 4}}_1(z,y)= z+y\, . }
We also have from \singtsym\ and \singthbps,
\eqn\quarsing{
Z^{{1\over 4},{1\over 4}}_{\rm s.t., sym.}(z,y)=
Z^{{1\over 4},{1\over 4}}_{\rm s.t., {1\over 2}-BPS}(z,y)
= \sum_{n=2}^\infty \chi_{(n,0)}(z,y) ={1\over (1-z)(1-y)}-z-y-1 \, .
}
This shows that all completely symmetric single trace operators are in
fact the $\half$-BPS ones which are related by $SU(2)$ lowering operators
to those in the $(\half,\half)$ sector.
For general single trace operators from \singt\ in this case
\eqn\quart{
Z^{{1\over 4},{1\over 4}}_{\rm s.t.}(z,y)= 
- \sum_{k=1}^\infty {\phi(k)\over k} \, \ln ( 1-z^k-y^k) - z - y \, .
}

For multi-trace operators the partition function becomes
\eqn\quarmt{ 
Z^{{1\over 4},{1\over 4}}_{\rm m.t.}(z,y)
=(1-z)(1-y) \prod_{k=1}^{\infty}{1\over 1-z^k-y^k}\,. }
Using from \quarsing\
$Z^{{1\over 4},{1\over 4}}_{\rm s.t., sym.}(z,y)=\sum_{k,l=0,
k+l\geq 2}^{\infty}z^{k}y^{l}$ we also get
\eqn\quarmtsym{
Z^{{1\over 4},{1\over 4}}_{\rm m.t., sym.}(z,y)=
Z^{{1\over 4},{1\over 4}}_{\rm sugra}(z,y)
= \prod_{k,l=0\atop k+l\geq 2}^{\infty} {1\over 1-z^{k}y^{l}}\,.
}

To determine the decomposition into contributions corresponding to $\quar$-BPS
multiplets $\B^{{1\over 4},{1\over4}}_{[q,p,q](0,0)}$ multiplets we use from 
\quarquarshv\ the character expressed in terms of $z,y$,
\eqn\charsnewpopo{
{\hat \chi}^{{1\over 4},{1\over 4}}_{(p+2q;q,p,q;0,0)}(z,y)=
\chi_{(p+q,q)}(z,y)=(z y)^q\, {z^{p+1}-y^{p+1}\over z-y}\,,
}
which is just a $U(2)$ character. The required expansion is then
\eqn\expquar{\eqalign{
Z^{{1\over 4},{1\over 4}}_{\rm m.t.}(z,y) = {}& \sum_{n,m=0}^\infty 
N^{{1\over 4},{1\over 4}}_{{\rm m.t.}, nm} \, \chi_{(n+m,m)}(z,y) \, , \cr
Z^{{1\over 4},{1\over 4}}_{\rm m.t.,sym.}(z,y) = {}& \sum_{n,m=0}^\infty 
N^{{1\over 4},{1\over 4}}_{{\rm m.t., sym.}, nm} \, \chi_{(n+m,m)}(z,y) \, , \cr}
}
where the coefficients determine the number of primary operators belonging to the
representation $\R^{(0,0)}_{[m,n,m]}$.
Expressions for the coefficients in some cases are obtained in appendix B.
These give the following results
\eqn\quarcoeff{\eqalign{
N^{{1\over 4},{1\over 4}}_{{\rm m.t.}, n 0} =
N^{{1\over 4},{1\over 4}}_{{\rm m.t., sym.}, n0} = {}& p(n) -p(n-1) \, , \quad
n=2,3,\dots \, , \cr
N^{{1\over 4},{1\over 4}}_{{\rm m.t.}, n 1} =
N^{{1\over 4},{1\over 4}}_{{\rm m.t., sym.}, n1} = {}& p(n)+p(n+1)-p(n+2) \, ,
\ \ n=3,4,\dots \, , \cr
N^{{1\over 4},{1\over 4}}_{{\rm m.t.}, n2} =
2 N^{{1\over 4},{1\over 4}}_{{\rm m.t., sym.}, n2} = {}& 2 \sum_{j=0}^{[{1\over 2}n]}
p(n-2j) \, , \quad n=3,4, \dots \, . \cr}
}
Clearly from \halfbps\ $N^{{1\over 4},{1\over 4}}_{{\rm m.t.}, n 0} =
N^{{1\over 2}{\rm -BPS}}_{{\rm m.t.},n}$ which reflects that for $m=0$ these
are just the $\half$-BPS operators in this sector. The operators corresponding
to $m=1$ are genuine $\quar$-BPS operators, for $m\ge 2$ these may combine with
other operators to form long multiplets as discussed further later.

For single trace operators we clearly have from \quarsing\ 
$N^{{1\over 4},{1\over 4}}_{{\rm s.t., sym.},n0} = 1$, $n=2,3,\dots$ as expected
since the relevant operators  are all part of $\half$-BPS multiplets. For
general single trace operators from \quart
\eqn\quarsingc{
N^{{1\over 4},{1\over 4}}_{{\rm s.t.}, n 0} = 1 \, , \quad
N^{{1\over 4},{1\over 4}}_{{\rm s.t.}, n 1} = 0 \, , \quad
N^{{1\over 4},{1\over 4}}_{{\rm s.t.}, n 2 } = 1+ [\half n] \, , \quad n=0,1,\dots \, .
}
The last case corresponds to operators $\tr(YZ^r Y Z^{n+2-r})$, 
$r=0,1,\dots , 1+ [\half n]$,
with the symmetric sum part of the $\half$-BPS multiplet formed from $\tr(Z^{n+4})$.

Corresponding to the results for the expansions \expquar, apart from the 
$\half$-BPS operators, we list the required 
$\quar$-BPS operators for the first few cases in Table 3.

\medskip
\vbox{
\hskip1.5cm Table 3
\nobreak

\hskip1.5cm
\vbox{\tabskip=0pt \offinterlineskip
\hrule
\halign{&\vrule# &\strut \ \hfil#\  \cr
height2pt&\multispan5 &\cr
&\multispan5 \hfil $(\quar,\quar)$ primary operators \hfil&\cr
height2pt&\multispan5 &\cr
&\multispan5\hrulefill& \cr
height2pt&\omit&&\omit&&\omit&\cr
&\ $\De$ \ \hfil   &&\  Symmetric operators \hfil &&
Remaining operators \ \ \hfil &\cr
height2pt&\omit&&\omit&&\omit&\cr
\noalign{\hrule}
height2pt&\omit&&\omit&&\omit&\cr
& \ 4  \ \hfil && \ $\R_{[2,0,2]}^{(0,0)}$ \hfil &
& \ $(1)\R_{[2,0,2]}^{(0,0)}$ \ \hfil &\cr
height1pt&\omit&&\omit&&\omit&\cr
& \ 5  \ \hfil&& \ $\R_{[1,3,1]}^{(0,0)}$,
$\R_{[2,1,2]}^{(0,0)}$ \hfil &
& \ $(1)\R_{[2,1,2]}^{(0,0)}$ \ \hfil &\cr
height1pt&\omit&&\omit&&\omit&\cr
& \ 6  \ \hfil&& \ $\R_{[1,4,1]}^{(0,0)}$,
$3\,\R_{[2,2,2]}^{(0,0)}$ \hfil &
& \ $3(2)\R_{[2,2,2]}^{(0,0)}$,
$(1)\R_{[3,0,3]}^{(0,0)}$ \ \hfil &\cr
height1pt&\omit&&\omit&&\omit&\cr
& \ 7  \ \hfil&& \ $3\,\R_{[1,5,1]}^{(0,0)}$,
$4\,\R_{[2,3,2]}^{(0,0)}$,\hfil  &
& \ $4(2)\R_{[2,3,2]}^{(0,0)}$,
$3(2)\R_{[3,1,3]}^{(0,0)}$\ \hfil &\cr
& \   \ && \
$2\,\R_{[3,1,3]}^{(0,0)}$\hfil && \ \ &\cr
height1pt&\omit&&\omit&&\omit&\cr
& \ 8  \ \hfil&& \ $4\, \R_{[1,6,1]}^{(0,0)}$,
$8\,\R_{[2,4,2]}^{(0,0)}$, \hfil &
& \ $8(3)\R_{[2,4,2]}^{(0,0)}$, $7(3)\R_{[3,2,3]}^{(0,0)}$,
\ \hfil &\cr
& \   \ && \ $3\, \R_{[3,2,3]}^{(0,0)}$,
$4\,\R_{[4,0,4]}^{(0,0)}$ \hfil && \ $7(3)\R_{[4,0,4]}^{(0,0)}$
\ \hfil &\cr
height2pt&\omit&&\omit&&\omit&\cr
}
\hrule}

{\eightpoint
{\parindent 1.5cm{\narrower
\noindent
$\quar$-BPS primary operators with 
conformal dimensions $\De$ belonging to representations $\R^{(0,0)}_{[q,p,q]}$, 
as obtained from expansion of partition function. When present numbers of single
trace operators are listed in parenthesis.

}}}}

The results here are in 
accord with the explicit construction of $\quar$-BPS operators in various cases
in \Doker.

\noindent
{\bf $t=\bt=\eight$ and $t=\quar, \, \bt = \eight$ semi-short operators}

The fields in these sectors are listed in \ZYd\ and \ZYfe.
In terms of the variables $z,y,a,b$ then 
the single particle partition function from \eeph\ in the $t=\bt=\eight$ sector is
\eqn\singpart{
Z^{{1\over 8},{1\over 8}}(z,y,a,b)=
\hchi^{{1\over 8},{1\over 8}}_{\lower 1pt \hbox{$\ss 1$}}(z,y;-a,-b)
= {z+y-a-b\over 1- \si }\, , \quad \si = {ab \over zy} \, .
}
Here expansion in powers of $\si$ counts derivatives in the basis \ZYd.
Results for $t= \quar $ are obtained by setting $b=0$, and hence $\si=0$,  and 
indeed for  $t=\bt = \quar$ by requiring $a=b=0$. 
The operators ${\tilde O}^{{1\over 8},{1\over 8}}_{[k,p,q](j,\bj)}$
corresponding to the expansion of this partition function 
satisfy $\Delta=j+\bj+k+p+q$, $j-\bj = \half (q-k)$, but as discussed in 
section 3 this is sufficient to count $(\eight,\eight)$ semi-short multiplets.
For $j$ or $ \bj$ zero these become $t=\quar$ or $\bt = \quar$ operators.

Applying \singt\ the partition function for single trace operators is
\eqn\quart{
Z^{{1\over 8},{1\over 8}}_{\rm s.t.}(z,y,a,b)=
- \sum_{k=1}^\infty {\phi(k)\over k} \, \ln \Big (  1-{z^k+y^k- a^k - b^k
\over 1-\si^k} \Big ) - {z + y  - a -b \over 1 - \si } \, .
}
Corresponding to \singthbps\ the single trace partition function for 
$\half$-BPS operators is 
\eqn\sssthalf{\eqalign{
Z^{{1\over 8},{1\over 8}}_{\rm s.t.,{1\over 2}-BPS}(z,y,a,b) = {}&
\sum_{p=2}^\infty 
\hchi^{\vphantom g}_p{\!\!}^{{1\over 8},{1\over 8}}(z,y;-a,-b) \cr
= {}& {1\over 1- \si} \bigg ( {(1-a)(1-b)\over (1-z)(1-y)} -1-z-y+a+b\bigg ) \, ,}
}
using \eeph.
With \singpart\ the
multi-trace partition function for this sector is given by
\eqn\zmtss{\eqalign{
Z^{{1\over 8},{1\over 8}}_{\rm m.t.}(z,y,a,b)=&{}
\prod_{n=0}^{\infty}{(1-z\, \si^n)(1-y\, \si^n)
\over (1-a\, \si^n)(1 -b\, \si^n)}\cr
&\times
\prod_{m=1}^{\infty} {1 - \si^m\over 1- \si^m - z^m - y^m +a^m + b^m }\,.
}}
The partition function corresponding to the
supergravity sector of the AdS dual theory corresponds to
operators which are simply those multi-trace operators obtained by products of
single trace $\half$-BPS operators. This partition function is obtained  from
\multigrav\ by using \sssthalf,
\eqn\sssugra{\eqalign{
Z_{\rm sugra}^{{1\over 8},{1\over 8}}(z,y,a,b) = {}&\exp\Big(\sum_{n=1}^{\infty}
{1\over n}
Z_{\rm s.t.,{1\over 2}-BPS}^{{1\over 8},{1\over 8}}(z^n,y^n,a^n,b^n)  \Big) \cr
= {}& \prod_{n=0}^\infty \,
{\prod_{k,l=0,k+l\ge 1}^\infty ( 1 - a\, z^ky^l\, \si^n) 
\, ( 1 - b\, z^ky^l\, \si^n) \over
\prod_{k,l=0,k+l\ge 2}^\infty ( 1 -  z^ky^l\, \si^n) \, 
\prod_{k,l=0}^\infty ( 1 - ab\, z^ky^l\, \si^n)} \, .}
}

In order to decompose these partition functions in terms of contributions
for different semi-short multiplets by a character expansion it is first
necessary to subtract the contribution of  $\half$-BPS multiplets. Using \halfbps\
this is given by
\eqn\zsshalf{\eqalign{
Z_{\rm m.t.}^{{1\over 2},{1\over 2}}(z,y,a,b) &=\sum_{n=0}^{\infty}
\big(p(n)-p(n-1)\big) \, 
\hchi^{\vphantom g}_n{\!\!}^{{1\over 8},{1\over 8}}(z,y;-a,-b) \cr
&= {(z-a)(z-b)\over (1-\si) z(z-y)}\prod_{m=2}^\infty {1\over 1-z^m}
+z\leftrightarrow y\,. \cr }
}
{}From \quart\ and \sssthalf
\eqn\quarz{
Z^{{1\over 8},{1\over 8}}_{\rm s.t.}(z,y,z,b)=
- \sum_{k=1}^\infty {\phi(k)\over k} \, \ln \big (  1-y^k \big ) -  y   
= Z^{{1\over 8},{1\over 8}}_{\rm s.t.,{1\over 2}-BPS}(z,y,z,b) = {y^2 \over 1-y} \, ,
}
with similar results for $z=b$ and $y=a,b$. In consequence we have
\eqn\zea{
Z^{{1\over 8},{1\over 8}}_{\rm m.t.}(z,y,z,b) =
Z_{\rm sugra}^{{1\over 8},{1\over 8}}(z,y,z,b) =
Z_{\rm m.t.}^{{1\over 2},{1\over 2}}(z,y,z,b) = \prod_{m=2}^\infty {1\over 1-y^m}\, ,
}
and analogously for $z=b$ and $y=a,b$. 

Initially we set $b=0$, so that no derivatives are present, and consider
expansions determining the contributions of $({1\over 4},{1\over 8})$ multiplets.
The relevant characters are obtained from \quarquarshvo\ and \charlim. After 
subtracting the $\half$-BPS contribution, given by \zsshalf\ for $b=0$, then
\zea\ shows that the result vanishes for $z,y=a$ and we can write
\eqnn\foure
$$\eqalignno{
& Z^{{1\over 8},{1\over 8}}_{\rm m.t.}(z,y,a,0) 
- Z_{\rm m.t.}^{{1\over 2},{1\over 2}}(z,y,a,0) \cr
& \! {}= {(1-z)(1-y)\over 1-a}\! \prod_{m=1}^\infty \! {1\over 1-z^m-y^m+a^m}
- {1\over z-y}\bigg ( \! (z-a)\! \prod_{k=2}^\infty {1\over 1-z^k} -
(y-a)\! \prod_{k=2}^\infty {1\over 1-y^k} \! \bigg ) \cr
&\hskip 2.5cm {}= 
(z-a)(y-a) \sum_{n,m= 0}^\infty\sum_{2\bj=-1}^\infty 
N^{{1\over 4},{1\over 8}}_{{\rm m.t.}, n m ,\bj}\,
(-a)^{2\bj+1} \chi_{(n+m,m)}(z,y) \, ,\cr
\noalign{\vskip 6pt}
& Z^{{1\over 8},{1\over 8}}_{\rm sugra}(z,y,a,0) 
- Z_{\rm m.t.}^{{1\over 2},{1\over 2}}(z,y,a,0) \cr
\noalign{\vskip 3pt}
& \! {}=  {\prod_{k,l=0,k+l\ge1}^\infty ( 1 - a\, z^ky^l)\over 
\prod_{k,l=0,k+l\ge 2}^\infty ( 1 - z^ky^l) }
- {1\over z-y}\bigg ( \! (z-a)\! \prod_{k=2}^\infty {1\over 1-z^k} -
(y-a)\! \prod_{k=2}^\infty {1\over 1-y^k} \! \bigg ) \cr
&\hskip 2.5cm {}= 
(z-a)(y-a) \sum_{n,m= 0}^\infty\sum_{2\bj=-1}^\infty
N^{{1\over 4},{1\over 8}}_{{\rm sugra}, n m ,\bj}\,
(-a)^{2\bj+1} \chi_{(n+m,m)}(z,y) \, . & \foure }
$$
Here $N^{{1\over 4},{1\over 8}}_{{\rm m.t.}, pq  ,\bj}$ and
$N^{{1\over 4},{1\over 8}}_{{\rm sugra}, p q ,\bj}$ are then the number of
$t=\quar, \, \bt= \eight$ multiplets which contribute to the relevant partition
functions and whose primary operators belong to the 
representation labelled by $\R^{(0,\bj)}_{[2+2\bj+q,p,q]}$ and has a scale dimension 
$\De= 3+3\bj + p +2q$.
It is easy to see, taking $a=0$ when only the $\bj=-\half$ term survives in the
sum, that by comparing with the expansion \expquar\ we have
\eqn\relN{
N^{{1\over 4},{1\over 8}}_{{\rm m.t.}, n m ,-{1\over 2}} =
N^{{1\over 4},{1\over 4}}_{{\rm m.t.}, n\, m+1} \, , \qquad
N^{{1\over 4},{1\over 8}}_{{\rm sugra}, n m ,-{1\over 2}} =
N^{{1\over 4},{1\over 4}}_{{\rm m.t.,sym.}, n\, m+1} \, ,
} 
in accord with \redt. 
For this case, as shown later, $ N^{{1\over 4},{1\over 8}}_{{\rm sugra}, n m ,\bj}
=  N^{{1\over 4},{1\over 8}}_{{\rm m.t.,sym.}, n m ,\bj}$ involving operators 
formed from symmetrised traces. Analytic formulae for the expansion coefficients
are harder to obtain
but for the lowest dimension operators results are given in Table 4.
\medskip

\vbox{Table 4
\nobreak

\hskip -1.2cm
\vbox{\tabskip=0pt \offinterlineskip
\hrule
\halign{&\vrule# &\strut \ \hfil#\  \cr
height2pt&\multispan7 &\cr
&\multispan7 \hfil $(\quar,\eight)$ primary operators \hfil&\cr
height2pt&\multispan7 &\cr
&\multispan7\hrulefill& \cr
height2pt&\omit&&\omit&&\multispan3&\cr
& $\De-\bj$  &&  Symmetric \hfil && \multispan3 \hfil Remaining operators \ \ \hfil &\cr
height1pt&\omit&&\omit&&\multispan3&\cr
&\omit&&\omit&&\multispan3\hrulefill&\cr
height1pt&\omit&&\omit&&\omit&&\omit&\cr
&\omit&& operators \hfil&&\ $(\quar,0)$ primary \hfil&&\ $(\quar,0)$ descendant\hfil&\cr
height1pt&\omit&&\omit&&\omit&&\omit&\cr
\noalign{\hrule}
& 3   \hfil && \  \hfil &
& \ $(1)\R_{[2,0,0]}^{(0,0)}$ \ \hfil &&&\cr
& 4   \hfil&& \  \hfil &
& \ $(1)\R_{[2,1,0]}^{(0,0)}$ \ \hfil &&&\cr
& 5   \hfil&& \  \hfil &
& $\, 3(2)R_{[2,2,0]}^{(0,0)}$,
$(1)\R_{[3,0,1]}^{(0,0)}$,
$\R_{[3,1,0]}^{(0,{1\over 2})}$,
$(1)\R_{[4,0,0]}^{(0,1)}$ \ \hfil &&&\cr
& 6   \hfil&& \  \hfil  &
& \ $4(2)\R_{[2,3,0]}^{(0,0)}$,
$3(2)\R_{[3,1,1]}^{(0,0)}$,
$4(2)\R_{[3,2,0]}^{(0,{1\over 2})}$,\ \hfil &
&\ $\R_{[3,1,1]}^{(0,0)}$, $(1)\R_{[4,0,1]}^{(0,{1\over 2})}$\hfil&\cr
& \hfil&& \ \hfil  && \ $\R_{[4,0,1]}^{(0,{1\over 2})}$,
$2(1)\R_{[4,1,0]}^{(0,1)}$ \ \hfil &&&\cr
&  7   \hfil&& \ $\R_{[3,2,1]}^{(0,0)}$ \hfil &
&  $8(3)\R_{[2,4,0]}^{(0,0)}$,$7(3)\R_{[4,0,2]}^{(0,0)}$,
$7(3)\R_{[3,2,1]}^{(0,0)}$,\ \hfil &
&\ $\R_{[4,0,2]}^{(0,0)}$, $4(2)\R_{[3,2,1]}^{(0,0)}$\hfil&\cr
&    \hfil&& \  \hfil &
& $7(2)\R_{[3,3,0]}^{(0,{1\over 2})}$, $10(4)\R_{[4,1,1]}^{(0,{1\over 2})}$,
$5(3)\R_{[4,2,0]}^{(0,1)}$,\ \hfil &&\ $2(1)\R_{[4,1,1]}^{(0,{1\over 2})}$\hfil&\cr
&  \hfil&& \  \hfil &
& \ $5(2)\R_{[5,0,1]}^{(0,1)}$,
$2\R_{[5,1,0]}^{(0,{3\over 2})}$, $(1)\R_{[6,0,0]}^{(0,2)}$\ \hfil &&&\cr
& 8  \hfil&&  $\, 2\R_{[3,3,1]}^{(0,0)}$, \hfil &
& $\, 11(3)\R_{[2,5,0]}^{(0,0)}$,$13(4)\R_{[4,1,2]}^{(0,0)}$,
$16(6)\R_{[3,3,1]}^{(0,0)}$,\hfil &
& $10(4)\R_{[4,1,2]}^{(0,0)}$, $7(2)\R_{[3,3,1]}^{(0,0)}$\hfil&\cr
& \hfil&&  $\, \R_{[4,1,2]}^{(0,0)}$ \hfil &
& $15(4)\R_{[3,4,0]}^{(0,{1\over 2})}$, $25(8)\R_{[4,2,1]}^{(0,{1\over 2})}$,
$17(6)\R_{[5,0,2]}^{(0,{1\over 2})}$, \hfil &
& $5(3)\R_{[4,2,1]}^{(0,{1\over 2})}$, $5(2)\R_{[5,0,2]}^{(0,{1\over 2})}$\hfil &\cr
&  \hfil&& \  \hfil &
& $14(5)\R_{[4,3,0]}^{(0,1)}$, $19(8)\R_{[5,1,1]}^{(0,1)}$, $7(2)\R_{[5,2,0]}^{(0,{3\over 2})}$,
\hfil &&\ $2\R_{[5,1,1]}^{(0,1)}$, $(1)\R_{[6,0,1]}^{(0,{3\over 2})}$\hfil &\cr
& \hfil && \hfil &
& $7(2)\R_{[6,0,1]}^{(0,{3\over 2})}$,
$2(1)\R_{[6,1,0]}^{(0,2)}$, $\R_{[7,0,0]}^{(0,{5\over 2})}$\ \hfil & \hfil && \cr
height2pt&\omit&&\omit&&\omit&&\omit&\cr
}
\hrule}
}

{\eightpoint
{\parindent 1.5cm{\narrower
\noindent
$(\quar,\eight)$ primary operators belonging to representations 
$\R^{(0,\bj)}_{[q+2\bj+2,p,q]}$ 
as obtained from expansion of partition function. When present numbers of single
trace operators are listed in parenthesis.

}}}

In the general case we consider expansions of the form
\eqn\ssspectrum{\eqalign{
Z^{{1\over 8},{1\over 8}}_{\rm m.t.}(z,y,a,b) 
- Z_{\rm m.t.}^{{1\over 2},{1\over 2}}(z,y,a,b) 
= {}& \sum_{n,m,j,\bj} N^{{1\over 8},{1\over 8}}_{{\rm m.t.}, n m ,j \bj}\,
\hchi_{nm,j\bj}(z,y,-a,-b) \, ,\cr
Z^{{1\over 8},{1\over 8}}_{\rm sugra}(z,y,a,b) 
- Z_{\rm m.t.}^{{1\over 2},{1\over 2}}(z,y,a,b) 
= {}& \sum_{n,m,j,\bj} N^{{1\over 8},{1\over 8}}_{{\rm sugra}, n m ,j\bj}\,
\hchi_{nm,j\bj}(z,y,-a,-b)  \, ,}
}
where, using the results \quarquarshvn\ and \shorchar\ for the characters in the
relevant limit,
\eqn\chss{
\hchi_{nm,j\bj}(z,y,a,b) = {(z+a) (y+a) (z+b)(y+b) \over 1- \si}\ a^{2\bj+1}b^{2j+1}
\, \chi_{(n+m,m)} (z,y) \, .
}
The left hand sides in \ssspectrum\ has a factor $(z-a) (y-a) (z-b)(y-b)$ as a 
consequence of \zea\ and the symmetry under $z\leftrightarrow y, \ 
a \leftrightarrow b$. 
In the summation $n=0,1,2,\dots$, $j,\bj=-\half,0,\half,\dots $ while $m$ may
take negative values, $m\ge-2j-2, -2\bj-2$. The expansion coefficients 
\ssspectrum\ $ N^{{1\over 8},{1\over 8}}_{{\rm m.t.}, n m ,j \bj}$ and
$N^{{1\over 8},{1\over 8}}_{{\rm sugra}, n m ,j\bj}$ then give
the numbers of semi-short primary operators in the representation 
$\R^{(j,\bj)}_{[m+2\bj+2,n,m+2j+2]}$, along with their descendants,
determined by the partition function. It is easy to see that
\eqn\symN{
N^{{1\over 8},{1\over 8}}_{{\rm m.t.}, n m ,j \bj}
= N^{{1\over 8},{1\over 8}}_{{\rm m.t.}, n m ,\bj j} \, , \qquad
N^{{1\over 8},{1\over 8}}_{{\rm sugra}, n m ,j\bj}
= N^{{1\over 8},{1\over 8}}_{{\rm sugra}, n m ,\bj j} \, ,
}
and, setting $b=0$,
\eqn\specN{
N^{{1\over 8},{1\over 8}}_{{\rm m.t.}, n m ,-{1\over 2}\, \bj} = 
N^{{1\over 4},{1\over 8}}_{{\rm m.t.}, n\, m+1 ,\bj} \, , \qquad
N^{{1\over 8},{1\over 8}}_{{\rm sugra}, n m ,-{1\over 2}\, \bj}
= N^{{1\over 4},{1\over 8}}_{{\rm sugra}, n \, m+1 , \bj} \, .
}

As emphasised in section 3 semi-short multiplets may combine to
form long multiplets for which there are no shortening conditions, as shown
diagrammatically in \diamond\ or \diamondt. It is crucial, in order to
determine when anomalous dimensions are allowed in the interacting theory,
to identify when semi-short multiplets may combine in this fashion.
All semi-short multiplets which are present in the expansion of
$Z^{{1\over 8},{1\over 8}}_{\rm m.t.}$ after subtracting those coming from
$Z^{{1\over 8},{1\over 8}}_{\rm sugra}$ do so combine.
To demonstrate this we note that (as first observed in~\beis)
\eqn\zstid{
Z^{{1\over 8},{1\over 8}}_{\rm s.t.}(z,y,zy,b)=
Z^{{1\over 8},{1\over 8}}_{\rm s.t.{1\over 2}-BPS}(z,y,zy,b)\,,
}
with a corresponding formula in the multi-trace case following directly from this,
\eqn\zmtid{
Z^{{1\over 8},{1\over 8}}_{\rm m.t.}(z,y,zy,b)=
Z^{{1\over 8},{1\over 8}}_{\rm m.t., sugra}(z,y,zy,b)\ .
}
Identical results to \zstid\ and \zmtid\ also follow for $b=zy$.
Following from \zmtid\ we may then write
\eqn\fourl{\eqalign{
Z^{{1\over 8},{1\over 8}}_{\rm m.t.}(z,y,a,0) 
- {}& Z^{{1\over 8},{1\over 8}}_{\rm sugra}(z,y,a,0) \cr
\noalign{\vskip -1pt}
= {}& \Big ( 1 - {zy\over a} \Big ) \,
(z-a)(y-a) \sum_{n,m,\bj} {\tilde N}^{{1\over 4},0}_{n m ,\bj}\,
(-a)^{2\bj+1} \chi_{(n+m,m)}(z,y) \, ,\cr}
}
and also
\eqn\eightl{\eqalign{
Z^{{1\over 8},{1\over 8}}_{\rm m.t.}(z,y,a,b)
- {}& Z^{{1\over 8},{1\over 8}}_{\rm sugra}(z,y,a,b) \cr
= {}& \Big ( 1 - {zy\over a} \Big ) \Big ( 1 - {zy\over b} \Big )
\sum_{n,m,j,\bj} {\tilde N}^{0,0}_{{\rm long},n m ,j\bj}\, 
\hchi_{nm,j\bj}(z,y,-a,-b)  \, .\cr}
}
Using the result \quarzero\ the right hand side in \fourl\ involves
only an expansion in terms of $t=\quar,\bt=0$ characters at the unitarity threshold
where $\bar \de$, $\de$, as defined in \defdd, are both zero.
As a consequence all $(\quar,\quar)$ operators
in the right-most column of Table 3, can combine with certain $(\quar,\eight)$
semi-short operators in the third column of Table 4 to form $(\quar,0)$ multiplets.
Then the remaining $(\quar,\eight)$ operators are in pairs, in the third and fourth
columns of Table 4, as  necessary 
to construct further $(\quar,0)$ $\eight$-BPS multiplets.
These combinations of multiplets are a reflection of the character identities,
following  from \fourch\ and \redt,
\eqn\decomp{\eqalign{
\chi^{{1\over 4},0}_{(\Delta;2+2\bj+q,p,q;0,\bj)} &
=\chi^{{1\over 4},{1\over 8}}_{(\Delta;2+2\bj+q,p,q;0,\bj)}
+\chi^{{1\over 4},{1\over 8}}_{(\Delta+{1\over 2};{5\over 2}+2\bj+q,p,q+1;0,
\bj-{1\over 2})}\cr
\chi^{{1\over 4},0}_{(\Delta;q+2,p,q;0,0)} &=
\chi^{{1\over 4},{1\over 8}}_{(\Delta;q+2,p,q;0,0)}
+\chi^{{1\over 4},{1\over 4}}_{(\Delta+1;q+2,p,q+2;0,0)}
\, ,}
}
where $\Delta=3+3\bj+p+2q$. In terms of the coefficients in \foure\ we have
\eqn\solN{
N^{{1\over 4},{1\over 8}}_{{\rm m.t.}, n m ,\bj} -
N^{{1\over 4},{1\over 8}}_{{\rm sugra}, n m ,\bj} =
{\tilde N}^{{1\over 4},0}_{n m ,\bj} + 
{\tilde N}^{{1\over 4},0}_{n \, m-1 ,\bj+{1\over 2}}\,.
}

In a similar fashion \longeight\ shows that in \eightl\ only characters for 
long multiplets with $\de=\bar \de=0$ are present in the expansion.
This indicates that all primary operators which are not supergravity dual states
can kinematically  join together to form long operators, as in \diamond.
Assuming, as discussed in the next section, such operators actually do combine 
dynamically and that they are the only ones that do so in this sector
we may then read off the number of long multiplets in
the interacting theory with anomalous dimensions for each representation. Hence
from \eightl
\eqn\solN{\eqalign{
& N^{{1\over 8},{1\over 8}}_{{\rm m.t.}, n m ,j\bj} -
N^{{1\over 8},{1\over 8}}_{{\rm sugra}, n m ,j\bj} \cr
&{} =
{\tilde N}^{0,0}_{{\rm long},n m ,j \bj} + 
{\tilde N}^{0,0}_{{\rm long},n \,m-1 ,j\, \bj+{1\over 2}} +
{\tilde N}^{0,0}_{{\rm long},n\, m-1 ,j+{1\over 2}\,\bj} +
{\tilde N}^{0,0}_{{\rm long},n\, m-2 ,j+{1\over 2}\,\bj+{1\over 2}} \,. \cr}
}

For $b=0$ setting $a=zy$ removes all contributions which may combine to 
form $(\quar,0)$ multiplets in \foure.
Thus
\eqn\indexe{\eqalign{
Z^{{1\over 8},{1\over 8}}_{\rm sugra}& (z,y,zy,0)
- Z_{\rm m.t.}^{{1\over 2},{1\over 2}}(z,y,zy,0)\cr
& {} = {1\over 1-zy} \prod_{k=2}^\infty {1\over (1-z^k)(1-y^k)} -
{z(1-y)\over z-y} \prod_{k=2}^\infty {1\over 1-z^k} +
{y(1-z)\over z-y} \prod_{k=2}^\infty {1\over 1-y^k} \cr
&{}= (1-z)(1-y)(zy)^2 \sum_{n,m} I^{{1\over 4},{1\over 8}}_{nm} \, \chi_{(n+m,m)}(z,y)\, ,
}
}
where, assuming \relN,
\eqn\Ieight{
I^{{1\over 4},{1\over 8}}_{nm} = \sum_{r = -1}^{m} (-1)^{r+1} 
N^{{1\over 4},{1\over 8}}_{{\rm sugra}, n\, m- r , {1\over 2}r} \, ,
\quad n\ge 0, \,  m\ge -1 \, .
}
Here $I^{{1\over 4},{1\over 8}}_{nm}$ is an index \mald\ whose magnitude determines 
a lower bound for the number of possible $t=\quar,\, \bt =\eight$ multiplets. For this
index we find
\eqn\Inm{
I^{{1\over 4},{1\over 8}}_{nm} = p(n+m+2)p(m+2) - 
\sum_{r=1}^{m+2} \big ( p(n+m+4-r) - p(n+m+2-r) \big ) p(m+2-r)  \, .} 
Particular results are easily found, for the lowest $n,m$,

\hskip 2cm
\vbox{\tabskip=0pt \offinterlineskip
\halign{& \vrule# &\strut \ \hfil#  \cr
\noalign{\vskip 3pt}
\omit &$I_{nm}^{{1\over 4},{1\over 8}}$\hfil && \ \ \hfil &\omit &
\hfil &\omit & \ \hfil \cr
\omit & $\quad n^{\dps \, m} $ \hfil &&\quad ${}^{\dps -1}$ \hfil
&\omit&\ ${}^{\dps 0}$ \hfil &\omit&\ ${}^{\dps 1}$ \hfil
&\omit&\ ${}^{\dps 2}$ \hfil &\omit&\ ${}^{\dps 3}$ \hfil
&\omit&\ ${}^{\dps 4}$ \hfil \cr
\noalign{\hrule}
\omit & \ 0  \ \hfil&& \quad 0  \hfil &
\omit & \ $1$ \ \hfil &\omit &\ $0$ \ \hfil
&\omit & \ $4$ \ \hfil &\omit &\ $-2$ \ \hfil&\omit & \ $14$ \ \hfil \cr
\omit & \ 1  \ \hfil&& \quad 0  \hfil &
\omit & \ $1$ \ \hfil &\omit &\ $2$ \ \hfil
&\omit & \ $4$ \ \hfil &\omit &\ $6$ \ \hfil&\omit & \ $17$ \ \hfil \cr
\omit & \ 2  \ \hfil&& \quad 0  \hfil &
\omit & \ $3$ \ \hfil &\omit &\ $2$ \ \hfil
&\omit & \ $12$ \ \hfil &\omit &\ $7$ \ \hfil&\omit & \ $27$ \ \hfil \cr
\omit & \ 3  \ \hfil&& \quad 1 \hfil &
\omit & \ $4$ \ \hfil &\omit &\ $7$ \ \hfil
&\omit & \ $16$ \ \hfil &\omit &\ $20$ \ \hfil&\omit & \ $70$ \ \hfil \cr
\omit & \ 4  \ \hfil&& \quad 1  \hfil &
\omit & \ $8$ \ \hfil &\omit &\ $9$ \ \hfil
&\omit & \ $29$ \ \hfil &\omit &\ $29$ \ \hfil&\omit & \ $92$ \ \hfil \cr
\omit & \ 5  \ \hfil&& \quad 3  \hfil &
\omit & \ $11$ \ \hfil &\omit &\ $17$ \ \hfil
&\omit & \ $41$ \ \hfil &\omit &\ $55$ \ \hfil&\omit & \ $176$ \ \hfil \cr
}
}

\noindent
The result here are in accord with the numbers of symmetric operators
listed in Tables 3 and 4.

To analyse those $(\eight,\eight)$ multiplets which cannot form long or
$(\eight,0)$ or $(0,\eight)$ multiplets we may set $a=b=zy$ when we obtain
\eqn\indexz{\eqalign{
& Z^{{1\over 8},{1\over 8}}_{\rm sugra}(z,y,zy,zy)
- Z_{\rm m.t.}^{{1\over 2},{1\over 2}}(z,y,zy,zy)\cr
& {} = {1\over 1-zy}\,  {\prod_{k=1}^\infty (1-z^{k+1}y^k)(1-z^ky^{k+1}) \over
\prod_{k=2}^\infty (1-z^k)(1-y^k)(1-z^ky^k)} \cr
& \quad {} -
{1\over (1-zy)(z-y)} \bigg (z(1-y)^2 \prod_{k=2}^\infty {1\over 1-z^k} -
y(1-z)^2 \prod_{k=2}^\infty {1\over 1-y^k} \bigg ) \cr
&{}= {(1-z)^2(1-y)^2\over 1-zy}\, (zy)^2 
\sum_{n,m} I^{{1\over 8},{1\over 8}}_{nm} \, \chi_{(n+m,m)}(z,y)\, ,
}
}
where
\eqn\Izero{
I^{{1\over 8},{1\over 8}}_{nm} = \sum_{r,s = -1}^{m} (-1)^{r+s}
N^{{1\over 8},{1\over 8}}_{{\rm sugra}, n\, m- r-s-2,{1\over 2}r\, {1\over 2}s}\, ,
\quad n\ge 0, \, m\ge-1 \, .
}
When $r,s=-1$ it is necessary to use \specN\ and \relN. For the special case $m=-1$,
from \Ieight\ and \Izero,
\eqn\Inone{
I^{{1\over 4},{1\over 8}}_{n\, -1} = I^{{1\over 8},{1\over 8}}_{n\, -1} 
= N^{({1\over4},{1\over 4})}_{{\rm m.t.,sym.,}n1} \, ,
}
which is the number of $\quar$-BPS multiplets belonging to the representation
$\R^{(0,0)}_{[1,n,1]}$, given explicitly in \quarcoeff. As remarked in \fadho\
such multiplets are always protected in that they cannot be combined to form
long multiplets. In other cases expansion of \indexz\ gives

\hskip 2cm
\vbox{\tabskip=0pt \offinterlineskip
\halign{& \vrule# &\strut \ \hfil#  \cr
\noalign{\vskip 3pt}
\omit &$I_{nm}^{{1\over 8},{1\over 8}}$\hfil   && \ \ \hfil &\omit &
\hfil &\omit & \ \hfil \cr
\omit & $\quad n^{\dps \, m} $ \hfil &&\quad ${}^{\dps 0}$ \hfil
&\omit&\ ${}^{\dps 1}$ \hfil &\omit&\ ${}^{\dps 2}$ \hfil
&\omit&\ ${}^{\dps 3}$ \hfil &\omit&\ ${}^{\dps 4}$ \hfil
&\omit&\ ${}^{\dps 5}$ \hfil \cr
\noalign{\hrule}
\omit & \ 0  \ \hfil&& \quad 1  \hfil &
\omit & \ $0$ \ \hfil &\omit &\ $7$ \ \hfil
&\omit & \ $-5$ \ \hfil &\omit &\ $29$ \ \hfil&\omit & \ $-13$ \ \hfil \cr
\omit & \ 1  \ \hfil&& \quad 1  \hfil &
\omit & \ $3$ \ \hfil &\omit &\ $6$ \ \hfil
&\omit & \ $8$ \ \hfil &\omit &\ $30$ \ \hfil&\omit & \ $14$ \ \hfil \cr
\omit & \ 2  \ \hfil&& \quad 4  \hfil &
\omit & \ $3$ \ \hfil &\omit &\ $21$ \ \hfil
&\omit & \ $7$ \ \hfil &\omit &\ $71$ \ \hfil&\omit & \ $21$ \ \hfil \cr
\omit & \ 3  \ \hfil&& \quad 5 \hfil &
\omit & \ $11$ \ \hfil &\omit &\ $27$ \ \hfil
&\omit & \ $26$ \ \hfil &\omit &\ $99$ \ \hfil&\omit & \ $85$ \ \hfil \cr
\omit & \ 4  \ \hfil&& \quad 11  \hfil &
\omit & \ $14$ \ \hfil &\omit &\ $49$ \ \hfil
&\omit & \ $36$ \ \hfil &\omit &\ $176$ \ \hfil&\omit & \ $84$ \ \hfil \cr
\omit & \ 5  \ \hfil&& \quad 15  \hfil &
\omit & \ $26$ \ \hfil &\omit &\ $69$ \ \hfil
&\omit & \ $74$ \ \hfil &\omit &\ $222$ \ \hfil&\omit & \ $208$ \ \hfil \cr
}
}

A third class of operators which are  interesting to examine are the operators 
formed from products of symmetrised traces. For the cases discussed previously
these were identical to the operators present in the supergravity limit.
For $(\eight,\eight)$ operators, which generate semi-short representations, there 
are symmetric operators which are not present in supergravity limit (the
simplest example being the Konishi operator). For completely symmetric
single trace operators the partition function in this sector is given by \singtsym
\eqn\zssstsym{\eqalign{
Z_{\rm s.t.,sym}^{{1\over 8},{1\over 8}}(z,y,a,b)
&=\exp\Big(\sum_{m=1}^{\infty} {1\over m}
Z^{{1\over 8},{1\over 8}}(z^m,y^m,a^m,b^m)
\Big)-Z^{{1\over 8},{1\over 8}}(z,y,a,b)-1\cr
&=\prod_{n=0}^{\infty}{(1- a\, \si^n)(1 - b\, \si^n\big) \over
(1- z\, \si^n)(1 - y \, \si^n\big)}
- {1\over 1- \si}(z+y-\si -a -b +1 )\,.}
}
The multi-trace operators then formed from symmetric single trace operators may 
then be counted from \multitrsym\ using the partition function,
\eqn\zssmtsym{
Z_{\rm m.t., sym.}^{{1\over 8},{1\over 8}}(z,y,a,b)
=\exp\Big(\sum_{m=1}^{\infty} {1\over m}
Z_{\rm s.t., sym.}^{{1\over 8},{1\over 8}}(z^m,y^m,a^m,b^m)\Big)\,,
}
Since, from \zssstsym, 
$Z_{\rm s.t.,sym}^{{1\over 8},{1\over 8}}(z,y,a,0) =
Z_{\rm s.t.,{1\over 2}-BPS}^{{1\over 8},{1\over 8}}(z,y,a,0)$, as given in 
\sssthalf, then also $Z_{\rm m.t., sym.}^{{1\over 8},{1\over 8}}(z,y,a,0) =
Z_{\rm sugra}^{{1\over 8},{1\over 8}}(z,y,a,0)$, so that the supergravity and symmetric
operator  partition functions are again identical in the $(\quar,\eight)$ sector. 
Furthermore as $Z_{\rm s.t.,sym}^{{1\over 8},{1\over 8}}(z,y,z,b) =
Z_{\rm s.t.,{1\over 2}-BPS}^{{1\over 8},{1\over 8}}(z,y,z,b)$, as in \quarz,
we may replace $ Z^{{1\over 8},{1\over 8}}_{\rm m.t.}(z,y,a,b)\to
Z_{\rm m.t., sym.}^{{1\over 8},{1\over 8}}(z,y,a,b)$ in~\ssspectrum\ with
$N^{{1\over 8},{1\over 8}}_{\rm m.t.}\to N^{{1\over 8},{1\over 8}}_{\rm sym.}$ 
to give the spectrum of symmetric primary operators, excluding those belonging to
$\half$-BPS multiplets, in the 
$(\eight,\eight)$ sector of $\N=4$ superconformal Yang Mills theory.

The representation content for the various highest operators that arise can be 
conveniently summarised by the following spaces
\eqn\defSSS{\eqalign{
\S_{\rm sugra} = {}& {\ts \bigoplus_{nm,j\bj}}\, 
N^{{1\over 8},{1\over 8}}_{{\rm sugra}, n m ,j\bj}\, 
\R_{[m+2\bj+2,n,m+2j+2]}^{(j,\bj)} \, , \cr
\S_{\rm sym.} = {}& {\ts\bigoplus_{nm,j\bj}}\, 
N^{{1\over 8},{1\over 8}}_{{\rm sym.}, n m ,j\bj}\, 
\R_{[m+2\bj+2,n,m+2j+2]}^{(j,\bj)} \, , \cr
\S_{\rm long} = {}& {\ts\bigoplus_{nm,j\bj}}\, 
{\tilde N}^{0,0}_{{\rm long},n m ,j\bj}\, 
\R_{[m+2\bj+2,n,m+2j+2]}^{(j,\bj)} \, , \cr
\S_{\rm free} = {}& {\ts\bigoplus_{nm,j\bj}}\, 
N^{{1\over 8},{1\over 8}}_{{\rm m.t.}, n m ,j\bj}\, 
\R_{[m+2\bj+2,n,m+2j+2]}^{(j,\bj)} \, , \cr}
}
where $\S_{\rm sugra}$ is then formed from the set of representations, with 
multiplicities, for highest weight states of supergravity dual multiplets
({\it i.e.} those formed from tensor products of $\half$-BPS states),
$\S_{\rm sym.}$  correspondingly from the  set of highest weight representations of 
supermultiplets formed from symmetric traces other than those in $\S_{\rm sugra}$,
$\S_{\rm long}$ from the set of highest weight representations
for long supermultiplets and $\S_{\rm free}$ from the set of all highest weight 
$(\eight,\eight)$ representations for all supermultiplets in the free
theory. We claim that these spaces of operators are all nested as 
follows,\foot{Previously discussed cases are much simpler. There we have
$\S_{\rm sugra}= \S_{\rm sym.} \subset \S_{\rm free}$}
\eqn\subs{
\S_{\rm sugra}\subset \S_{\rm sym.} \, , \qquad 
\S_{\rm sym.}/\S_{\rm sugra} \subset \S_{\rm long}\subset
\S_{\rm free}\ ,
}
requiring $0\le N^{{1\over 8},{1\over 8}}_{{\rm sym.}, n m ,j\bj} - 
N^{{1\over 8},{1\over 8}}_{{\rm sugra}, n m ,j\bj}\le 
{\tilde N}^{0,0}_{{\rm long},n m ,j\bj}$. The assumption here that all symmetric 
operators, other than those dual to the supergravity sector, are primary operators 
for long multiplets is discussed further in the next section.

The results,  obtained using the above formulae for character expansions, 
in the simplest cases are presented  in appendix C. 
Table 6 lists those operators dual to the supergravity sector obtained from
$N^{{1\over 8},{1\over 8}}_{{\rm sugra}, n m ,j\bj}$ in \ssspectrum, Tables
7 and 8 list those obtained from \eightl\ which form primary operators
for long multiplets, divided into those which are formed from symmetric traces 
other than those in the supergravity sector and also those which are not,
while Table 9 lists all remaining $(\eight,\eight)$ operators which are
descendants for those in Tables 7 and 8. Tables 10 and 11 give the same information
as in Tables 7 and 8 but for single trace operators.

\noindent
{\bf $t=\quar, \ \bt=0$ operators}

In this sector the operators are constructed from the fields in \Zquar,
with associated letters $z,y,x,a,{\bar a}$.
The single particle partition function is given, from \qzph, by
\eqn\eigthsp{ Z^{{1\over 4},0}(z,y,x,a,{\bar a})=
\hchi{}_1^{{1\over 4},0}(z,y,x;-a,-\ba)= z+y+x-a-{\bar a}\, . }
The multi-trace partition function then takes the form
\eqn\eigthmt{ Z^{{1\over 4},0}_{\rm m.t.}(z,y,x,a,{\bar a})
= {(1-z)(1-y)(1-x)\over (1-a)(1-{\bar a})}\prod_{m=1}^\infty
{1\over 1-z^m-y^m-x^m+a^m+{\bar a}^{m}}\,.} 

We also find from \singtsym\ and \singthbps,
\eqn\eigthstsym{\eqalign{
Z^{{1\over 4},0}_{\rm s.t., sym.}(z,y,x,a,{\bar a}) &=
Z^{{1\over 4},0}_{\rm s.t., {1\over 2}-BPS}(z,y,x,a,{\bar a})
= \sum_{p=2}^\infty 
\hchi^{\vphantom g}_p{\!\!}^{{1\over 4},0}(z,y,x;-a,-\ba) \cr 
&={(1-a)(1-{\bar a})\over (1-z)(1-y)(1-x)}-z-y-x+a+{\bar a}-1 \cr
& =\sum_{k,l,m\ge0,r,s=0,1 \atop k+l+m+r+s \ge 2}z^ky^lx^m(-a)^r(-{\bar a})^s \,,}
}
using \qzph\ for the $\half$-BPS characters. In this sector
symmetric single trace operators are identical with $\half$-BPS operators.
It follows that for the multi-trace partition functions
$Z^{{1\over 4},0}_{\rm m.t., sym.}(z,y,x,a,{\bar a})=
Z^{{1\over 4},0}_{\rm sugra}(z,y,x,a,{\bar a})$, with explicitly
\eqn\eigthmtsym{
Z^{{1\over 4},0}_{\rm m.t.,sym.}(z,y,x,a,{\bar a}) 
= {\prod_{k,l,m\ge 0,k+l+m\ge 1} (1- a\, z^ky^lx^m)
(1- {\bar a}\,z^ky^lx^m)\over
\prod_{k,l,m\ge 0,k+l+m\ge 2}(1-z^ky^lx^m)
\prod_{k,l,m\ge 0}(1- a{\bar a}\, z^ky^lx^m)}\,.
}

To disentangle the contributions of various supermultiplets we require
expansions of the form
\eqn\expZ{\eqalign{
Z^{{1\over 4},0}_{\rm m.t.}(z,y,x,a,{\bar a}) = {}& \sum_{p =2}^\infty
N_{{\rm m.t.},p}^{{1\over 2}{\rm -BPS}} \, 
\hchi{}_{\lower 1pt \hbox{$\ss p$}}^{{1\over 4},0}(z,y,x;-a,-\ba) \cr
&{} + \sum_{p,q= 0}^\infty\sum_{2\bj=-1}^\infty
N^{{1\over 4},{1\over 8}}_{{\rm m.t.}, pq ,\bj}\,
\hchi^{{1\over 4},0}_{{\rm semi},pq,\bj}(z,y,x;-a,-\ba) \cr
&{} + \sum_{k,p,q= 0\atop k-q \ge 2 + 2\bj }^\infty\sum_{2\bj=0}^\infty
N^{{1\over 4},0}_{{\rm m.t.},[k,p,q] ,\bj}\,
\hchi^{{1\over 4},0}_{[k,p,q],\bj}(z,y,x;-a,-\ba) \, , 
}}
involving $\half$-BPS characters given in  \qzph\ as well as 
\eqn\charqr{\eqalign{
\hchi^{{1\over 4},0}_{[k,p,q],\bj}(z,y,x;a,\ba) = {}&
(xyz)^{{1\over 2}(k-q-2\bj-4)}\, \chi_{(p+q,q,0)}(z,y,x)\, \chi_{(2\bj,0)}(a,\ba) \cr
&{}\times (z+a)(y+a)(x+a)(z+\ba)(y+\ba)(x+\ba)\, ,}
}
which is obtained from \quarzeshv\ and \charsu\ involving $U(3)$ and $U(2)$
characters $\chi_{(p+q,q,0)}$ and $\chi_{(2\bj,0)}$, and also the reduced characters 
calculated from \charmixed,
\eqn\charqmix{\eqalign{
\hchi^{{1\over 4},0}_{{\rm semi},pq,\bj}(z,y,x;a,\ba) = 
{\frak W}^{\S_3}_{xyz} {\overline \frak{W}}{}^{\S_2}_{a\ba} & \big ( 
(yz)^{-1} C_{(p+q,q,0)}(z,y,x)\, C_{(2\bj,0)}(a,\ba) \cr
&{}\times (z+a)(y+a)(x+a)(z+\ba)(y+\ba) \big ) \, .}
}
In \charqmix\ the Weyl symmetrisers over $x,y,z$ and $a,\ba$ act so as to
generate an expression for this semi-short character in terms of a sum over
standard $U(3)$ and $U(2)$ characters. In the limit $x,\ba\to 0$, with
$\ba/x = yz/a$, the expansion reduces to that in \foure\ and \fourl. With
a similar expansion to \expZ\ for $Z^{{1\over 4},0}_{\rm m.t.,sym.}(z,y,x,a,{\bar a})$,
in addition to the $(\quar,\quar)$ and $(\quar,\eight)$ operators tabulated 
previously, the first few
necessary primary $(\quar,0)$ $\eight$-BPS operators are listed in Table 5.

\medskip
\vbox{
\hskip 1cm
Table 5 
\nobreak

\hskip 1cm
\vbox{\tabskip=0pt \offinterlineskip
\hrule
\halign{&\vrule# &\strut \ \hfil#\  \cr
height2pt&\multispan5 &\cr
&\multispan5 \hfil\ \ $(\quar,0)$ primary operators\ \ \hfil&\cr
height2pt&\multispan5 &\cr
&\multispan5\hrulefill& \cr
height2pt&\omit&&\omit&&\omit&\cr
&\ $\De-\bj$ \   && \ Symmetric operators \hfil && Remaining operators\ \ \hfil &\cr
height2pt&\omit&&\omit&&\omit&\cr
\noalign{\hrule}
height2pt&\omit&&\omit&&\omit&\cr
& \ 6  \ \hfil && \ $\R_{[4,0,0]}^{(0,0)}$ \hfil &
& \ $4(2)\R_{[4,0,0]}^{(0,0)}$ \ \hfil &\cr
& \ 7  \ \hfil&& \ $\R_{[4,1,0]}^{(0,0)}$ \hfil &
& \ $7(3)\R_{[4,1,0]}^{(0,0)}$, $4(2)\R_{[5,0,0]}^{(0,{1\over 2})}$,\ \hfil &\cr
& \ 8  \ \hfil&& \
$3 \R_{[4,2,0]}^{(0,0)}$ \hfil &
& \
$24(8)\R_{[4,2,0]}^{(0,0)}$,
$12(5)\R_{[5,0,1]}^{(0,0)}$, \ \hfil &\cr
& \   \ \hfil&& \
\hfil  &
& \ $22(8)\R_{[5,1,0]}^{(0,{1\over 2})}$, $8(3)\R_{[6,0,0]}^{(0,1)}$
 \ \hfil &\cr
height2pt&\omit&&\omit&&\omit&\cr
}
\hrule}
}

{\eightpoint
{\parindent 1.5cm{\narrower
\noindent
$(\quar,0)$ primary operators belonging to representations 
$\R^{(0,\bj)}_{[k,p,q]}$, with $k-q>2+2\bj$, 
as obtained from expansion of partition function. When present numbers of single
trace operators are listed in parenthesis.

}}}

\noindent
{\bf $t=\eight,\ \bt=0$ semi-short operators}

The fields for this sector are listed in \ZYX. All the various $(t,\bt)$ considered 
above are subsumed as special cases by setting various letters to zero.
We have not attempted in this paper to extend at the same level of detail
the previous discussion to this sector, it is considered in \mald. Of course
the results already obtained are a necessary corollary of the expansions of 
the partition functions for this sector in terms of appropriate characters although
the algebraic complexity increases significantly. 

We list below a few 
salient formulae in the notation of this paper for future reference.

The basic single particle partition function is given by \ezph, with $xyz=a\ba$
and $\lam=b/yz$,
\eqn\singpartsixteenth{\eqalign{
&Z^{{1\over 8},0}(z,y,x,a,{\bar a},\lam) 
= \hchi_{\lower 1pt \hbox{$\ss 1$}}^{{1\over 8},0}(z,y,x;-a,-\ba,-b) \cr
&{} = { 1 \over (1 - \lam a)(1-\lam \ba)} \, \big ( z+y+x - a - \ba 
- \lam(yz+zx+xy) + xyz(\lam^2+\lam) \big ) \, .
}}
The corresponding single trace partition function for $\half$-BPS operators is
\eqn\sssthalfsh{
Z_{\rm s.t., {1\over 2}-BPS}^{{1\over 8},0}(z,y,x,a,{\bar a},\lam)
= \sum_{p=2}^\infty
\hchi^{\vphantom g}_p{\!\!}^{{1\over 8},0}(z,y,x;-a,-\ba,-b) \, .
}
To calculate this starting from \ezph\ we may use
\eqn\calW{\eqalign{
& \sum_{p=1}^\infty
\hchi^{\vphantom g}_p{\!\!}^{{1\over 8},0}(z,y,x;-a,-\ba,-b)\cr
\noalign{\vskip -2pt}
&{} = { 1 \over (1 - \lam a)(1-\lam \ba)}  \,
{\frak W}^{\S_3}_{xyz} \bigg ( {
(1- z^{-1} a)(1- z^{-1} \ba)(1 - \lam x)(1- \lam y) \over (z-y)(z-x)(y-x)}\, {z^3 y \over 1-z}
\bigg ) \cr
&{}= {1\over (1 - \lam a)(1-\lam \ba)(1-x)(1-y)(1-z)}\cr
&\quad {}\times \Big (
(x+y+z)(1-\lam xyz) - (\lam+1)(xy+yz+zx-2xyz) + \lam^2 xyz (1+  xyz)\cr
\noalign{\vskip -2pt}
&\qquad\quad {}- (a+\ba)\big (1 - \lam (xy+yz+zx-xyz)+ \lam^2 xyz \big )\Big ) \, . \cr}
}
With this we may obtain
\eqn\Zeight{\eqalign{
Z_{\rm s.t., {1\over 2}-BPS}^{{1\over 8},0}(z,y,x,a,{\bar a},1)
={}&  {x\over 1-x} + {y\over 1-y} + {z\over 1-z} - {a\over 1-a} - {\ba\over 1-\ba}\cr
&{} + {(1-x)(1-y)(1-z)\over (1-a)(1-\ba)} - 1 \, ,}
}
agreeing with comparable results in \mald.

\newsec{Further Remarks}

The results of the analysis of partition functions in the previous section
and also in \mald\ may be summarised in part as follows

\item{$\bullet$} In the $(\half,\half)$, $(\quar,\quar)$, $(\quar,0)$
sectors, which are composed of what are generally referred to as $\half$-, $\quar$-, 
$\eight$-BPS operators, as well as the $(\quar,\eight)$ sector, the set of operators 
obtained from symmetrised traces is identical with the supergravity dual operators 
formed by products of corresponding $\half$-BPS single trace operators.

\item{$\bullet$} In the $(\eight,\eight)$ sector supergravity dual
operators are a subset of the symmetric operators.

\item{$\bullet$} All operators in the $(\eight,\eight)$ and $(\quar,\quar)$, 
$(\quar,\eight)$ sectors, other than those dual to supergravity fields,
are potentially part of long multiplets and may gain an anomalous dimension in 
the interacting theory. Highest weight symmetric operators in the $(\eight,\eight)$ 
sector, other than those which are supergravity dual operators, are primary operators
for potential long multiplets.

In general in the analysis of $\N=4$ SYM a central issue is whether particular 
operators are protected in the interacting theory or gain an anomalous  scale
dimension for non zero coupling. Our discussion in the previous section
is essentially kinematic, determining when there are appropriate combinations
to form long multiplets which may then have an anomalous dimension for 
$\lambda\ne0$. It is then a dynamical question as to whether this occurs
in each case. The results for the different sectors formed from the basic elementary
fields of $\N=4$ SYM, described in section 6 depends on the superconformal
transformations given for free field in \QSfree. In an interacting theory these
are modified and in the $(\eight,\eight)$ sector we expect\foot{In \QSfree\ for 
the interacting theory the algebra is modified by
{\vskip -12pt \vbox{$$
\{ Q^i{}_{\! \alpha} , \lambda_{j\beta} \} = \de^i{}_{\! j}
F_{\alpha\beta} +  \half g \, \vep_{\alpha\beta}\, \vep^{iklm} [
\vphi_{jk}, \vphi_{lm} ] \, , \qquad
\{ \bQ_{\smash {j\dal}} , {\bar \lambda}^i{}_{\! \smash \dbe} \}
= \de^i{}_{\! j} {\bar F}_{\smash{\dal\dbe}} + \half g \, \vep_{\smash{\dal\dbe}}\, 
\vep^{iklm}[ \vphi_{jk}, \vphi_{lm} ] \, .
$$} \vskip-12pt} \noindent
and otherwise derivatives are replaced by covariant derivatives.}
\eqn\intQ{
\big \{ Q^1{}_{\! 2} , \lambda_{41} \big \} \sim \big [ Z , Y \big ] \, , \qquad
\big \{ \bQ_{41} , {\bar \lambda}{}^1{}_{\! 2 } \big \} \sim \big [ Z , Y \big ] \, .
}
For $t=\quar, \, \bt=0$ we have also
$\big \{ Q^1{}_{\! 2} , \lambda_{31} \big \} \sim \big [ X , Z \big ]$,
$\big \{ Q^1{}_{\! 2} , \lambda_{21} \big \} \sim \big [ Y , X \big ]$. In 
consequence the supercharges $Q^1{}_{\!2}, \, \bQ_{41}$ no longer
annihilate the $(\eight,\eight)$ primary operators, as in the free theory for a
semi-short multiplet, but generate the operators necessary to complete a long 
multiplet.
The descendants which are formed, assuming the supercharges act as in  \intQ\ 
and also when acting on covariant derivatives of fields inside a trace, involve 
commutators and so exclude operators formed by symmetrised traces in accord
with the kinematic observations made above.

All operators other than those dual to supergravity fields are therefore expected 
to combine into long multiplets not only kinematically but dynamically implying
that they then have non zero anomalous dimensions. 
Furthermore a natural conjecture is that those `supergravity' operators
which may also combine to form long operators kinematically nevertheless 
remain protected dynamically. These conclusions follow from an analytic 
superspace analysis \Aspects\ for the classical interacting theory and 
a recent paper \biswas\ gives further evidence for this claim in the 
$\eight$-BPS sector.

For $\half$-, $\quar$- and $\eight$-BPS,
operators constructed from chiral fields as in \ZZZ, \ZY\ and \ZYfe, 
the statements concerning protected operators
follow from the notion of the chiral ring \Chiral\ composed of
operators formed by multiple traces of the basic chiral fields modulo relations 
requiring that  commutators such as $[Z,Y]=[Z,{\bar\lambda}{}^1{}_{\! \dal}]=0$ or 
anti-commutators $\{{\bar \lambda}{}^1{}_{\!1} ,{\bar \lambda}{}^1{}_{\!2}\}=0$.
The counting of such operators is identical to that for symmetrised traces
of $Z,Y,X, {\bar \lambda}{}^1{}_{\!\dal}$ and these operators form a closed set 
under multiplication with scale dimensions additive. 

To justify these claims we review some evidence from perturbative calculations
and also from application of the AdS/CFT correspondence for the large $N$
strong coupling limit.

Historically the first protected operator to be identified in this context \OPEN\
belonged to the representation $\R^{(0,0)}_{[0,2,0]}$ with canonical 
conformal dimension $4$, denoted by ${\cal D}_{\bf 20'}$ in \opemix. 
In free theory this gives a highest weight state for a semi-short multiplet but,
as in \diamond, it may combine potentially with others to form a long multiplet. 
If we identify the primary $\half$-BPS operators $\O_{rs}=\tr (X_r X_s)
- {1\over 6} \de_{rs} \tr(X_t X_t)$ then this operator  is a double trace
operator of the form $\O_{rt}\O_{st} - {1\over 6} \de_{rs} \O_{tu}\O_{tu}$.
The  anomalous dimension was shown to vanish at one-loop \opemix. In addition 
a procedure for counting states to determine whether long multiplets may
be formed was also described in \BMStao\ (whimsically called the Eratosthenes' super 
sieve by analogy to finding prime numbers, this procedure of successively
removing superconformal descendants by hand is equivalent to an expansion
in terms of supermultiplet characters).
The authors of \mald\ devoted some attention to this example and were able to show
there are no available operators of higher conformal dimensions that 
${\cal D}_{\bf 20'}$ can be combined with to become long and therefore allow it
to acquire an anomalous dimension. 

These results for ${\cal D}_{\bf 20'}$, and other related operators, may also
be arrived at by applying the
index $I^{{1\over 8},{1\over 8}}_{nm}$ for the simple values of $m,n$. For $m=-1$ 
the index, from \Inone, just counts the number of 
$\quar$-BPS operators transforming in $\R^{(0,0)}_{[1,n,1]}$ representation
which, as  stressed in \fadho, cannot combine into long multiplets and must remain
short. For $m=0$ from \Izero, considering only operators dual to
supergravity fields for which the index may be non zero,
\eqn\Ibps{
I^{{1\over 8},{1\over 8}}_{n0} = N^{{1\over 4},{1\over 4}}_{{\rm m.t. sym.}, n2}
\big (\R^{(0,0)}_{[2,n,2]}\big )
- 2 N^{{1\over 4},{1\over 8}}_{{\rm sugra}, n0,0}\big (\R^{(0,0)}_{[2,n,0]}\big ) + 
N^{{1\over 8},{1\over 8}}_{{\rm sugra}, n 0, 00}\big (\R^{(0,0)}_{[0,n,0]}\big )\, ,
}
where we list in parenthesis the representations of the primary 
operators of each short/semi-short multiplet and we have used the symmetry
under conjugation $N^{{1\over 4},{1\over 8}} = N^{{1\over 8},{1\over 4}}$.
The non zero numbers for the first few cases are as follows,
\eqn\Inum{\eqalign{
I^{{1\over 8},{1\over 8}}_{00} ={}& N^{{1\over 4},{1\over 4}}_{{\rm m.t. sym.}, 02}
= 1 \, , \qquad  \
I^{{1\over 8},{1\over 8}}_{10} = N^{{1\over 4},{1\over 4}}_{{\rm m.t. sym.}, 12} 
= 1 \, , \cr
I^{{1\over 8},{1\over 8}}_{20} = {}& 4 \, , \ \
N^{{1\over 4},{1\over 4}}_{{\rm m.t. sym.}, 22} = 3 \, ,  \  \
N^{{1\over 8},{1\over 8}}_{{\rm sugra}, 2 0 , 00} = 1 \, ,\cr
I^{{1\over 8},{1\over 8}}_{30} = {}& 5 \, , \ \
N^{{1\over 4},{1\over 4}}_{{\rm m.t. sym.}, 32} = 4 \, , \ \
N^{{1\over 8},{1\over 8}}_{{\rm sugra}, 2 0 , 00} = 1 \, ,\cr
I^{{1\over 8},{1\over 8}}_{40} = {}& 11 \, , \
N^{{1\over 4},{1\over 4}}_{{\rm m.t. sym.}, 42} = 8 \, , \ \
N^{{1\over 8},{1\over 8}}_{{\rm sugra}, 40 , 00} = 3 \, ,\cr
I^{{1\over 8},{1\over 8}}_{50} = {}& 15 \, , \
N^{{1\over 4},{1\over 4}}_{{\rm m.t. sym.}, 22} = 11 \, , \
N^{{1\over 8},{1\over 8}}_{{\rm sugra}, 50 , 00} = 4 \, ,\cr
I^{{1\over 8},{1\over 8}}_{60} = {}& 25 \, , \
N^{{1\over 4},{1\over 4}}_{{\rm m.t. sym.}, 62} = 19 \, , \
N^{{1\over 4},{1\over 8}}_{{\rm sugra}, 60,0} = 1 \, , \
N^{{1\over 8},{1\over 8}}_{{\rm sugra}, 60, 00} = 8 \, .\cr}
}
For $n=2,3,4,5$ there are no negative contributions to the index so that
the $\quar$-BPS operators listed must all be protected as well as the
associated $(\eight,\eight)$ primary operators for semi-short multiplets, 
$n=2$ corresponds to the operator ${\cal D}_{\bf 20'}$ mentioned above.
The results for $n=2,3,4$  are in agreement with calculations which take into account
non-planar operator mixing of multi-trace operators formed by the fields
$Z,Y$ by applying a long-range version of the $SU(2)$ spin chain \Moraletal.
In each case as expected there are the required number of operators
with protected scale dimensions. Taking $n=6$ would however be a less trivial
case to consider in future investigation since then it is possible to form
a long multiplet.
For $m=1$ the index chain is longer,
\eqn\Ibpsone{\eqalign{
I^{{1\over 8},{1\over 8}}_{n1} = {}& N^{{1\over 4},{1\over 4}}_{{\rm m.t. sym.}, n3}
\big (\R^{(0,0)}_{[3,n,3]}\big )
- 2 N^{{1\over 4},{1\over 8}}_{{\rm sugra}, n1,0}\big (\R^{(0,0)}_{[3,n,1]}\big ) + 
2 N^{{1\over 4},{1\over 8}}_{{\rm sugra}, n0,{1\over 2}}
\big (\R^{(0,{1\over 2})}_{[3,n,0]}\big )\cr
& \!\!\! {}  + N^{{1\over 8},{1\over 8}}_{{\rm sugra}, n \,-1, 00}
\big (\R^{(0,0)}_{[1,n,1]}\big ) -
2 N^{{1\over 8},{1\over 8}}_{{\rm sugra}, n \,-2, {1\over 2}0}
\big (\R^{({1\over 2},0)}_{[0,n,1]}\big ) +
N^{{1\over 8},{1\over 8}}_{{\rm sugra}, n \,-3,{1\over 2}{1\over 2}}
\big (\R^{({1\over 2},{1\over 2})}_{[0,n,0]}\big ) \, . \cr}
}
For $n=0$ all contributions are zero whereas for $n=1,2$
\eqn\Inumt{\eqalign{
I^{{1\over 8},{1\over 8}}_{11} = {}& 3 \, , \
N^{{1\over 4},{1\over 4}}_{{\rm m.t. sym.}, 13} = 2 \, ,  \
N^{{1\over 8},{1\over 8}}_{{\rm sugra},1\,-1 , 00} = 1 \, ,\cr
I^{{1\over 8},{1\over 8}}_{21} = {}& 3 \, , \
N^{{1\over 4},{1\over 4}}_{{\rm m.t. sym.}, 23} = 3 \, ,
N^{{1\over 4},{1\over 8}}_{{\rm sugra}, 21,0} = 1 \, ,
N^{{1\over 8},{1\over 8}}_{{\rm sugra}, 2\,-1, 00} = 2 \, .\cr}
}
The second case allows for the possibility of forming a long multiplet.

However the corresponding $\quar$-BPS operators,
of scaling dimension $8$ and belonging to the  $\R^{(0,0)}_{[3,2,3]}$ representation,
have been analysed by Morales \Morales, following the methods of \Moraletal.
It transpires that three symmetric operators
corresponding to $\O^{{1\over 4},{1\over 4}}_{[3,2,3](0,0)}$ have vanishing
anomalous dimension at one-loop, including for non-planar interactions as
stressed above, although only two would be necessary to satisfy the index
(along with one $(\eight,\eight)$ operator belonging to $\R^{(0,0)}_{[1,2,1]}$). 
The analysis is non  trivial. It is necessary to include in the basis for the
spin chain
all multi-trace operators formed by fields $Z^5Y^3$ and also from $Z^8$,
which yields 7 protected $\half$-BPS operators, $Z^7Y$ giving additionally only 
protected $\quar$-BPS operators, 4 in all, and  $Z^6Y^2$  producing 8 more protected 
(for a total of 19 protected including the previous ones) as well as 8 unprotected 
operators. The operators formed from $Z^5Y^3$ then give 10 operators beyond those 
which are $SU(2)$ partners of the ones already discussed.
In order to resolve the mixing, it is necessary to diagonalise a $37 \times
37$ matrix or, equivalently, find the zeroes of the degree 37
characteristic polynomial! Luckily there are 22 zero eigenvalues
which corresponds to the 19 protected operators counted above plus
3 more corresponding to all three $\O^{{1\over 4},{1\over 4}}_{[3,2,3](0,0)}$
operators formed from symmetric traces.

Although tentative this seems to be an indication that the dynamics of $\N=4$ 
superconformal Yang Mills theory is more strictly constrained
than required just by $PSU(2,2|4)$ representation theory. 
Of course one loop results
for one operator are not sufficient to demonstrate such conclusions.
In \BRSta\ a specific triple trace scalar operator
of canonical conformal dimension $6$ (called ${\cal T}_6=\O_{rs}\O_{st}\O_{tr}$) 
was shown to have vanishing anomalous dimension at one loop despite its being 
part of a long unprotected multiplet. 
In both these one loop examples, attempting a two-loop calculation may
help sort out such issues but this is a formidable task.
Our expectation is that the vanishing of the anomalous dimension of 
${\cal T}_6$ is a one-loop accident, while all three 
$\O^{{1\over 4},{1\over 4}}_{[3,2,3](0,0)}$ operators 
remain {\it dynamically} protected.

\noindent
{\it Beyond the chiral ring}

If we look beyond the chiral ring at semi-short $(\eight,\eight)$ operators, or
$({1\over 8},0)$ operators the situation is more complicated. However, 
in \refs{\HesH,\Aspects} all operators which remain semi-short in the classical 
interacting
theory were classified with the aid of analytic superspace. This classification, 
which is expected to be valid in the quantum theory at non zero coupling assuming
no quantum anomalies in the action of supercharges, 
is straightforward: an operator is short/semi-short in the classical
interacting theory if and only if it is short/semi-short in the free
theory and is constructed from $\half$-BPS operators.\foot{Mixing in the
quantum theory will mean the precise definition of the protected
operators may be different from this, but the counting should remain
the same.}

With this classification of operators for weakly interacting
$\N=4$ SYM, inherited from the classical superconformal theory, it is trivial
that the partition function, restricted to short/semi-short  operators, is the
same at weak and strong coupling, since they are
calculated in exactly the same way.    At weak
coupling the partition function is defined as in \multigrav. At
strong coupling, on the other hand, according to the AdS/CFT
correspondence,  the partition function is that of a gas of free
gravitons together with superpartners on AdS/CFT. The single particle
partition function of these states is represented by \singthbps, which is
equal to the single particle partition function of $\half$-BPS operators
in the field theory, and the partition function of a free gas of such
states is defined precisely as in \multigrav.

Applying these arguments independently shows that  ${\cal D}_{\bf 20'}$ is
non-renormalised: it is semi-short in the free theory
and is constructed from two $\half$-BPS operators (up to mixing). 
More generally for $\O^{(n)}_{r_1 \dots r_n} = \tr(X_{(r_1} \dots 
X_{r_n)}) - \hbox{traces}$, the single trace $\half$-BPS primary
operators formed from the basic six component scalars $X_r$ for 
representation $[0,n,0]$, then all
semi-short supergravity dual primary operators belonging to the
representations $\R^{(0,0)}_{[0,p,0]}$, given by symmetric traceless tensors
of rank $p$ $\O_{r_1\dots r_{p}}$, are formed from products of two or
more $\O^{(n)}$, $\sum n=p+2$,  with one pair of indices contracted and all 
remaining $p$ free indices symmetrised and  traces subtracted. 
For representations $\R^{(0,0)}_{[1,p-2,1]}$, described by tensor fields
$\O_{r,r_1\dots r_{p-1}}=\O_{r,(r_1\dots r_{p-1})}$ where
$\O_{(r,r_1\dots r_{p-1})}=0$ and contractions of any pair of indices are zero,
a similar construction of all corresponding semi-short operators is possible. 
Starting from a product $\prod_n \O^{(n)}$, $\sum n=p+2$, one 
pair of indices is again contracted and another pair is antisymmetrised 
before $p-1$ free indices are symmetrised and traces subtracted. The 
indices which are contracted or antisymmetrised must come from different $\O^{(n)}$.
This procedure gives the two semi-short operators
$\O^{{1\over 8},{1\over 8}}_{[1,2,1](0,0)}$, formed from $\O^{(2)\, 3}$ and
$\O^{(2)}\O^{(4)}$, which  are therefore protected.

There is in fact an alternative way to show that some semi-short operators are
non-renormalised at non zero coupling.
This involves considering three point functions for
two $\half$-BPS operators $\O^{{1\over 2},{1\over 2}}_{[0,p,0](0,0)}\equiv \O^{(p)}$
and an operator $\O$ of the form $\langle \O^{(p)} \O^{(q)} \O\rangle$. 
This was analysed in \sok\ and shown to be non zero only for
\eqnn\nonz
$$\eqalignno{
\O = {}& \O^{{1\over 2},{1\over 2}}_{[0,p+q-2r,0](0,0)} \, , \ r=0,\dots, q \, , 
\quad \O^{{1\over 4},{1\over 4}}_{[s,p+q-2r,s](0,0)}  \, , \ 
r=0,\dots, q, \, s=1,\dots q-r\,,\cr
{}& \O^{{1\over 8},{1\over 8}}_{[s,p+q-2r-2s,s](j,j)}\, , \ 
r=1,\dots, q, \, s=0,\dots q-r\, , \cr
{}& \O^{0,0}_{[s,p+q-2r-2s,s](j,j)} \, , \ r=2,\dots, q, \, s=0,\dots q-r\, , &\nonz\cr}
$$
assuming $p\ge q$ and for the long operator $\O^{0,0}$ we must have 
$\De\ge 2+ p+q-2r +2j$. For the semi-short operators $O^{{1\over 8},{1\over 8}}$,
$r=1$ is special so that
if $\O$ has twist $\De-2j= p+q$, the associated three point function is non zero
only if $\O$ is semi-short and has vanishing anomalous dimension. 
This was used in \sok\ to prove the non-renormalisation of the
operator ${\cal D}_{\bf 20'}$ and further demonstrates 
that any short/semi-short operator of twist $p+q$ in the operator product of
$\O^{(p)} \O^{(q)}$ is protected. 
Equivalent results may also be obtained using superconformal Ward identities
for the operator product expansion applied to the four point function
$\langle\O^{(p)}\O^{(q)} \O^{(p)} \O^{(q)} \rangle$ \refs{\NO,\OPEexp}. 
Hence performing a conformal partial wave analysis on this four point function
in the free theory implies that all semi-short operators of twist $p+q$ 
which are present in the operator product expansion for $\O^{(p)} \O^{(q)}$
must be protected in the interacting theory. This analysis
demonstrates that semi-short operators belonging to the representations
$\R^{(j,j)}_{[q,p,q]}$ must be present for any $j,p,q$ although it is harder
to disentangle the number of such operators in each case in this fashion.

\noindent
{\it Construction of semi-short operators}

As  has been discussed in section 6 the construction of $(\eight,\eight)$
semi-short operators can be reduced to operators formed from the fields
in \ZYd. In terms of the letters in Table 1 these correspond to words
$a^{s+1}b^{t+1}z^uy^v$, with $u\ge v$, where the associated operator belongs to
the representation $\R^{({1\over 2}t,{1\over 2}s)}_{[v+s,u-v,v+t]}$. To
illustrate some examples we adopt the notation
\eqn\notS{
\Z^i=(Z,Y) \, , \ \  \lambda = \lambda_{4 1} \, , \ \ 
\blam = {\bar \lambda}^1{}_{\! 2} \, , \ \  \pr=\pr_{12} \, , \quad
S_i = (S_2{}^{\!1}, S_3{}^{\!1}) \, , \ \ \bS^i = (\bS^{22}, \bS^{32}) \, , 
}
so that from \QSfree
\eqn\SZY{\eqalign{
\big [ S_i , \pr^n \Z^j \big ] = {}& 2in \, \de^i{}_{\! j}\, \pr^{n-1} \blam \,, 
\qquad\qquad \ \big [ \bS^i , \pr^n \Z^j \big ] = 
- 2in \, \vep^{ij}\, \pr^{n-1} \lambda \,,\cr
\big \{ S_i , \pr^n \lambda \big \} = {}& - 4(n+1) \, \vep_{ij} \, \pr^n \Z^j \, , 
\qquad \big \{ \bS^i , \pr^n \blam \big \} = 4(n+1) \, \pr^n \Z^i \, , \cr}
}
as well as $\{ S_i , \pr^n \blam  \}= \{ \bS^i , \pr^n \lambda \} =0$.
For each word $a^{s+1}b^{t+1}z^uy^v$ there are various possible operators
formed by one or more traces which are annihilated by $ S_i , \bS^j$ which
may also be required to be $SU(2)$ highest weight states. If only
$Z,Y$, without any derivatives,  are present these are just $\quar$- or 
$\half$-BPS operators. To discuss the simplest operators in this sector we
consider
\eqn\trial{
\sum_{s=0}^{n+1} \Big ( \alpha_s^{(n)} \, \pr^s \blam \otimes_\pm \pr^{n-s} \lambda
- \beta_s^{(n)} \, i\vep_{ij} \, \pr^s \Z^i \otimes_\pm \pr^{n+1-s}\Z^j \Big ) \, , 
\quad \alpha_{n+1}= 0 \, , \  \beta_{n+1-s}^{(n)} = \mp \beta_s^{(n)} \, ,
}
where $\otimes_\pm$ represents the symmetric/antisymmetric tensor product for $SU(N)$
matrices.
Imposing the conditions that this commutes with $S_i, \bS^i$ gives the relations
$(n-s+1)\alpha_s^{(n)} = - (s+1)\beta_{s+1}^{(n)}$, $(s+1)\alpha_s^{(n)} =
(n-s+1)\beta_{s+1}^{(n)}$ and hence we may take
\eqn\solab{
\alpha_s^{(n)} = {n+1\choose s} {n+1\choose s+1}(-1)^s \, , \qquad
\beta_s^{(n)} = {n+1\choose s}^2(-1)^s \, ,
}
with $n$ even, odd according to the two signs in \trial. 

Hence for the word $ab$ there are semi-short singlet operators of twist 2,
corresponding to the representations $\R^{({1\over 2}n,{1\over 2}n)}_{[0,0,0]}$ 
in Table 7, represented in this sector by
\eqn\Kon{
\sum_{s=0}^{n+1}  
\Big ( \alpha_s^{(n)} \, \tr \big (\pr^s \blam  \, \pr^{n-s} \lambda \big ) 
- \beta_s^{(n)} \, i\vep_{ij} \, \tr \big ( \pr^s \Z^i \, \pr^{n+1-s}\Z^j \big )\Big )
\, , \quad n=0,2,\dots \, .
}
For $n=0$ this corresponds to the well known Konishi scalar. 
The result \Kon\ is equivalent to the results for twist 2 operators in  \BHR. 
Similarly for the semi-short operators in Table 7 for the representations
$\R^{({1\over 2}n,{1\over 2}n)}_{[0,1,0]}$ we have
\eqn\Kone{
\sum_{s=0}^{n+1}
\Big ( \alpha_s^{(n)} \, \tr \big (Z \,[\pr^s \blam  , \pr^{n-s} \lambda]_{\mp} \big )
- \beta_s^{(n)} \, i\vep_{ij} \, \tr \big ( Z \, [ \pr^s \Z^i , \pr^{n+1-s}\Z^j ]_{\pm}
\big )\Big ) \, , 
}
with $[X,Y]_\mp = X Y \mp Y X$ and the two cases in \Kone\ require $n$ even/odd.
Other multiple trace operators may be similarly found. Thus there are four
$\Delta=4$ semi-short operators for $\R^{(0,0)}_{[0,2,0]}$
represented by, 
\eqn\Ofour{\eqalign{
{}& \tr \big ( Z \blam) \, \tr(Z \lambda\big ) - 2i \vep_{ij} \, \tr\big (Z\Z^i)\,  
\tr(Z \pr \Z^j\big ) \, ,
\quad \tr \big ( ZZ\big ) \, \tr\big 
(\blam \lambda - 2i \vep_{ij} \, \Z^i  \pr \Z^j \big ) \, , \cr
{}& \tr \big ( ZZ \,([\blam ,\lambda] - 2i \vep_{ij} \, \{\Z^i , \pr \Z^j \}) \big ) 
\, ,  \quad \quad \qquad \
\tr \big ( Z\blam Z \lambda - 2i \vep_{ij} \, Z \Z^i Z  \pr \Z^j\big ) \, , \cr}
}
in accord with Tables 6,7,8. The remaining 
$\Delta=4$ semi-short operator necessary according to  Table 8 for the representation
$\R^{(0,0)}_{[1,0,1]}$ corresponds to the single trace operator $ \vep_{kl}\, \tr 
\big ( \Z^k \Z^l \,([\blam, \lambda] - 2i\vep_{ij} \, \{\Z^i , \pr \Z^j \}) \big )$.

A privileged set of multi-trace operators are those constructed from $\half$-BPS
single trace operators. The simplest example in the $(\eight,\eight)$ sector
are those double trace operators composed of $[0,2,0]$ $\half$-BPS 
operators formed from the descendants of $\tr(ZZ)$. There are two cases to 
consider, an $SU(2)$ singlet
\eqn\Osing{\eqalign{
\O^{(n)}_0 = \sum_{s=0}^{n+2}
\Big ( {}& \half \alpha_{0,s}^{(n)} \, 
\pr^s\tr \big ( \blam \lambda - i\vep_{ij} \, \pr\Z^i \, \Z^j \big ) \,
\pr^{n-s} \tr \big ( \blam \lambda - i\vep_{kl} \, \pr\Z^k \, \Z^l \big ) \cr
\noalign{\vskip -6pt}
&{} + \beta_{0,s}^{(n)} \, i\vep_{ij} \, \pr^s \tr \big ( \blam \Z^i \big )
\, \pr^{n+1-s} \tr \big ( \lambda \Z^j \big )\cr 
\noalign{\vskip -3pt}
&{} + \quar \gamma_{0,s}^{(n)} \, \vep_{ik}\vep_{jl} \, \pr^s 
\tr\big ( \Z^i \Z^j \big )\, \pr^{n+2-s} \tr \big ( \Z^k\Z^l\big ) \Big ) \, , \cr}
}
where $\alpha_{0,s}^{(n)} = \alpha_{0,n-s}^{(n)}. \ 
\gamma_{0,s}^{(n)} = \gamma_{0,n+2-s}^{(n)}$ and the highest weight state for
a $SU(2)$ triplet
\eqnn\Otrip
$$\eqalignno{
\O^{(n)}_1 = \sum_{s=0}^{n+1}
\Big ( {}& \half \alpha_{1,s}^{(n)} \,
\pr^s\tr \big ( ZZ \big ) \,
\pr^{n-s} \tr \big ( \blam \lambda - i\vep_{kl} \, \pr\Z^k \, \Z^l \big ) 
- \beta_{1,s}^{(n)} \, \pr^s \tr \big ( \blam Z \big )
\, \pr^{n-s} \tr \big ( \lambda Z \big )\cr
\noalign{\vskip -3pt}
&{} + \quar \gamma_{1,s}^{(n)} \, i\vep_{ij}\, \pr^s
\tr\big ( \Z^i Z \big )\, \pr^{n+1-s} \tr \big ( \Z^j Z \big ) \Big ) \, , &\Otrip\cr}
$$
with $\gamma_{1,s}^{(n)} = - \gamma_{1,n+1-s}^{(n)}$. The operators in
\Osing\ and \Otrip\ correspond to the representations 
$\R^{({1\over 2}(n+1),{1\over 2}(n+1))}_{[1,0,1]}$ and
$\R^{({1\over 2}n,{1\over 2}n)}_{[0,2,0]}$ in Table 6. By using
\eqn\commS{\eqalign{
&\big [ S_i ,\pr^n \tr \big ( \blam \lambda - i\vep_{kl} \, \pr\Z^k \, \Z^l \big )
\big ] = 2(n+3) \, \vep_{ij} \pr^n \tr \big ( \blam \Z^j \big )  \, , \cr
& \big [ \bS^i ,\pr^n \tr \big ( \blam \lambda - i\vep_{kl} \, \pr\Z^k \, \Z^l \big )
\big ] = 2(n+3) \,  \pr^n \tr \big ( \lambda \Z^i \big ) \, , \cr
& \big \{ S_i , \pr^n \tr \big ( \lambda \Z^j \big ) \} =
2in \, \de_i{\!}^j \, \pr^{n-1} 
\tr \big ( \blam \lambda - i\vep_{kl} \, \pr\Z^k \, \Z^l \big ) - 2(n+2)\, \vep_{ik}
\pr^n \tr \big ( \Z^k\Z^j\big )  \, , \cr
& \big \{ \bS^i , \pr^n \tr \big ( \blam \Z^j \big ) \}
= 2in \, \vep^{ij} \, \pr^{n-1}
\tr \big ( \blam \lambda - i\vep_{kl} \, \pr\Z^k \, \Z^l \big ) + 2(n+2)\,
\pr^n \tr \big ( \Z^i\Z^j\big )  \, , \cr
&\big [ S_i ,\pr^n \tr \big ( \Z^j \Z^k \big ) \big ] = 2in \, \pr^{n-1}
\big ( \de_i{\!}^j \, \tr \big ( \blam \Z^k \big ) +
\de_i{\!}^k \, \tr \big ( \blam \Z^j \big )  \big ) \, , \cr
&\big [ \bS^i ,\pr^n \tr \big ( \Z^j \Z^k \big ) \big ] = - 2in \, \pr^{n-1}
\big ( \vep^{ij}\,  \tr \big ( \lambda \Z^k \big ) +
\vep^{ik}\,   \tr \big ( \lambda \Z^j \big )  \big ) \, , \cr
}
}
we may determine the conditions for $\O^{(n)}_0$ and $\O^{(n)}_1$ to be
superconformal primary operators. These give
\eqn\solabc{\eqalign{
\alpha_{0,s}^{(n)} ={}& {n+3\choose s}{n+3\choose s+3}(-1)^s \, , \quad
\beta_{0,s}^{(n)} = {n+3\choose s}{n+3\choose s+2}(-1)^s \, , \cr
\gamma_{0,s}^{(n)} = {}& {n+3\choose s}{n+3\choose s+1}(-1)^s \, , \cr
\alpha_{1,s}^{(n)} ={}& {n+3\choose s}{n+1\choose s+1}(-1)^s \, , \quad
\gamma_{1,s}^{(n)} = {n+3\choose s+1}{n+1\choose s}(-1)^s \, ,  \cr
\beta_{1,s}^{(n)} = {}& {n+3\over n+2} {n+2\choose s}{n+2\choose s+2}(-1)^s \, ,
}}
where to satisfy the symmetry conditions we must take $n$ to be even. It is easy
to verify that $\O^{(0)}_1$ is a linear combination of the double trace
operators in \Ofour.

The operator $\O^{(0)}_1$ represents the protected operator ${\cal D}_{\bf 20'}$.
All the operators $\O^{(n)}_0,\, \O^{(n)}_1$ are still annihilated by 
$Q^1{}_{\! 2} , \bQ_{41}$ in the interacting theory since they are formed 
from products of $\half$-BPS operators and their descendants and the action of
the supercharges on these short multiplets cannot be changed (although the 
detailed form of the the $\half$-BPS operators in terms of elementary fields 
may modified) and hence are protected.
In \Osing\ and \Otrip\ the operators corresponding to the $[0,2,0]$ short
multiplet in this sector are unaffected by interactions except for the 
descendant
$\tr \big ( \blam \lambda - i\vep_{ij} \, D\Z^i \, \Z^j \big )$ where it is
necessary to take $\pr \to D$, a
$SU(N)$ covariant derivative (derivatives outside the trace are not changed).
Extending \intQ\ to 
\eqn\intQ{\eqalign{
\big \{ Q^1{}_{\! 2} , \lambda \big \} ={}& g\, \vep_{ij} \big [ \Z^i , \Z^j \big ] \, ,
\qquad\quad
\big [  Q^1{}_{\! 2} , D \Z^i \big ] = g \, i \big [\blam ,  \Z^i  \big ]\, , \cr
\big \{ \bQ_{41} , \blam \big \} = {}& - g\, \vep_{ij} \big [ \Z^i , \Z^j \big ]\, ,\qquad
\big [ \bQ_{41} , D\Z_i \big ] = g \, i \big [\lambda,  \Z^i  \big ] \, ,\cr}
}
it is easy to see that $\tr \big ( \blam \lambda - i\vep_{ij} \, D\Z^i \, \Z^j \big )$ 
commute with $Q^1{}_{\! 2} , \bQ_{41}$ for non zero $g$. This is in contrast to 
operators such as those in \Kon\ and \Kone\ which are therefore part of long
multiplets.

\bigskip

\noindent{\bf Acknowledgements}
\medskip

M.B. would like to thank J.~Maldacena, Ya.~Stanev, E. Sokatchev
for useful discussions and especially to thank J.F. Morales for
collaboration on the resolution of the operator mixing discussed
in the concluding section. During completion of this work M.B. was
visiting the Galileo Galilei Institute of Arcetri (FI); INFN is
acknowledged for hospitality. This work
was supported in part by INFN, by the MIUR-COFIN contract
2003-023852, by the EU contracts MRTN-CT-2004-503369 and
MRTN-CT-2004-512194, by the INTAS contract 03-516346, by the NATO
grant PST.CLG.978785 and by the National Science Foundation under
Grant No. PHY99-07949 while at KITP in Santa Barbara.

P.H. was visiting LAPP, Annecy-le-Vieux during the preparation 
of this paper and is grateful for support and discussions there and also
to Paul Howe for email correspondence.
F.D. is grateful for hospitality at The School of Theoretical Physics
at the Institute for Advanced Studies, Dublin, during completion of this work.

\vfil\eject

\appendix{A}{$PSU(2,2|4)$ Superconformal Algebra and Subalgebras}

The generators of the $\N=4$ superconformal group $PSU(2,2|4)$ consist of 
the usual Lorentz transformations $M_{ab}$, translations $P_a$,
special conformal translations $K_a$, dilatations $D$, and
$SU(4)_R$ $R$-symmetry generators $R^i{}_{\! j}$
together with supercharges $Q^i{}_{\! \alpha}, \, \bQ_{i\dal}$ and
their superconformal partners $S_i{}^{\!\alpha}, \, \bS{}^{i\dal}$,
for $i,j=1,2,3,4$. In a spinorial basis $P_{\alpha\dal} = 
(\si^a)_{\alpha\dal} P_a$, 
${\tilde K}{}^{\dal\alpha} = (\bsi^a)^{\dal\alpha} K_a$, 
$ M_\alpha{}^{\! \beta} = - \quar i ( \si^a \bsi^b)_\alpha{}^{\! \beta}
M_{ab}$, ${\bar M}{}^\dal{}_{\!\smash{\dbe}} = - \quar i
( \bsi^a \si^b)^\dal{}_{\!\smash{\dbe}} M_{ab}$ we may write
\eqn\defMab{
\M_\A{}^{\! \B} = \pmatrix{ M_\alpha{}^{\! \beta} + \half 
\de_\alpha{}^{\! \beta} D & \half \, P_{\smash{\alpha\dbe}} \cr
\half \, {\tilde K}{}^{\dal\beta} & {\bar M}{}^\dal{}_{\!\smash{\dbe}}
- \half \de{}^\dal{}_{\!\smash{\dbe}} D \cr} \, , \ \ 
\Q^i{\!}_\A = \pmatrix{Q^i{}_{\! \alpha}\cr  \bS{}^{i\dal}} \, , \ \
{\bar \Q}_i{\!}^\B = \pmatrix{ S_i{}^{\!\beta}& \bQ_{\smash {i\dbe}} }\, ,
}
and their various commutators, anticommutators are then determined by
\eqn\comm{\eqalign{
\big [ \M_\A{}^{\! \B}, \M_\C{}^{\! \D} \big ] = {}& \de_\C{}^{\! \B} 
\M_\A{}^{\! \D} - \de_\A{}^{\! \D} \M_\C{}^{\! \B} \, , \quad
\big [ R^i{}_{\! j} , R^k{}_{\! l} \big ] = \de^k{}_{\! j}R^i{}_{\! l} -
\de^i{}_{\! l}R^k{}_{\! j} \, , \cr
\big [ \M_\A{}^{\! \B}, \Q^i{\!}_\C \big ] = {}& \de_\C{}^{\! \B}
\Q^i{\!}_\A - \quar \de_\A{}^{\! \B} \Q^i{\!}_\C \, , \qquad
\big [ \M_\A{}^{\! \B}, {\bar \Q}_i{\!}^\C \big ] = - \de_\A{}^{\! \C} 
{\bar \Q}_i{\!}^\B + \quar \de_\A{}^{\! \B} {\bar \Q}_i{\!}^\C \, , \cr
\big [ R^i{}_{\! j} ,  \Q^k{\!}_\A \big ] = {}& \de^k{}_{\! j}
\Q^i{\!}_\A  - \quar  \de^i{}_{\! j}  \Q^k{\!}_\A \, , \qquad \, 
\big [ R^i{}_{\! j} , {\bar \Q}_k{\!}^\A \big ] = - \de^i{}_{\! k}
{\bar \Q}_j{\!}^\A + \quar  \de^i{}_{\! j} {\bar \Q}_k{\!}^\A \, , \cr
\big \{ \Q^i{\!}_\A , {\bar \Q}_i{\!}^\B \big \} = {}& 4 \big ( \de^i{}_{\! j}
\M_\A{}^{\! \B} - \de_\A{}^{\! \B} R^i{}_{\! j} \big ) \, , \quad
\big \{ \Q^i{\!}_\A , \Q^j{\!}_\B  \big \} = 0 \, , \quad
\{ {\bar \Q}_i{\!}^\A , {\bar \Q}_j{\!}^\B \big \} = 0 \, ,\cr}
}
for $\de_\A{}^{\! \B} = \pmatrix{\de_\alpha{}^{\! \beta}&0\cr 0& 
\de{}^\dal{}_{\!\smash{\dbe}}}$. 
In terms of the usual angular momentum generators we have
\eqn\MJ{
\big [M_\alpha{}^{\! \beta}\big ] = \pmatrix{ J_3 & J_+ \cr J_- & -J_3 } \, ,
\qquad \big [{\bar M}{}^\dbe{}_{\!\smash{\dal}}\big ] =
\pmatrix{ \bJ_3 & \bJ_+ \cr \bJ_- & -\bJ_3 } \, ,
}
and for $SU(4)$ there is the decomposition
\eqn\Rfour{\hskip -12pt{
\big [ R^i{}_{\! j} \big ] = 
\pmatrix{\quar(3H_1{+2H_2}{+H_3})&\!  E_1{}^{\!+} & \!
[E_1{}^{\!+},E_2{}^{\!+}]& \! [E_1{}^{\!+},[E_2{}^{\!+},E_3{}^{\!+}]]\cr
E_1{}^{\!-} & \!\!\! \quar({-H_1}{+2H_2}{+H_3})& \! E_2{}^{\!+} & \!
[E_2{}^{\!+},E_3{}^{\!+}] \cr \!
-[E_1{}^{\!-},E_2{}^{\!-}]&\! E_2{}^{\!-} & \!\!\! -\quar(H_1{+2H_2}{-H_3}) &
\! E_3{}^{\!+} \cr
\! [E_1{}^{\!-},[E_2{}^{\!-},E_3{}^{\!-}]] &\! -[E_2{}^{\!-},E_3{}^{\!-}]&
\! E_3{}^{\!-} & \!\!\! - \quar(H_1{+2H_2}{+3H_3}) \cr}  ,}
}
with $E_i{}^{\!\pm}$ the raising/lowering operators for the simple roots.

The standard hermiticity requirements are
\eqn\herm{
\big ( \M_\A{}^{\! \B} \big )^\dagger = (\tau \M \tau)_\B{}^{\! \A} \, , \quad
\big (R^i{}_{\! j}\big )^\dagger = R^j{}_{\! i} \, , \quad 
\big (\Q^i{\!}_\A\big )^\dagger = ({\bar \Q}_i \tau)^\A \, , \quad \tau =
\pmatrix{0&1\cr 1&0} \, .
}
Thus $D^\dagger =-D$ and $(M_\alpha{}^{\! \beta})^\dagger =
{\bar M}{}^\dbe{}_{\!\smash{\dal}}$, interchanging $SU(2)_J$ and $SU(2)_\bJ$.
However for states formed by the action of local field
operators on the vacuum the two point function defines a scalar product
leading to modified hermiticity conditions \fadho. 
For an operator $\O$ we define an alternative conjugation operation by
\eqn\hermO{
\O^+ = U^{-1} \O^\dagger U \, , \qquad U=U^\dagger = \exp\big ({\ts {\pi\over 2}
(P_0-K_0)}\big )  \, .
}
With this definition dotted and undotted indices are still interchanged but
\eqn\Oplus{\eqalign{
& D^+ = D \, , \quad P_{\al\dal}{}^{\! +} = - (\si_0)_{\smash{\al\dbe}}
{\tilde K}^{\dbe \be} (\si_0)_{\be \dal} \, , \quad
\big (M_\alpha{}^{\! \beta}\big )^+ = (\bsi_0)^{\dbe\ga} M_\ga{}^{\! \de}
(\si_0)_{\de\dal} \, , \cr
& \big ({\bar M}{}^\dbe{}_{\!\smash{\dal}}\big )^+ = (\si_0)_{\al\dga}
{\bar M}{}^\dga{}_{\!\smash{\dde}} (\bsi_0)^{\dde \be} \, , \quad
Q^i{}_{\! \al}{}^+ = S_i{}^{\! \be}(\si_0)_{\be \dal} \, , \quad
\bQ_{i\al}{}^+ = - (\si_0)_{\smash{\al \dbe}} \bS^{i\dbe} \, , \cr}
}
where $\si_0\bsi_0=1$ and we may take $\si_0=\bsi_0=1$.

Corresponding to the various shortening conditions labelled by
$t,\bt$, described in section 3, there are corresponding reductions to 
subgroups which we list in turn.

\noindent
$PSU(2,2|4) \supset \big ( SU(1|1) \otimes PSU(1,2|3) \big ) \ltimes 
U(1)_{R}$, $t= {1\over 8}$.

The generators of $SU(1|1)$ are $Q^1{}_{\! 2}, S_1{}^{\!2},\H$ with
\eqn\algH{
\big \{ Q^1{}_{\! 2} , S_1{}^{\!2} \big \} = 2\H \, , \quad
\big [ \H, Q^1{}_{\! 2} \big ] = \big [ \H, S_1{}^{\!2} \big ] = 0 \, , \qquad
\H = D - 2J_3 - 2R^1{}_{\! 1} \, .
}
The conditions \Oplus\ require $\H$ to have a positive spectrum.
For $PSU(1,2|3)$ we may write the generators in a similar fashion to \defMab
\eqn\defMabt{\eqalign{
{\tilde \M}_\A{}^{\! \B} = {}& \pmatrix{ {\ts {2\over 3}} \hD 
& \half \, P_{\smash{1\dbe}} \cr
\half \, {\tilde K}{}^{\dbe 1} & {\bar M}{}^\dal{}_{\!\smash{\dbe}}
- \thir \de{}^\dal{}_{\!\smash{\dbe}} \hD \cr} \, , \qquad 
{\tilde R}^i{}_{\! j} = R^i{}_{\! j} + \thir \de^i{}_{\! j} R^1{}_{\! 1}\, , \cr
\Q^i{\!}_\A = {}& \pmatrix{Q^i{}_{\! 1}\cr  \bS{}^{i\dal}} \, , \ \
{\bar \Q}_j{\!}^\B = \pmatrix{ S_j{}^{\! 1}& \bQ_{\smash {j\dbe}} }\, ,
\qquad i,j=2,3,4 \, , \quad \hD = D + J_3 \, , \cr}
}
and the algebra is as in \comm, with the obvious modification $\quar \to \thir$, 
except for
\eqn\cent{
\big \{ \Q^i{\!}_\A , {\bar \Q}_j{\!}^\B \big \} = 4 \big ( \de^i{}_{\! j}
{\tilde \M}_\A{}^{\! \B} - \de_\A{}^{\! \B} {\tilde R}^i{}_{\! j} \big ) 
- {\ts {2\over 3}} \de^i{}_{\! j} \de_\A{}^{\! \B} \H \, ,
}
where $\H$ is here a central extension. The $U(1)$ with generator $R=4R^1{}_{\! 1}$
and plays the role of an external automorphism with the action
\eqn\Uone{
\big [ R,   Q^1{}_{\! 2} \big ] =  3 Q^1{}_{\! 2} \, , \
\big [ R ,  S_1{}^{\!2} \big ] = - 3 S_1{}^{\!2} \, , 
\qquad \big [ R , \Q^i{\!}_\A \big ] =  \Q^i{\!}_\A \, , \
\big [ R , {\bar \Q}_i{\!}^\A  \big ] =  - {\bar \Q}_i{\!}^\A \, .
}

\noindent
$PSU(2,2|4) \supset \big ( SU(1|1) \otimes  SU(1|1) \otimes PSU(1,1|2) \big ) 
\ltimes \big ( U(1)_{H_+} \otimes U(1)_{H_-} \big )$,
$t= \bt = {1\over 8}$.

In this case first $SU(1|1)$ is as above in \algH\ but the second $SU(1|1)$
has generators $\bQ_{41}, \bS^{41}, {\bar \H}$ with
\eqn\algbH{
\big \{ \bQ_{41} , \bS^{41} \big \} = -2{\bar\H} \, , \quad
\big [ {\bar \H} , \bQ_{41} \big ] = \big [  {\bar \H}, \bS^{41} \big ] = 0 \, , \qquad
{\bar \H} = D - 2\bJ_3 + 2R^4{}_{\! 4} \, ,
}
with $ {\bar \H} $ also positive.
The generators of $PSU(1,1|2)$ can be expressed as
\eqnn\defMabtt
$$\eqalignno{
{\tilde \M}_\A{}^{\! \B} = {}& \pmatrix{ \half \hD
& \half \, P_{\smash{1 2}} \cr
\half \, {\tilde K}{}^{2 1} & - \half \hD \cr} \, , \qquad
{\tilde R}^i{}_{\! j} = R^i{}_{\! j} + \half \de^i{}_{\! j}( R^1{}_{\! 1}
+  R^4{}_{\! 4} )\, , \
[ {\tilde R}^i{}_{\! j} ] = \pmatrix{ \!\! \half H_2 & E_2{}^{\!+} \cr
E_2{}^{\!-} & \!\! -\half H_2 \cr} \, , \cr
\Q^i{\!}_\A = {}& \pmatrix{Q^i{}_{\! 1}\cr  \bS{}^{i 2}} \, , \ \
{\bar \Q}_j{\!}^\B = \pmatrix{ S_j{}^{\! 1}& \bQ_{\smash {j \, 2}} }\, ,
\qquad i,j=2,3 \, , \quad \hD = D + J_3 + \bJ_3 \, . & \defMabtt \cr}
$$
The corresponding algebra is evident except for
\eqn\centt{
\big \{ \Q^i{\!}_\A , {\bar \Q}_j{\!}^\B \big \} = 4 \big ( \de^i{}_{\! j}
{\tilde \M}_\A{}^{\! \B} - \de_\A{}^{\! \B} {\tilde R}^i{}_{\! j} \big )
- \de^i{}_{\! j} \de_\A{}^{\! \B} (\H - {\bar \H} )  \, .
}
The two $U(1)$ automorphisms are generated by $ R^1{}_{\! 1}$ and 
$R^4{}_{\! 4}$ or equivalently to
\eqn\defHH{
H_+ =  R^1{}_{\! 1} - R^4{}_{\! 4} = H_1+H_2+H_3 \, , \qquad H_- = 
2( R^1{}_{\! 1} +  R^4{}_{\! 4})  = H_1- H_3 \, .
}
The non zero commutators are then 
\eqn\HH{\eqalign{
\big [ H_\pm ,   Q^1{}_{\! 2} \big ] = {}&  Q^1{}_{\! 2} \, , \
\big [ H_\pm ,  S_1{}^{\!2} \big ] = - S_1{}^{\!2} \, , \
\big [ H_\pm ,  \bQ_{41} \big ] =  \pm \bQ_{41}  \, , \
\big [ H_\pm ,  \bS^{41}  \big ] = \mp  \bS^{41} \, , \cr
\big [ H_+ , \Q^i{\!}_\A \big ] =  {}& 
\big [ H_+ ,  {\bar \Q}_i{\!}^\A  \big ] = 0 \, ,
\qquad \big [ H_- , \Q^i{\!}_\A \big ] = -  \Q^i{\!}_\A \, , \
\big [ H_- , {\bar \Q}_j{\!}^\A  \big ] =  {\bar \Q}_j{\!}^\A \, .\cr}
}

\noindent
$PSU(2,2|4) \supset SU(2|1) \otimes SU(2|3) $,
$t= {1\over 4}$.

Here the generators of $SU(2|1)$ are $M_\alpha{}^{\! \beta}, \H_0,\, 
Q^1{}_{\! \alpha}, S_1{}^{\!\beta}$ with
\eqn\tquar{\eqalign{
\big \{ Q^1{}_{\! \alpha} , S_1{}^{\! \beta} \big \} = {}&
4 M_\alpha{}^{\! \beta}
+ 2 \de_\alpha{}^{\! \beta}\H_0 \, ,  \qquad \H_0 = D - 2 R^1{}_{\! 1} \, , \cr
\big [ \H_0 , Q^1{}_{\! \alpha}] = {}& 
- Q^1{}_{\! \alpha}\, , \qquad  \big [ \H_0 ,  S_1{}^{\! \beta} \big ] =
S_1{}^{\! \beta}  \, . \cr}
}
The generators for $SU(2|3)$ are then ${\bar M}{}^\dal{}_{\!\smash{\dbe}},
\hD, {\tilde R}^i{}_{\! j},\bS{}^{i\dal} ,\bQ_{\smash {j\dbe}}$ 
for $i,j=2,3,4$ and
\eqn\tquars{\eqalign{
\big \{ \bS{}^{i\dal} , \bQ_{\smash {j\dbe}} \big \} = {}& 4 \big (
\de^i{}_{\! j} {\bar M}{}^\dal{}_{\!\smash{\dbe}} - \de{}^\dal{}_{\!\smash{\dbe}}
{\tilde R}^i{}_{\! j} \big ) - {\ts {4\over 3}} \de{}^\dal{}_{\!\smash{\dbe}} 
\de^i{}_{\! j} \hD\, , \qquad
{\tilde R}^i{}_{\! j} = R^i{}_{\! j} + \thir \de^i{}_{\! j} R^1{}_{\! 1}\, , \cr
\big [ \hD , \bS{}^{i\dal} \big ] = {}& - \half  \bS{}^{i\dal} \, , \quad
\big [ \hD ,  \bQ_{\smash {j\dbe}} \big ] = \half  \bQ_{\smash {j\dbe}} \, , \qquad
\quad \hD = {\ts {3\over 2}} D -  R^1{}_{\! 1} \, . \cr}
}

\noindent
$PSU(2,2|4) \supset \big ( SU(2|1) \otimes  SU(1|1) \otimes SU(1|2) \big )
\ltimes U(1)_{H_+}$, $t= {1\over 4}$, $\bt = {1\over 8}$.

The first $SU(2|1)$ is as in \tquar\ and $SU(1|1)$ as in \algbH. $SU(1|2)$
has generators ${\tilde R}^i{}_{\! j},\,  \hD,\,  \bS^{i2}, \, \bQ_{j2}$,
$i,j =2,3$, where $H_+$ is as in \defHH, ${\tilde R}^i{}_{\! j}$ as in \defMabtt\  
and 
\eqn\qeight{\eqalign{
\big \{ \bS{}^{i2} , \bQ_{j2} \big \} = {}& - 4 {\tilde R}^i{}_{\! j} 
- 2 \de^i{}_{\! j} \hD \, , \quad \hD = D + 2 \bJ_3 - \half H_- \, , \cr
\big [ \hD , \bS{}^{i2} \big ] = {}& - \bS{}^{i2} \, , \quad
\big [ \hD ,  \bQ_{j2} \big ] = \bQ_{j2} \, , \quad i,j =2,3 \, .
\cr}
}
The action of $H_+$ can be determined from \HH.

\noindent
$PSU(2,2|4) \supset \big ( SU(2|1) \otimes  SU(2|1) \otimes SU(2) \big )
\ltimes U(1)_{H_+}$, $t= \bt = {1\over 4}$.

The first $SU(2|1)$ is as previously in \tquar\ whereas the second has 
generators ${\bar M}{}^\dal{}_{\!\smash{\dbe}},{\bar \H}_0, \bS{}^{4\dal}$, 
$\bQ_{\smash {4\dbe}}$ with
\eqn\ttquar{\eqalign{
\big \{ \bS{}^{4\dal} , \bQ_{\smash {4\dbe}} \big \} = {}& 
4 {\bar M}{}^\dal{}_{\!\smash{\dbe}}- 2 \de {}^\dal{}_{\!\smash{\dbe}} 
{\bar \H}_0 \, ,   \qquad {\bar \H}_0 = D + 2 R^4{}_{\! 4} \, , \cr
\big [ {\bar \H}_0 , \bQ_{\smash {4\dbe}} \big ] = {}& - \bQ_{\smash {4\dbe}}\, , 
\qquad  \big [ {\bar \H}_0 , \bS{}^{4\dal} \big ] = \bS{}^{4\dal} \, , \cr}
}
and the $SU(2)$ corresponds to ${\tilde R}^i{}_{\! j}$ which is defined in
\defMabtt\ and  corresponds to the generators $H_2,E_2{}^{\!\pm}$. The 
additional $U(1)$ automorphism generated by  $H_+$ in \defHH\ has the action
\eqn\HHH{
\big [ H_+ ,  Q^1{}_{\! \alpha} \big ] = Q^1{}_{\! \alpha} \, , \
\big [ H_+ ,  S_1{}^{\! \alpha} \big ] = - S_1{}^{\!\alpha} \, , \
\big [ H_+ ,  \bQ_{\smash {4\dal}} \big ] =  \bQ_{\smash {4\dal}} \, , \ 
\big [ H_+ ,  \bS{}^{4\dal}   \big ] = - \bS{}^{4\dal} \, .
}

\noindent
$PSU(2,2|4) \supset \big ( PSU(2|2) \otimes  PSU(2|2) \otimes U(1)_{\tilde \H} \big )
\ltimes U(1)_D$, $t=\half$ and $t= \bt = {1\over 2}$.

The generators of the $PSU(2|2)$ factors are $M_\alpha{}^{\! \beta},
{\tilde R}^i{}_{\! j},Q^i{}_{\! \alpha}, S_j{}^{\! \alpha}$ for $i,j=1,2$
and ${\bar M}{}^\dal{}_{\!\smash{\dbe}}, {\bar R}^k{}_{\! l}, \bS{}^{k\dal},
\bQ_{\smash {l\dal}}$ for $k,l=3,4$, where
\eqn\QSa{\eqalign{
\big \{ Q^i{}_{\! \alpha} , S_j{}^{\! \beta} \big \} = {}&4 \big (
\de ^i{}_{\! j} M_\alpha{}^{\! \beta} - \de_\alpha{}^{\! \beta} 
{\tilde R}^i{}_{\! j}
\big ) + 2 \de ^i{}_{\! j} \de_\alpha{}^{\! \beta} {\tilde \H} \, , \ \ \
{\tilde R}^i{}_{\! j} = R^i{}_{\! j} + \half \de^i{}_{\! j}( R^3{}_{\! 3}
+  R^4{}_{\! 4} )\, , \cr
\big \{ \bS{}^{k\dal} , \bQ_{\smash {l\dbe}} \big \} = {}&
4 \big ( \de^k{}_{\! l} {\bar M}{}^\dal{}_{\!\smash{\dbe}}- 
\de ^\dal{}_{\!\smash{\dbe}} {\bar R}^k{}_{\! l} \big ) -
2 \de ^k{}_{\! l} \de^\dal{}_{\!\smash{\dbe}} {\tilde \H} \, , \ \
{\bar R}^k{}_{\! l} = R^k{}_{\! l} + \half \de^k{}_{\! l}( R^1{}_{\! 1}
+  R^2{}_{\! 2} )\, , \cr}
}
with $\tilde \H$ a central charge
\eqn\tilH{
\tilde \H = D - \half ( H_1+ 2H_2+H_3) \, , \quad
\big [ \tilde \H , Q^i{}_{\! \alpha} \big ] = \big [ \tilde \H ,  S_j{}^{\! \beta}
\big ] = \big [ \tilde \H , \bS{}^{k\dal} \big ] = \big [ \tilde \H , 
\bQ_{\smash {l\dbe}}  \big ] =  0 \, .
}
For this case $D$ generates an external automorphism with
\eqn\Dcom{
\big [ D , Q^i{}_{\! \alpha} \big ] = \half Q^i{}_{\! \alpha} \, , \ \
\big [ D , \bQ_{\smash {l\dbe}}  \big ] = \half \bQ_{\smash {l\dbe}} \, , \ \
\big [ D ,  S_j{}^{\! \beta} \big ] = - \half S_j{}^{\! \beta} \, , \ \
\big [ D ,  \bS{}^{k\dal} \big ] = - \half  \bS{}^{k\dal} \, .
}

\appendix{B}{Expansions in Schur Polynomials}

Here for application in the decomposition of partition functions we discuss
how to expand a general symmetric function $f(z,y)=f(y,z)$ in terms of 2 
variable Schur polynomials $\chi_{(n+m,m)}(z,y)$, defined as in \charsnewpopo,
\eqn\expSc{
f(z,y) = \sum_{n,m\ge 0} N_{n,m} \, \chi_{(n+m,m)}(z,y) \, ,
}
where we require $f(z,y)$ to have a power series expansion in $z,y$. It is 
important to note the symmetry relations
\eqn\symSc{
\chi_{(n+m,m}(z,y)=- \chi_{(m-1,n+m+1)}(z,y) \quad \Rightarrow \quad
N_{n,m} = - N_{-n-2,n+m+1} \, , \ N_{-1,m} = 0 \, .
}
Writing
\eqn\exy{
f(z,y) = \sum_{r=0}^\infty f_r(z) \, y^r \, , \qquad 
\chi_{(n+m,m)}(z,y) = (zy)^m \sum_{r=0}^n y^r z^{n-r} \, ,
}
we easily get
\eqn\powerF{
\sum_{n=0}^\infty \sum_{m\ge 0,s-n}^s  N_{n,m} z^{n+2m-s} = f_s(z) \, .
}
Using \symSc\ this can be rearranged as
\eqn\finSc{
\sum_{n=-s-1}^\infty N_{n,s} \, z^{n+s} = \hf_s(z) \, ,
}
for
\eqn\fhat{
\hf_0(z)=f_0(z) \, , \quad
\hf_s(z) = f_s(z) - {1\over z} f_{s-1}(z) \,, \,  s=1,2,\dots \, .
}

For application in section 7 we consider first, as in \quarmt,
\eqn\fdef{ 
f(z,y) =(1-z)(1-y) \prod_{k=1}^{\infty}{1\over 1-z^k-y^k}\,. }
The definitions \exy\ and \finSc\ give
\eqn\fnn{\eqalign{
f_0(z) = {}& (1-z) \prod_{k=1}^\infty {1\over 1-z^k} \, , \quad
\hf_1(z) = \bigg ( z+1-{1\over z} \bigg ) \prod_{k=1}^\infty {1\over 1-z^k} \, , \cr
\hf_2(z) = {}& {2z^2 \over 1-z^2}  \prod_{k=1}^\infty {1\over 1-z^k} \, , \cr
\hf_3(z) = {}& \bigg ({z\over (1-z)^2} - {z\over 1-z^2} + {1-z\over 1-z^3}-{1\over z}\bigg ) 
\prod_{k=1}^\infty {1\over 1-z^k} \, . \cr}
}
Similarly with, as in \quarmtsym,
\eqn\gfun{
g(z,y) = \prod_{k,l=0\atop k+l\geq 2}^{\infty} {1\over 1-z^{k}y^{l}}\,.
}
we get
\eqn\gnn{\eqalign{
g_0(z) = {}& (1-z) \prod_{k=1}^\infty {1\over 1-z^k} \, , \quad
\hg_1(z) = \bigg ( z+1-{1\over z} \bigg ) \prod_{k=1}^\infty {1\over 1-z^k} \, , \cr
\hg_2(z) = {}& {z^2 \over 1-z^2}  \prod_{k=1}^\infty {1\over 1-z^k} \, , \cr
\hg_3(z) = {}& \bigg ({1\over (1-z)(1-z^6)} - {z\over1-z^6} -{1\over z}\bigg ) 
\prod_{k=1}^\infty {1\over 1-z^k} \, . \cr}
}
Expansion of these gives the results in \quarcoeff.

If we consider the corresponding expression for single trace operators as in
\quart
\eqn\hfun{
h(z,y)= 
- \sum_{k=1}^\infty {\phi(k)\over k} \, \ln ( 1-z^k-y^k) - z - y \, ,
}
we find
\eqn\hnn{\eqalign{
h_0(z) = {}& {z^2\over 1-z} \, , \quad \hh_1(z)=0 \, , \quad 
\hh_2(z)= {z^2\over (1-z)(1-z^2)}\, \cr
\hh_3(z) = {}& {2z^3 \over (1-z)(1-z^2)(1-z^3)} - {z\over 1-z^2} - {1\over z} \, .}
}
Expansion gives \quarsingc.

For the purposes of counting $1\over 4$-BPS operators for large $R$-symmetry 
charges $p,q$ a different approach may be more relevant.
An alternative expression for $N_{n,m}$ in \expSc\ is obtained by employing 
the orthogonality relation,
\eqn\orth{\eqalign{
{1\over 8 \pi^2}& \oint\oint \chi_{(n+m,m)}(z,y)\, \chi_{(p+q,q)}(z^{-1},y^{-1})
(z^{-1}-y^{-1})^2 \,{\rm d}z\,{\rm d}y\cr
& = \de_{n\, p}\,\de_{m\, q}-\de_{n\,-p{-2}}\,\de_{m\,p{+}q{+}1}\,,}
} 
with contours encircling the origin. This relation is consistent with \symSc\
and follows from an orthogonality relation given in \dunkl\ for 
Jack polynomials for which Schur polynomials are a special case.
Hence
\eqn\npq{
N_{nm}={1\over 8 \pi^2}\oint\oint \chi_{(n+m,m)}(z,y)f(z^{-1},y^{-1})
(z^{-1}-y^{-1})^2 \,{\rm d}z\,{\rm d}y\,.
}

\vfil\eject
\appendix{C}{Tables}

\noindent
Here we list primary $(\eight,\eight)$
semi-short operators obtained from partition functions in terms
of their $SU(4)_R\otimes SU(2)_J\otimes SU(2)_{\bar J}$
representations $\R^{(j,\bj)}_{[k,p,q]}$ and scale dimension $\De$.

Table 6

\vskip -4pt
\hskip0cm
\vbox{\tabskip=0pt \offinterlineskip
\halign{&\vrule# &\strut \ \hfil#\  \cr
height0pt&\omit &&\omit &&\omit&\cr
&\multispan3\hrulefill &\cr
&\multispan3 \hfil Semi-short ($\eight$,$\eight$) primary operators \hfil&\cr
height0pt&\omit &&\omit &\cr
\noalign{\vskip-0pt}
&\multispan3\hrulefill &\cr
\noalign{\vskip-0pt}
& $\ \Delta$ \hfil  && Operators in $\S_{\rm sugra}$  \hfil &\cr
height2pt&\omit&&\omit&\cr
&\multispan3\hrulefill &\cr
&4\hfil&& $\rC_{[0,2,0]}^{(0,0)}$\ \hfil &\cr
&5\hfil&& $\rC_{[1,0,1]}^{({1\over2},{1\over2})},\rC_{[1,1,1]}^{(0,0)},
\rC_{[0,3,0]}^{(0,0)}$\ \hfil &\cr
&6\hfil&& $\rC_{[0,2,0]}^{(1,1)},\rC_{[1,1,1]}^{({1\over2},
{1\over2})},\rC_{[0,3,0]}^{({1\over2},{1\over2})},2
\rC_{[2,0,2]}^{(0,0)},2 \rC_{[1,2,1]}^{(0,0)},3
\rC_{[0,4,0]}^{(0,0)}$\ \hfil &\cr
&${13\over 2}$\hfil&&$\rC_{[1,1,2]}^{({1\over2},0)},
\rC_{[2,1,1]}^{(0,{1\over2})}$\ \hfil &\cr
&7\hfil&&$\rC_{[1,0,1]}^{({3\over2},{3\over2})},
\rC_{[1,1,1]}^{(1,1)},\rC_{[0,3,0]}^{(1,1)},4
\rC_{[1,2,1]}^{({1\over2},{1\over2})},
2 \rC_{[0,4,0]}^{({1\over2},{1\over2})},4 \rC_{[2,1,2]}^{(0,0)},
5 \rC_{[1,3,1]}^{(0,0)},4 \rC_{[0,5,0]}^{(0,0)} $\ \hfil &\cr
&${15\over 2}$\hfil&&$\rC_{[1,1,2]}^{(1,{1\over2})},
\rC_{[2,1,1]}^{({1\over2},1)},\rC_{[2,0,3]}^{({1\over2},0)},
\rC_{[3,0,2]}^{(0,{1\over2})},
2 \rC_{[1,2,2]}^{({1\over2},0)},2 \rC_{[2,2,1]}^{(0,{1\over2})},
\rC_{[0,4,1]}^{({1\over2},0)},\rC_{[1,4,0]}^{(0,{1\over2})}$\ \hfil &\cr
 &8\hfil&& $\rC_{[1,1,1]}^{({3\over2},{3\over2})},\rC_{[0,2,0]}^{(2,2)},
3\rC_{[2,0,2]}^{(1,1)},\rC_{[0,3,0]}^{({3\over2},{3\over2})},
3\rC_{[1,2,1]}^{(1,1)},3 \rC_{[0,4,0]}^{(1,1)},
5\rC_{[2,1,2]}^{({1\over2},{1\over2})},7
\rC_{[1,3,1]}^{({1\over2},{1\over2})},$\ \hfil &\cr
&&&$2\rC_{[3,0,3]}^{(0,0)},4 \rC_{[0,5,0]}^{({1\over2},{1\over2})},
11 \rC_{[2,2,2]}^{(0,0)},9 \rC_{[1,4,1]}^{(0,0)},
8 \rC_{[0,6,0]}^{(0,0)} \ $ \hfil &\cr
&\multispan3\hrulefill &\cr}
}

\vskip -2pt

\hskip - 1cm Table 7
\vskip 1pt

\hbox spread 2cm{\hskip -1cm
\vbox{\tabskip=0pt \offinterlineskip
\halign{&\vrule# &\strut \ \hfil#\  \cr
height0pt&\omit &&\omit &&\omit&\cr
&\multispan3\hrulefill &\cr
&\multispan3 \hfil Semi-short ($\eight$,$\eight$) primary operators \hfil&\cr
height0pt&\omit &&\omit &\cr
\noalign{\vskip-0pt}
&\multispan3\hrulefill &\cr
\noalign{\vskip-0pt}
& $\ \Delta$ \hfil  && $(0,0)$ primary operators in $\S_{\rm sym.}/ \S_{\rm sugra}$
\hfil &\cr
height2pt&\omit&&\omit&\cr
&\multispan3\hrulefill &\cr
&2\hfil&& $\rC_{[0,0,0]}^{(0,0)}$ \ \hfil &\cr
&3\hfil&& $\rC_{[0,1,0]}^{(0,0)}$ \ \hfil &\cr
&4\hfil&& $ \rC_{[0,0,0]}^{(1,1)}, \rC_{[0,1,0]}^{({1\over2},{1\over2})},
2\rC_{[0,2,0]}^{(0,0)}$\ \hfil & \cr
&${9\over 2}$\hfil&& $\rC_{ [0,1,1]}^{({1\over2},0)}, 
\rC_{[1,1,0]}^{(0,{1\over2})}$ \ \hfil & \cr
&5\hfil&&$\rC_{[0,1,0]}^{(1,1)}, 3\rC_{[1,0,1]}^{({1\over2},{1\over2})},
2\rC_{[0,2,0]}^{({1\over2},{1\over2})},
\rC_{[1,1,1]}^{(0,0)}, 3\rC_{[0,3,0]}^{(0,0)}$\ \hfil &  \cr
&${11\over 2}$\hfil&&$ \rC_{ [0,0,1]}^{({3\over2},1)},
\rC_{[1,0,0]}^{(1,{3\over2})}, \rC_{[0,1,1]}^{(1,{1\over2})},
\rC_{[1,1,0]}^{({1\over2},1)}, \rC_{[1,0,2]}^{({1\over2},0)},
\rC_{[2,0,1]}^{(0,{1\over2})}, 2\rC_{[0,2,1]}^{({1\over2},0)}, 
2\rC_{[1,2,0]}^{(0,{1\over2})}  $\ \hfil &\cr
&6\hfil&&$ \rC_{[0,0,2]}^{({3\over2},{1\over2})},
2\rC_{[1,0,1]}^{(1,1)}, \rC_{[2,0,0]}^{({1\over2},{3\over2})},
\rC_{[0,0,0]}^{(2,2)}, \rC_{[0,1,0]}^{({3\over2},{3\over2})},
5\rC_{[0,2,0]}^{(1,1)}, 6\rC_{[1,1,1]}^{({1\over2},{1\over2})},
4\rC_{[0,3,0]}^{({1\over2},{1\over2})},2\rC_{[2,0,2]}^{(0,0)},
2\rC_{[1,2,1]}^{(0,0)}, 5\rC_{[0,4,0]}^{(0,0)}  $\ \hfil &\cr
&${13\over 2}$\hfil&&$ 4\rC_{ [0,1,1]}^{({3\over2},1)},
4\rC_{ [1,1,0]}^{(1,{3\over2})}, 3\rC_{[1,0,2]}^{(1,{1\over2})},
3\rC_{[2,0,1]}^{({1\over2},1)}, 4\rC_{[0,2,1]}^{(1,{1\over2})}, 
4\rC_{[1,2,0]}^{({1\over2},1)}, 3\rC_{[1,1,2]}^{({1\over2},0)}, 
3\rC_{[2,1,1]}^{(0,{1\over2})}, 4\rC_{ [0,3,1]}^{({1\over2},0)},
4\rC_{ [1,3,0]}^{(0,{1\over2})} $\ \hfil &\cr
&7\hfil&&$ 3\rC_{[0,2,0]}^{({3\over2},{3\over2})},
2\rC_{[0,1,0]}^{(2,2)}, 7\rC_{[1,0,1]}^{({3\over2},{3\over2})},
\rC_{[0,1,2]}^{({3\over2},{1\over2})}, 
11\rC_{[1,1,1]}^{(1,1)}, \rC_{[2,1,0]}^{({1\over2},{3\over2})}, 
8\rC_{[0,3,0]}^{(1,1)}, \rC_{[1,0,3]}^{(1,0)}$,\ \hfil &\cr
&&& $5\rC_{[2,0,2]}^{({1\over2},{1\over2})},
\rC_{[3,0,1]}^{(0,1)}, 15\rC_{[1,2,1]}^{({1\over2},{1\over2})},
7\rC_{[0,4,0]}^{({1\over2},{1\over2})}, 4\rC_{[2,1,2]}^{(0,0)},
5\rC_{[1,3,1]}^{(0,0)}, 7\rC_{[0,5,0]}^{(0,0)}  $\ \hfil &\cr
&${15\over 2}$\hfil&&$6\rC_{ [1,0,2]}^{({3\over2},1)},
6\rC_{[2,0,1]}^{(1,{3\over2})}, \rC_{[0,0,1]}^{({5\over2},2)},
\rC_{[1,0,0]}^{(2,{5\over2})}, 5\rC_{[0,1,1]}^{(2,{3\over2})},
5\rC_{[1,1,0]}^{({3\over2},2)}, 10\rC_{[0,2,1]}^{({3\over2},1)},
10\rC_{[1,2,0]}^{(1,{3\over2})}, 11\rC_{[1,1,2]}^{(1,{1\over2})}$,\ \hfil &\cr
&&& $11\rC_{[2,1,1]}^{({1\over2},1)}, 8\rC_{[0,3,1]}^{(1,{1\over2})}, 
8\rC_{[1,3,0]}^{({1\over2},1)}, 3\rC_{[2,0,3]}^{({1\over2},0)},
3\rC_{[3,0,2]}^{(0,{1\over2})}, 7\rC_{[1,2,2]}^{({1\over2},0)},
7\rC_{[2,2,1]}^{(0,{1\over2})}, 7\rC_{ [0,4,1]}^{({1\over2},0)}, 
7\rC_{ [1,4,0]}^{(0,{1\over2})}$\ \hfil &\cr
&8\hfil&&$3\rC_{[0,1,2]}^{(2,1)},
22\rC_{[1,1,1]}^{({3\over2},{3\over2})}, 3\rC_{[2,1,0]}^{(1,2)},
\rC_{[0,0,0]}^{(3,3)} ,\rC_{[0,1,0]}^{({5\over2},{5\over2})},
2\rC_{[0,0,2]}^{({5\over2},{3\over2})}, 6\rC_{[1,0,1]}^{(2,2)},
2\rC_{[2,0,0]}^{({3\over2},{5\over2})}, 8\rC_{[0,2,0]}^{(2,2)} , 
\rC_{[1,0,3]}^{({3\over2},{1\over2})}$,\ \hfil &\cr
&&&
$22\rC_{[2,0,2]}^{(1,1)}, \rC_{[3,0,1]}^{({1\over2},{3\over2})},
10\rC_{[0,3,0]}^{({3\over2},{3\over2})}, 
4\rC_{[0,2,2]}^{({3\over2},{1\over2})}, 29\rC_{[1,2,1]}^{(1,1)}, 
4\rC_{[2,2,0]}^{({1\over2},{3\over2})}, 
17\rC_{[0,4,0]}^{(1,1)}, 2\rC_{[1,1,3]}^{(1,0)},
20\rC_{[2,1,2]}^{({1\over2},{1\over2})} $,\ \hfil &\cr
&&& $2\rC_{[3,1,1]}^{(0,1)},
\rC_{[0,3,2]}^{(1,0)}, 28\rC_{[1,3,1]}^{({1\over2},{1\over2})},
\rC_{[2,3,0]}^{(0,1)}, 2\rC_{[3,0,3]}^{(0,0)},
12\rC_{[0,5,0]}^{({1\over2},{1\over2})}, 10\rC_{[2,2,2]}^{(0,0)},
8\rC_{[1,4,1]}^{(0,0)}, 11\rC_{[0,6,0]}^{(0,0)}  $\ \hfil &\cr
&\multispan3\hrulefill &\cr}
}}

\vfil

\hskip -1cm Table 8
\vskip 1pt

\hbox spread 2cm{\hskip -1.6cm
\vbox{\tabskip=0pt \offinterlineskip
\halign{&\vrule# &\strut \ \hfil#\  \cr
height0pt&\omit &&\omit &&\omit&\cr
&\multispan3\hrulefill &\cr
&\multispan3 \hfil Semi-short ($\eight$,$\eight$) primary operators \hfil&\cr
height0pt&\omit &&\omit &\cr
\noalign{\vskip-0pt}
&\multispan3\hrulefill &\cr
\noalign{\vskip-0pt}
& $\ \Delta$  \hfil  && $(0,0)$ primary 
operators in $\S_{\rm long}/(\S_{\rm sym.}/ \S_{\rm sugra})$ \hfil &\cr
height2pt&\omit&&\omit&\cr
&\multispan3\hrulefill &\cr
\cr
&4\hfil&&$ \rC_{[0,1,0]}^{({1\over2},{1\over2})}, \rC_{[1,0,1]}^{(0,0)},
\rC_{[0,2,0]}^{(0,0)}  $\ \hfil &\cr
&5\hfil&&$\rC_{[0,0,2]}^{(1,0)}, 3\rC_{[1,0,1]}^{({1\over2},{1\over2})},
\rC_{[2,0,0]}^{(0,1)}, \rC_{[0,2,0]}^{({1\over2},{1\over2})},
2\rC_{[1,1,1]}^{(0,0)}, \rC_{[0,3,0]}^{(0,0)}  $\ \hfil &\cr
&${11\over 2}$\hfil&&
$\rC_{ [0,0,1]}^{({3\over2},1)}, \rC_{[1,0,0]}^{(1,{3\over2})}, 
2\rC_{[0,1,1]}^{(1,{1\over2})}, 2\rC_{[1,1,0]}^{({1\over2},1)}, 
2\rC_{[0,2,1]}^{({1\over2},0)}, 2\rC_{[1,2,0]}^{(0,{1\over2})}  $\ \hfil &\cr
&6\hfil&&$2\rC_{[1,0,1]}^{(1,1)} ,\rC_{[0,1,0]}^{({3\over2},{3\over2})},
3\rC_{[0,2,0]}^{(1,1)}, 2\rC_{[0,1,2]}^{(1,0)},
10\rC_{[1,1,1]}^{({1\over2},{1\over2})}, 2\rC_{[2,1,0]}^{(0,1)}, 
4\rC_{[0,3,0]}^{({1\over2},{1\over2})}, 5\rC_{[2,0,2]}^{(0,0)}, 
5\rC_{[1,2,1]}^{(0,0)}, 3\rC_{[0,4,0]}^{(0,0)}$\ \hfil &\cr
&${13\over 2}$\hfil&&$
3\rC_{ [0,1,1]}^{({3\over2},1)}, 3\rC_{[1,1,0]}^{(1,{3\over2})}, 
8\rC_{[1,0,2]}^{(1,{1\over2})}, 8\rC_{[2,0,1]}^{({1\over2},1)}, 
5\rC_{[0,2,1]}^{(1,{1\over2})}, 5\rC_{[1,2,0]}^{({1\over2},1)},
7\rC_{[1,1,2]}^{({1\over2},0)}, 7\rC_{[2,1,1]}^{(0,{1\over2})}, 
3\rC_{[0,3,1]}^{({1\over2},0)}, 3\rC_{[1,3,0]}^{(0,{1\over2})}$\ \hfil &\cr
&7\hfil&&$3\rC_{[0,2,0]}^{({3\over2},{3\over2})}, \rC_{[0,1,0]}^{(2,2)},
2\rC_{[0,0,2]}^{(2,1)}, 8\rC_{[1,0,1]}^{({3\over2},{3\over2})},
2\rC_{[2,0,0]}^{(1,2)}, 5\rC_{[0,1,2]}^{({3\over2},{1\over2})}, 
22\rC_{[1,1,1]}^{(1,1)}, 5\rC_{[2,1,0]}^{({1\over2},{3\over2})}, 
6\rC_{[0,3,0]}^{(1,1)}, 4\rC_{[1,0,3]}^{(1,0)},$\ \hfil &\cr
&&& $ 15\rC_{[2,0,2]}^{({1\over2},{1\over2})}, 4\rC_{[3,0,1]}^{(0,1)},
5\rC_{[0,2,2]}^{(1,0)}, 29\rC_{[1,2,1]}^{({1\over2},{1\over2})},
5\rC_{[2,2,0]}^{(0,1)}, 8\rC_{[0,4,0]}^{({1\over2},{1\over2})}, 
9\rC_{[2,1,2]}^{(0,0)}, 11\rC_{[1,3,1]}^{(0,0)}, 
4\rC_{[0,5,0]}^{(0,0)}$\ \hfil &\cr
&${15\over 2}$\hfil&&$\rC_{ [0,0,3]}^{(2,{1\over2})}, 
15\rC_{[1,0,2]}^{({3\over2},1)}, 15\rC_{[2,0,1]}^{(1,{3\over2})},
\rC_{[3,0,0]}^{({1\over2},2)}, \rC_{[0,0,1]}^{({5\over2},2)},
\rC_{[1,0,0]}^{(2,{5\over2})}, 5\rC_{[0,1,1]}^{(2,{3\over2})},
5\rC_{[1,1,0]}^{({3\over2},2)}, 15\rC_{[0,2,1]}^{({3\over2},1)} ,$\ \hfil &\cr
&&& $15\rC_{[1,2,0]}^{(1,{3\over2})}, 2\rC_{[0,1,3]}^{({3\over2},0)}, 
37\rC_{[1,1,2]}^{(1,{1\over2})}, 37\rC_{[2,1,1]}^{({1\over2},1)}, 
2\rC_{[3,1,0]}^{(0,{3\over2})}, 17\rC_{[0,3,1]}^{(1,{1\over2})}, 
17\rC_{[1,3,0]}^{({1\over2},1)}, 
14\rC_{[2,0,3]}^{({1\over2},0)}, 14\rC_{[3,0,2]}^{(0,{1\over2})},$\ \hfil &\cr
&&& $ 18\rC_{[1,2,2]}^{({1\over2},0)}, 18\rC_{[2,2,1]}^{(0,{1\over2})},
8\rC_{ [0,4,1]}^{({1\over2},0)}, 8\rC_{ [1,4,0]}^{(0,{1\over2})} $\ \hfil &\cr
&8\hfil&&$12\rC_{[0,1,2]}^{(2,1)}, 44\rC_{[1,1,1]}^{({3\over2},{3\over2})},
12\rC_{[2,1,0]}^{(1,2)}, \rC_{[0,1,0]}^{({5\over2},{5\over2})}, 
2\rC_{[0,0,2]}^{({5\over2},{3\over2})}, 8\rC_{[1,0,1]}^{(2,2)},
2\rC_{[2,0,0]}^{({3\over2},{5\over2})}, 5\rC_{[0,2,0]}^{(2,2)}, 
\rC_{[0,0,4]}^{(2,0)}, 19\rC_{[1,0,3]}^{({3\over2},{1\over2})},$\ \hfil &\cr
&&& $69\rC_{[2,0,2]}^{(1,1)}, 19\rC_{[3,0,1]}^{({1\over2},{3\over2})}, 
\rC_{[4,0,0]}^{(0,2)}, 13\rC_{[0,3,0]}^{({3\over2},{3\over2})}, 
18\rC_{[0,2,2]}^{({3\over2},{1\over2})}, 77\rC_{[1,2,1]}^{(1,1)},
18\rC_{[2,2,0]}^{({1\over2},{3\over2})}, 17\rC_{[0,4,0]}^{(1,1)}, 
17\rC_{[1,1,3]}^{(1,0)}, 79\rC_{[2,1,2]}^{({1\over2},{1\over2})},$ &\cr
&&& $17\rC_{[3,1,1]}^{(0,1)}, 13\rC_{[0,3,2]}^{(1,0)},
67\rC_{[1,3,1]}^{({1\over2},{1\over2})}, 13\rC_{[2,3,0]}^{(0,1)}, 
10\rC_{[3,0,3]}^{(0,0)}, 15\rC_{[0,5,0]}^{({1\over2},{1\over2})},
28\rC_{[2,2,2]}^{(0,0)}, 19\rC_{[1,4,1]}^{(0,0)}, 8\rC_{[0,6,0]}^{(0,0)} $\ \hfil &\cr
&\multispan3\hrulefill &\cr}
}}

\vskip -2pt
\hskip -1.3cm Table 9
\vskip 1pt

\hbox spread 2cm{\hskip -1.6cm
\vbox{\tabskip=0pt \offinterlineskip
\halign{&\vrule# &\strut \ \hfil#\  \cr
height0pt&\omit &&\omit &&\omit&\cr
&\multispan3\hrulefill &\cr
&\multispan3 \hfil Semi-short ($\eight$,$\eight$) primary operators \hfil&\cr
height0pt&\omit &&\omit &\cr
&\multispan3\hrulefill &\cr
&  $\ \Delta  $ \hfil  && $(0,0)$ descendant operators in 
$\S_{\rm free}/\S_{\rm long}$ \hfil &\cr
height2pt&\omit&&\omit&\cr
&\multispan3\hrulefill &\cr
&${9\over 2}$\hfil&&$\rC_{ [0,0,1]}^{(1,{1\over2})}, \rC_{ [1,0,0]}^{({1\over2},1)},
2\rC_{ [0,1,1]}^{({1\over2},0)}, 2\rC_{ [1,1,0]}^{(0,{1\over2})}$\ \hfil &\cr
&5\hfil&&$\rC_{[1,0,1]}^{({1\over2},{1\over2})}, 4\rC_{[1,1,1]}^{(0,0)} $\ \hfil &\cr
&${11\over 2}$\hfil&&$\rC_{ [0,1,1]}^{(1,{1\over2})}, \rC_{[1,1,0]}^{({1\over2},1)}, 
7\rC_{[1,0,2]}^{({1\over2},0)}, 7\rC_{[2,0,1]}^{(0,{1\over2})}, 
3\rC_{[0,2,1]}^{({1\over2},0)}, 3\rC_{[1,2,0]}^{(0,{1\over2})}    
$\ \hfil &\cr
&6\hfil&& $2\rC_{[0,0,2]}^{({3\over2},{1\over2})}, 4\rC_{[1,0,1]}^{(1,1)}, 
2\rC_{[2,0,0]}^{({1\over2},{3\over2})},
3\rC_{[0,1,2]}^{(1,0)}, 7 \rC_{[1,1,1]}^{({1\over2},{1\over2})},
3\rC_{[2,1,0]}^{(0,1)}, 8\rC_{[2,0,2]}^{(0,0)} , 11\rC_{[1,2,1]}^{(0,0)}$\ \hfil &\cr
&$13\over 2$\hfil&& $2\rC_{ [0,1,1]}^{({3\over2},1)}, 2\rC_{ [1,1,0]}^{(1,{3\over2})},
\rC_{ [0,0,1]}^{(2,{3\over2})}, \rC_{ [1,0,0]}^{({3\over2},2)},
\rC_{ [0,0,3]}^{({3\over2},0)}, 7\rC_{ [1,0,2]}^{(1,{1\over2})},
7\rC_{ [2,0,1]}^{({1\over2},1)},
\rC_{ [3,0,0]}^{(0,{3\over2})}, 8\rC_{ [0,2,1]}^{(1,{1\over2})},$\ \hfil &\cr
&&& $8\rC_{ [1,2,0]}^{({1\over2},1)},
21\rC_{ [1,1,2]}^{({1\over2},0)}, 21\rC_{ [2,1,1]}^{(0,{1\over2})},
8\rC_{ [0,3,1]}^{({1\over2},0)}, 8\rC_{[1,3,0]}^{(0,{1\over2})}$\ \hfil &\cr
&7\hfil&&$\rC_{[1,0,1]}^{({3\over2},{3\over2})}, 
7\rC_{[0,1,2]}^{({3\over2},{1\over2})},
16\rC_{[1,1,1]}^{(1,1)}, 7\rC_{[2,1,0]}^{({1\over2},{3\over2})}, 
12\rC_{[1,0,3]}^{(1,0)}, 26\rC_{[2,0,2]}^{({1\over2},{1\over2})}, 
12\rC_{[3,0,1]}^{(0,1)}, 9\rC_{[0,2,2]}^{(1,0)},
26\rC_{[1,2,1]}^{({1\over2},{1\over2})},$\ \hfil &\cr
&&& $9\rC_{[2,2,0]}^{(0,1)}, 36\rC_{[2,1,2]}^{(0,0)},
22\rC_{[1,3,1]}^{(0,0)}  $\ \hfil &\cr
&$15\over 2$\hfil&&$ 2\rC_{ [0,0,3]}^{(2,{1\over2})}, 
17\rC_{ [1,0,2]}^{({3\over2},1)}, 
17\rC_{[2,0,1]}^{(1,{3\over2})}, 2\rC_{[3,0,0]}^{({1\over2},2)}, 
3\rC_{[0,1,1]}^{(2,{3\over2})}, 3\rC_{[1,1,0]}^{({3\over2},2)}, 
6\rC_{[0,2,1]}^{({3\over2},1)}, 6\rC_{[1,2,0]}^{(1,{3\over2})}, 
6\rC_{[0,1,3]}^{({3\over2},0)}, 46\rC_{[1,1,2]}^{(1,{1\over2})},$\ \hfil &\cr
&&& $46\rC_{[2,1,1]}^{({1\over2},1)}, 6\rC_{[3,1,0]}^{(0,{3\over2})}, 
14\rC_{[0,3,1]}^{(1,{1\over2})}, 14\rC_{[1,3,0]}^{({1\over2},1)}, 
36\rC_{[2,0,3]}^{({1\over2},0)}, 36\rC_{[3,0,2]}^{(0,{1\over2})}, 
58\rC_{[1,2,2]}^{({1\over2},0)}, 58\rC_{[2,2,1]}^{(0,{1\over2})}, 
15\rC_{ [0,4,1]}^{({1\over2},0)}, 15\rC_{ [1,4,0]}^{(0,{1\over2})}$\ \hfil &\cr
&8\hfil&&$10\rC_{[0,1,2]}^{(2,1)}, 23\rC_{[1,1,1]}^{({3\over2},{3\over2})},
10\rC_{[2,1,0]}^{(1,2)}, 2\rC_{[0,0,2]}^{({5\over2},{3\over2})}, 
4\rC_{[1,0,1]}^{(2,2)}, 
2\rC_{[2,0,0]}^{({3\over2},{5\over2})}, \rC_{[0,0,4]}^{(2,0)},
24\rC_{[1,0,3]}^{({3\over2},{1\over2})}, 57\rC_{[2,0,2]}^{(1,1)},$\ \hfil &\cr
&&& $24\rC_{[3,0,1]}^{({1\over2}, {3\over2})}, \rC_{[4,0,0]}^{(0,2)},
25\rC_{[0,2,2]}^{({3\over2},{1\over2})}, 56\rC_{[1,2,1]}^{(1,1)}, 
25\rC_{[2,2,0]}^{({1\over2},{3\over2})}, 56\rC_{[1,1,3]}^{(1,0)},
129\rC_{[2,1,2]}^{({1\over2},{1\over2})},
56\rC_{[3,1,1]}^{(0,1)}, 25\rC_{[0,3,2]}^{(1,0)},$\ \hfil &\cr
&&& $64\rC_{[1,3,1]}^{({1\over2},{1\over2})}, 25\rC_{[2,3,0]}^{(0,1)} ,
54\rC_{[3,0,3]}^{(0,0)}, 94\rC_{[2,2,2]}^{(0,0)}, 45\rC_{[1,4,1]}^{(0,0)}$\ \hfil &\cr
&\multispan3\hrulefill &\cr}
}}

\hskip 2cm Table 10
\vskip 1pt

\hbox spread 2cm{\hskip 2cm
\vbox{\tabskip=0pt \offinterlineskip
\halign{&\vrule# &\strut \ \hfil#\  \cr
&\multispan3\hrulefill &\cr
height1pt&\multispan3 &\cr
&\multispan3 \hfil Semi-short ($\eight$,$\eight$) operators \hfil&\cr
height1pt&\multispan3 &\cr
&\multispan3\hrulefill &\cr
& $\ \Delta$ \hfil  && $(0,0)$ primary single trace operators in 
$\S_{\rm sym.}/\S_{\rm sugra}$ 
 \hfil
&\cr
height2pt&\omit&&\omit&\cr
&\multispan3\hrulefill &\cr
&2\hfil&& $\rC_{[0,0,0]}^{(0,0)}$ \ \hfil &\cr
&3\hfil&& $\rC_{[0,1,0]}^{(0,0)}$ \ \hfil &\cr
&4\hfil&& $ \rC_{[0,0,0]}^{(1,1)},\, \rC_{[0,1,0]}^{({1\over2},{1\over2})},
\,\rC_{[0,2,0]}^{(0,0)}$\ \hfil & \cr
&5\hfil&&$\rC_{[0,1,0]}^{(1,1)},
\rC_{[1,0,1]}^{({1\over2},{1\over2})}, \,
\rC_{[0,2,0]}^{({1\over2},{1\over2})},\, \rC_{[0,3,0]}^{(0,0)}$ \ \hfil &  \cr
&${11\over 2}$\hfil&&$ \rC_{ [0,0,1]}^{({3\over2},1)},\,
\rC_{[1,0,0]}^{(1,{3\over2})} $\ \hfil &\cr
&6\hfil&&$
\rC_{[0,0,0]}^{(2,2)}, \, \rC_{[0,1,0]}^{({3\over2},{3\over2})},\,
2\rC_{[0,2,0]}^{(1,1)},\, \rC_{[1,1,1]}^{({1\over2},{1\over2})}, \,
\rC_{[0,3,0]}^{({1\over2},{1\over2})},\, \rC_{[0,4,0]}^{(0,0)}  $\ \hfil &\cr
&${13\over 2}$\hfil&&$\rC_{ [0,1,1]}^{({3\over2},1)},\,
\rC_{ [1,1,0]}^{(1,{3\over2})}$\ \hfil &\cr
&7\hfil&&$ \rC_{[0,2,0]}^{({3\over2},{3\over2})}, \,
2\rC_{[0,1,0]}^{(2,2)}, \, 2\rC_{[1,0,1]}^{({3\over2},{3\over2})}, \,
\rC_{[1,1,1]}^{(1,1)},\, 2\rC_{[0,3,0]}^{(1,1)}$,\ \hfil &\cr
&&& $\rC_{[1,2,1]}^{({1\over2},{1\over2})},\, \rC_{[0,4,0]}^{({1\over2},{1\over2})}, \, 
\rC_{[0,5,0]}^{(0,0)}  $\ \hfil &\cr
&${15\over 2}$\hfil&&$ \rC_{[0,0,1]}^{({5\over2},2)},\,
\rC_{[1,0,0]}^{(2,{5\over2})}, \, \rC_{[0,1,1]}^{(2,{3\over2})}, 
\, \rC_{[1,1,0]}^{({3\over2},2)}, \, \rC_{[0,2,1]}^{({3\over2},1)},\,
\rC_{[1,2,0]}^{(1,{3\over2})}$\ \hfil &\cr 
&8\hfil&&$2\rC_{[1,1,1]}^{({3\over2},{3\over2})}, \,
\rC_{[0,0,0]}^{(3,3)}, \, \rC_{[0,1,0]}^{({5\over2},{5\over2})},
\rC_{[1,0,1]}^{(2,2)},  \, 3\rC_{[0,2,0]}^{(2,2)}$,\ \hfil &\cr
&&&
$\rC_{[2,0,2]}^{(1,1)}, \, 2\rC_{[0,3,0]}^{({3\over2},{3\over2})}, \, 
\rC_{[1,2,1]}^{(1,1)}, \, 2\rC_{[0,4,0]}^{(1,1)}$,\ \hfil &\cr
&&& $ \rC_{[1,3,1]}^{({1\over2},{1\over2})} ,
\rC_{[0,5,0]}^{({1\over2},{1\over2})}, \, \rC_{[0,6,0]}^{(0,0)}  $\ \hfil &\cr
&\multispan3\hrulefill &\cr}
}}

\vfil
\hskip -1cm Table 11
\vskip 1pt

\hbox spread 2cm{\hskip -1cm
\vbox{\tabskip=0pt \offinterlineskip
\halign{&\vrule# &\strut \ \hfil#\  \cr
&\multispan3\hrulefill &\cr
height1pt&\multispan3 &\cr
&\multispan3 \hfil Semi-short ($\eight$,$\eight$) operators \hfil&\cr
height1pt&\multispan3 &\cr
\noalign{\vskip-0pt}
&\multispan3\hrulefill &\cr
\noalign{\vskip-0pt}
& $\ \Delta$  \hfil  && $(0,0)$ primary 
single-trace operators in $\S_{\rm int.}/(\S_{\rm sym.}/\S_{\rm sugra})$ \hfil
&\cr
height2pt&\omit&&\omit&\cr
&\multispan3\hrulefill &\cr
\cr
&4\hfil&&$ \rC_{[0,1,0]}^{({1\over2},{1\over2})}, \rC_{[1,0,1]}^{(0,0)},
\rC_{[0,2,0]}^{(0,0)} $\ \hfil &\cr
&5\hfil&&$\rC_{[0,0,2]}^{(1,0)}, 3\rC_{[1,0,1]}^{({1\over2},{1\over2})},
\rC_{[2,0,0]}^{(0,1)}, \rC_{[0,2,0]}^{({1\over2},{1\over2})},
2\rC_{[1,1,1]}^{(0,0)}, \rC_{[0,3,0]}^{(0,0)}  $\ \hfil &\cr
&${11\over 2}$\hfil&&$\rC_{ [0,0,1]}^{({3\over2},1)},
\rC_{[1,0,0]}^{(1,{3\over2})}, 2\rC_{[0,1,1]}^{(1,{1\over2})},
2\rC_{[1,1,0]}^{({1\over2},1)}, 2\rC_{[0,2,1]}^{({1\over2},0)}, 
2\rC_{[1,2,0]}^{(0,{1\over2})}  $\ \hfil &\cr
&6\hfil&&$2\rC_{[1,0,1]}^{(1,1)}, \rC_{[0,1,0]}^{({3\over2},{3\over2})},
3\rC_{[0,2,0]}^{(1,1)}, \rC_{[0,1,2]}^{(1,0)}, 7\rC_{[1,1,1]}^{({1\over2},{1\over2})},
\rC_{[2,1,0]}^{(0,1)}, 3\rC_{[0,3,0]}^{({1\over2},{1\over2})}, 
3\rC_{[2,0,2]}^{(0,0)}, 3\rC_{[1,2,1]}^{(0,0)}, 2\rC_{[0,4,0]}^{(0,0)}$\ \hfil &\cr
 &${13\over 2}$\hfil&&$3\,\rC_{ [0,1,1]}^{({3\over2},1)},
3\rC_{[1,1,0]}^{(1,{3\over2})}, 6\rC_{[1,0,2]}^{(1,{1\over2})}, 
6\rC_{[2,0,1]}^{({1\over2},1)}, 4\rC_{[0,2,1]}^{(1,{1\over2})},
4\rC_{[1,2,0]}^{({1\over2},1)}, 4\rC_{[1,1,2]}^{({1\over2},0)}, 
4\rC_{[2,1,1]}^{(0,{1\over2})}, 2\rC_{[0,3,1]}^{({1\over2},0)}, 
2\rC_{[1,3,0]}^{(0,{1\over2})}$\ \hfil &\cr
&7\hfil&&$3\rC_{[0,2,0]}^{({3\over2},{3\over2})}, \rC_{[0,1,0]}^{(2,2)},
2\rC_{[0,0,2]}^{(2,1)}, 8\rC_{[1,0,1]}^{({3\over2},{3\over2})},
2\rC_{[2,0,0]}^{(1,2)}, 4\rC_{[0,1,2]}^{({3\over2},{1\over2})}, 
17\rC_{[1,1,1]}^{(1,1)}, 4\rC_{[2,1,0]}^{({1\over2},{3\over2})}, 5\rC_{[0,3,0]}^{(1,1)},
2\rC_{[1,0,3]}^{(1,0)},$\ \hfil &\cr
&&& $ 8\rC_{[2,0,2]}^{({1\over2},{1\over2})}, 2\rC_{[3,0,1]}^{(0,1)},
3\rC_{[0,2,2]}^{(1,0)}, 17\rC_{[1,2,1]}^{({1\over2},{1\over2})},
3\rC_{[2,2,0]}^{(0,1)}, 5\rC_{[0,4,0]}^{({1\over2},{1\over2})}, 
4\rC_{[2,1,2]}^{(0,0)}, 6\rC_{[1,3,1]}^{(0,0)}, 2\rC_{[0,5,0]}^{(0,0)}$\ \hfil &\cr
&${15\over 2}$\hfil&&$10\rC_{[1,0,2]}^{({3\over2},1)}, 
10\rC_{[2,0,1]}^{(1,{3\over2})}, \rC_{[0,0,1]}^{({5\over2},2)}, 
\rC_{[1,0,0]}^{(2,{5\over2})}, 5\rC_{[0,1,1]}^{(2,{3\over2})}, 
5\rC_{[1,1,0]}^{({3\over2},2)}, 11\rC_{[0,2,1]}^{({3\over2},1)},$\ \hfil &\cr
&&& 
$11\rC_{[1,2,0]}^{(1,{3\over2})}, 18\rC_{[1,1,2]}^{(1,{1\over2})}, 
18\rC_{[2,1,1]}^{({1\over2},1)}, 10\rC_{[0,3,1]}^{(1,{1\over2})}, 
10\rC_{[1,3,0]}^{({1\over2},1)}, 6\rC_{[2,0,3]}^{({1\over2},0)}, 
6\rC_{[3,0,2]}^{(0,{1\over2})},$\ \hfil &\cr
&&& $ 8\rC_{[1,2,2]}^{({1\over2},0)}, 8\rC_{[2,2,1]}^{(0,{1\over2})},
4\rC_{ [0,4,1]}^{({1\over2},0)}, 4\rC_{[1,4,0]}^{(0,{1\over2})} $\ \hfil &\cr
&8\hfil&&$8\rC_{[0,1,2]}^{(2,1)},
32\rC_{[1,1,1]}^{({3\over2},{3\over2})}, 8\rC_{[2,1,0]}^{(1,2)},
\rC_{[0,1,0]}^{({5\over2},{5\over2})}, 
2\rC_{[0,0,2]}^{({5\over2},{3\over2})}, 8\rC_{[1,0,1]}^{(2,2)},
2\rC_{[2,0,0]}^{({3\over2},{5\over2})}, 5\rC_{[0,2,0]}^{(2,2)}, 
\rC_{[0,0,4]}^{(2,0)}, 10\rC_{[1,0,3]}^{({3\over2},{1\over2})},$\ \hfil &\cr
&&& $ 37\rC_{[2,0,2]}^{(1,1)}, 10\rC_{[3,0,1]}^{({1\over2},{3\over2})},
\rC_{[4,0,0]}^{(0,2)}, 10\rC_{[0,3,0]}^{({3\over2},{3\over2})},
10\rC_{[0,2,2]}^{({3\over2},{1\over2})}, 44\rC_{[1,2,1]}^{(1,1)}, 
10\rC_{[2,2,0]}^{({1\over2},{3\over2})}, 10\rC_{[0,4,0]}^{(1,1)},
8\rC_{[1,1,3]}^{(1,0)}, $ &\cr
&&& $ 36\rC_{[2,1,2]}^{({1\over2},{1\over2})}, 8\rC_{[3,1,1]}^{(0,1)},
5\rC_{[0,3,2]}^{(1,0)}, 29\rC_{[1,3,1]}^{({1\over2},{1\over2})},
5\rC_{[2,3,0]}^{(0,1)}, 4\rC_{[3,0,3]}^{(0,0)},
7\rC_{[0,5,0]}^{({1\over2},{1\over2})}, 10\rC_{[2,2,2]}^{(0,0)}, 
7\rC_{[1,4,1]}^{(0,0)}, 3\rC_{[0,6,0]}^{(0,0)} $\ \hfil &\cr
&\multispan3\hrulefill &\cr}
}}

\appendix{D}{Use of Characters for Product of Fundamental Representation}

For many cases characters are very convenient for determining the decomposition
of products of representations into irreducible components. We show here how
this can be achieved for the product of two fundamental representations
$\F \otimes \F, \ \F\equiv \B^{{1\over 2},{1\over 2}}_{[0,1,0](0,0)}$. The character 
in this case was obtained in \elemp, it can be written more compactly in the form 
\eqn\elemsim{\eqalign{
\chi_\F (s;\uu;x,\by)={}& 
\chi^{{1\over 2},{1\over 2}}_{(1;0,1,0;0,0)}(s;\uu;x,\by) \cr
= {}& {\ts \sum_{n=0}^4} \D_{1-{1\over 2}n}(s,x,\by) \,
\chi_{(1^{4-n}0^n)}(\uu) \, , \quad \hbox{if} \ \D_{-j} = \oD_j \, , \
j=0,\half,1 \, . \cr}
}
The relations $\oD_{1\over 2} = \D_{- {1\over 2}}, \oD_0 = \D_0$ follow directly
from \charfree\ using \chartwo, otherwise for $\D_{-1}$ this is a just a notational
convention. Nevertheless we may then write the product of two $\D_j$ for
$j=0,\pm \half ,\pm 1$ in the convenient form \fadl
\eqn\DD{
\D_j \D_{j'} = \sum_{q\ge 0} \D_{{1\over 2}q+j_>,{1\over 2}q-j_<} +
\de_{j,1}\de_{j',1} \, \D_1 + \de_{j,-1}\de_{j',-1} \, \oD_1 \, ,
}
for $j_>,j_<= j,j'$ if $j\ge j'$ otherwise $j_>,j_<=j',j$, and also
using \conserved\ with the special cases
\eqnn\speceq
$$\eqalignno{
& \D_{j,0} = \A_{j+2,j,0} \, , \quad \D_{0,j} = \A_{j+2,0,j} \, , \quad
\D_{j,-{1\over 2}} = \A_{j+{5\over 2},j-{1\over 2},0} \, , \quad 
\D_{-{1\over 2},j} = \A_{j+{5\over 2},0,j-{1\over 2}} \, , \cr
& \D_1 + \D_{1,-1} = \oD_1 + \D_{-1,1} = \A_{2,0,0} \, , \quad
\D_{0,-{1\over 2}} = \D_{-{1\over 2},0} = 0 \, , & \speceq \cr}
$$
where $\A_{\De,j,\bj}$ is defined in \long.
Applying \DD\ we easily obtain
\eqn\FF{\eqalign{
\chi_\F^{\, 2} ={}& 
\sum_{q\ge 0} \bigg ( 
\sum_{0\le n<m \le 4}  2 \, \D_{{1\over 2}(q-n)+1, {1\over 2}(q+m)-1} \, 
\chi_{(1^{4-n} 0^n)} \chi_{(1^{4-m}0^m)} \cr 
\noalign{\vskip -4pt}
& \ \qquad {} + \sum_{0\le n \le 4} \D_{{1\over 2}(q-n)+1, {1\over 2}(q+n)-1} \,
\chi_{(1^{4-n} 0^n)} \chi_{(1^{4-n}0^n)} \bigg ) + \D_1 + \oD_1  \, . }
}
Using \oneone\ we have
\eqn\Fexp{\eqalign{
\chi_\F^{\, 2}  - {} & 
\sum_{q\ge 0} \chi^{\C^{1,1}}_{(q;0,0,0;{1\over 2}q-1,{1\over 2}q-1)} \cr
={}& \D_{1,1} - \D_{-{1\over 2},-{1\over 2}} - \D_{0,0} - \D_{{1\over 2},{1\over 2}}
+ \D_1 + \oD_1 \cr
&{}+ \D_{0,0} \, \chi_{(1^2 0^2)}^{\ 2} + \D_{{1\over 2},0} \, \chi_{(1^2 0^2)}
\chi_{(1^3 0)} + \D_{0,{1\over 2}} \, \chi_{(1^2 0^2)} \chi_{(1 0^3)} \cr
&{}+ \big ( \D_{{1\over 2},{1\over 2}} - \D_{0,0}\big  ) \chi_{(1^3 0)} \chi_{(1 0^3)}
+ \D_{{1\over 2},-{1\over 2}}\, \chi_{(1^3 0)}^{\, 2} + \D_{-{1\over 2},{1\over 2}} \,
\chi_{(1 0^3)}^{\, 2} \cr
&{}+ \big ( \D_{1,0} + \D_{0,1} - \D_{{1\over 2},-{1\over 2}} - 
\D_{-{1\over 2},{1\over 2}} \big ) \chi_{(1^2 0^2)}\cr
&{}+ \big ( \D_{1,{1\over 2}} - \D_{{1\over 2},0} \big ) \chi_{(1 0^3)} +
\big ( \D_{{1\over 2},1} - \D_{0,{1\over 2}} \big ) \chi_{(1^3 0)}  \cr
={}& \chi^{{1\over 2},{1\over 2}}_{(2;0,2,0;0,0)} \, , \cr}
}
where this may be shown to be equal to \zerotwozero\  by using \speceq\ and 
the $SU(4)$ decomposition into irreducible representations given by
\eqn\prodfour{\eqalign{
\chi_{(1,1,0,0)}\chi_{(1,1,0,0)} = {}& \chi_{(2,2,0,0)} + \chi_{2,1,1,0)} + 1 \, , \cr
\chi_{(1,0,0,0)}\chi_{(1,1,1,0)} = {}& \chi_{2,1,1,0)} + 1 \, , \cr
\chi_{(1,1,0,0)}\chi_{(1,0,0,0)} = {}& \chi_{(2,1,0,0)} + \chi_{1,1,1,0)} \, , \cr
\chi_{(1,1,0,0)}\chi_{(1,1,1,0)} = {}& \chi_{(2,2,1,0)} + \chi_{1,1,1,0)} \, , \cr
\chi_{(1,0,0,0)}\chi_{(1,0,0,0)} = {}& \chi_{(2,0,0,0)} + \chi_{1,1,0,0)} \, , \cr
\chi_{(1,1,1,0)}\chi_{(1,1,1,0)} = {}& \chi_{(2,2,2,0)} + \chi_{1,1,0,0)} \, . \cr}
}

The result \Fexp\ is in accord with the tensor product \beiserto
\eqn\prodF{
\F \otimes \F = \B^{{1\over 2},{1\over 2}}_{[0,2,0](0,0)} \oplus
\B^{{1\over 4},{1\over 4}}_{[1,0,1](0,0)} \oplus
\bigoplus_{q\ge 1} \, {\frak V}_q \, , \qquad
{\frak V}_q = \C^{1,1}_{[0,0,0]({1\over 2}(q-1),{1\over 2}(q-1))} \, .
}
where we note that ${\frak V}_{0} \simeq \B^{{1\over 4},{1\over 4}}_{[1,0,1](0,0)}$,
as in \idCB.

An interesting extension in this context is to consider the perturbative expansion
of partition functions where $D=D_0+\lambda D_2+ \rO(\lambda^2)$, 
in the notation of \stau. 
The first correction to $\rO(\lambda)$ is given by
\eqn\Dtwo{
\langle D_2 \rangle= \tr_{\F\otimes \F}
\big (D_2 \, x^{2D_0}\, u_1{}^{H_1+H_2+H_3}\,  u_2{}^{H_2+H_3}\, u_3{}^{H_3}\, x^{2J_3}
\, \by^{2{\bar J}_3} \big ) \, ,
}
restricting to $\F\otimes \F$. With the decomposition in \prodF\ and noting that 
$D_2 {\frak V}_q=h(q){\frak V}_q$ where
$h(q)=\sum_{r=1}^q1/r$ are harmonic numbers, we may obtain that from
\oneoness\ and using $\sum_{q\ge 1} h(q)\, x^q = - \ln(1-x)/(1-x)$,
\eqnn\expdtw
$$\eqalignno{
\langle D_2 \rangle ={} & \sum_{q=1}^{\infty}h(q) \,
\chi^{\CC^{1,1}}_{(q;0,0,0;{1\over 2}q-1,{1\over 2}q-1)}(s,\uu,x,\by)\cr
\noalign{\vskip -6pt}
= {}& - \frak{W}^{\S_2}{\overline \frak{W}}{}^{\S_2}
\prod_{i=1}^4 (1+s \ux_i x)(1+s \ux_i{}^{-1}\by) & \expdtw \cr
\noalign{\vskip -8pt}
& \hskip 2cm {}\times {\ln(1-s^2x \by)
\over (1-x^2)(1-\by^2)(1-s^2 x \by)^2(1-s^2 x^{-1}\by)(1-s^2 x \by^{-1})} \, ,
\cr}
$$
where $\frak{W}^{\S_2}{\overline \frak{W}}{}^{\S_2}$ imposes symmetry
under $x\leftrightarrow x^{-1}, \by\leftrightarrow \by^{-1}$.
This can be evaluated for $x,\by=1$ in the form
\eqn\resD{\eqalign{
\langle D_2 \rangle \big |_{x,\by=1} ={}&
- {1\over (1- s^2)^6} \sum_{n,m=0}^4 s^{n+m} \chi_{(1^n 0^{4-n})}(\uu)
\chi_{(1^{4-m} 0^m)}(\uu) \cr
&\quad {}\times \Big ( \big ( 1-n +(n-3)s^2 \big ) \big ( 1-m +(m-3)s^2 \big )
\ln(1-s^2) \cr
\noalign{\vskip -2pt}
& \qquad\qquad {} + \big (1-n-m + (n+m-6) s^2 \big ) s^2 \Big ) \, .
}
}
Similar results to \resD\ were given in \refs{\pendant,\har}
where $\langle D_2 \rangle$ was of prime importance in 
the calculation of one-loop corrections to the Hagedorn temperature,
although the calculations there were restricted to particular sectors.

\listrefs \bye